\documentclass[10pt]{article}

\usepackage{authblk}
\usepackage{natbib}
\usepackage{mathtools}
\usepackage{amsmath}
\usepackage{amsthm}
\usepackage{amssymb}
\usepackage{amsbsy}
\usepackage{amsfonts}
\usepackage{amscd}
\usepackage{mathrsfs}
\usepackage{bm}
\usepackage{breqn}
\usepackage{color, soul}
\usepackage{enumerate}
\usepackage{empheq}
\usepackage{rotating}
\usepackage{ftnxtra}
\usepackage{fnpos}
\usepackage[titletoc,toc,title]{appendix}
\usepackage{euscript}
\usepackage{graphicx}
\usepackage{epsfig}
\usepackage{epstopdf}
\DeclareGraphicsExtensions{.pdf,.png,.jpg,.eps}
\usepackage{pstool}
\usepackage{upgreek}
\usepackage{mathrsfs}

\usepackage{tikz}

\usepackage{psfrag}

\numberwithin{equation}{section}

\theoremstyle{plain}	
\newtheorem{thm}{Theorem}[section]

\newtheorem{prop}[thm]{Proposition}
\newtheorem*{prop*}{Proposition} 
\theoremstyle{definition}	
\newtheorem{definition}[thm]{Definition}
\newtheorem{remark}[thm]{Remark}
\newtheorem{example}[thm]{Example}

\usepackage{graphicx}
\usepackage{lipsum}

\usepackage{caption}
\usepackage{subcaption}

\usepackage{hyperref}
\hypersetup{colorlinks=true, linkcolor=blue}
\hypersetup{colorlinks=true,citecolor=blue}

\DeclareMathAlphabet{\mathpzc}{OT1}{pzc}{m}{it}

\usepackage{amsmath, amsthm, amssymb}

\usepackage{cleveref}

\DeclarePairedDelimiter\abs{\lvert}{\rvert}

\makeatletter
\newsavebox{\@brx}
\newcommand{\llangle}[1][]{\savebox{\@brx}{\(\m@th{#1\langle}\)}%
  \mathopen{\copy\@brx\mkern2mu\kern-0.9\wd\@brx\usebox{\@brx}}}
\newcommand{\rrangle}[1][]{\savebox{\@brx}{\(\m@th{#1\rangle}\)}%
  \mathclose{\copy\@brx\mkern2mu\kern-0.9\wd\@brx\usebox{\@brx}}}%
\let\oldabs\abs
\def\abs{\@ifstar{\oldabs}{\oldabs*}}
\makeatother

\usepackage{accents}

    {\end{bmatrix}}%

\usepackage{enumitem}

\usepackage[utf8]{inputenc}
\usepackage{dutchcal}
\usepackage{multirow}

\usepackage{xcolor}
\usepackage{float}
\usepackage{setspace}

\newcommand{\beq}{\begin{equation}}
\newcommand{\eeq}{\end{equation}}
\newcommand{\beqs}{\begin{eqnarray}}
\newcommand{\eeqs}{\end{eqnarray}}
\newcommand{\beql}{\begin{equation} \label}

\usetikzlibrary{arrows.meta,calc}

\begin{document}

\title{\textbf{Complementary Cauchy Elasticity}}

\author[1]{Arash Yavari\thanks{e-mail: arash.yavari@ce.gatech.edu}}
\affil[1]{\small \textit{School of Civil and Environmental Engineering and The George W. Woodruff School of Mechanical Engineering, Georgia Institute of Technology, Atlanta, GA 30332, USA,}}

\maketitle

\begin{abstract}
We develop a general nonlinear theory of complementary Cauchy elasticity, in which the right Cauchy--Green strain is prescribed as a function of the second Piola--Kirchhoff stress, without assuming constitutive invertibility or the existence of a complementary strain-energy function. Complementary Cauchy elasticity is a distinct explicit constitutive representation within implicit elasticity, with its own stress-space geometry and integrability condition. The strain response defines a complementary stress-work $1$-form whose exterior derivative characterizes the obstruction to complementary hyperelasticity through the failure of major symmetry of the tangent compliance tensor. For general implicit elasticity, we show that the stress-work and complementary stress-work $1$-forms are simultaneously closed on the constitutive manifold, a condition characterized by its Lagrangian structure. We formulate complementary versions of classical work postulates, examine material symmetry, and use the Darboux classification of differential $1$-forms to obtain local canonical representations of anisotropic and isotropic complementary Cauchy elastic solids, including explicit non-hyperelastic isotropic representatives of the complementary Ericksen and Edelen classes.
We study kinematic and stress constraints, showing that stress constraints arise naturally in complementary Cauchy elasticity and provide a framework for extremal composite materials, including pentamode solids. For internally constrained materials, the constitutively relevant stress variable is an equivalence class modulo reactive stresses, rather than an individual total stress. This quotient structure provides the natural formulation of constrained implicit and complementary Cauchy elastic constitutive relations.
We formulate the balance laws and thermodynamics covariantly and establish a fundamental distinction in the derivation of the balance of angular momentum. In Cauchy elasticity, covariance of the energy balance yields the balance of angular momentum only when covariance of the generalized energy functions is additionally imposed. In complementary Cauchy elasticity, the balance of angular momentum follows directly from covariance of the energy balance, without requiring this additional assumption.
Finally, we linearize implicit, Cauchy, and complementary Cauchy elasticity about finitely deformed and prestressed configurations. We show that a nonlinear non-hyperelastic solid may have a hyperelastic linearized response at a given constitutive state, with the obstruction to hyperelasticity appearing only at higher orders. The resulting framework accommodates non-hyperelastic, internally constrained, and stress-constrained elastic responses within implicit elasticity.
\end{abstract}

\begin{description}
\item[Keywords:] Cauchy elasticity, complementary Cauchy elasticity, complementary stress-work $1$-form.
\end{description}

\tableofcontents

\section{Introduction}

The classical theory of nonlinear elasticity is formulated by assuming that stress at a material point is determined by strain at that point. In material description, such a constitutive equation may be written as $\mathbf{S}=\mathbf{S}(\mathbf{C}^\flat,\mathbf{G})$, where $\mathbf{S}$ is the second Piola--Kirchhoff stress and $\mathbf{C}$ is the right Cauchy--Green strain ($\mathbf{G}$ is the material metric and $\mathbf{C}^\flat=\mathbf{G}\mathbf{C}$). Following Truesdell, we refer to this general theory as \emph{Cauchy elasticity} \citep{Cauchy1828,Truesdell1952}. A particular class of Cauchy elastic solids consists of those for which stress derives from a strain energy function. These materials are called Green elastic or, more commonly, hyperelastic solids \citep{Green1838,Green1839,Truesdell1952,Spencer2015}.

There is no reason to expect that, in an elastic response, stress must always be a function of strain. The complementary constitutive description reverses the roles of stress and strain and assumes that strain is determined by stress, namely $\mathbf{C}^\flat=\mathbf{C}^\flat(\mathbf{S},\mathbf{G})$. We refer to this theory as \emph{complementary Cauchy elasticity}. A complementary Cauchy elastic solid need not also be Cauchy elastic: the strain response function may fail to be invertible, either locally or globally. Similarly, a Cauchy elastic solid does not necessarily admit a complementary representation. The two material classes overlap when the stress and strain response functions are local inverses of one another, but neither material class is a subset of the other.

Complementary Cauchy elasticity also belongs naturally to the broader theory of implicit elasticity. An implicit elastic solid is described by a constitutive relation of the form $\boldsymbol{\mathcal{F}}(\mathbf{S},\mathbf{C}^\flat,\mathbf{G})=\mathbf{0}$ \citep{Morgan1966,Rajagopal2003,Rajagopal2007,Rajagopal2011,Bustamante2009,Bustamante2011,Yavari2024ImplicitElasticity}. When this relation can be solved locally for stress, it defines a Cauchy elastic solid. When it can be solved locally for strain, it defines a complementary Cauchy elastic solid. At constitutive states where both representations exist, they are locally equivalent. At states where one of the relevant derivatives becomes singular, one explicit representation may fail while the implicit constitutive equation remains meaningful.

The distinction between a general elastic constitutive equation and one generated by an energy function has been the subject of a long-standing discussion in nonlinear elasticity. Rivlin questioned the physical admissibility of Cauchy elasticity in the discussion of \citet{Truesdell1964} and subsequently expressed similar objections in \citep{Rivlin1986}. \citet{Carroll2009} posed the question explicitly: must elastic materials be hyperelastic? Carroll discussed the non-zero work that may be induced in a closed cycle of deformation by a non-hyperelastic Cauchy elastic solid and regarded this behavior as a serious difficulty for the physical interpretation of such materials. The laws of thermodynamics, however, do not by themselves require every elastic constitutive equation to be hyperelastic \citep{GreenLaws1967,GreenNaghdi1971}. As discussed in our earlier work, the apparent difficulty is particularly relevant to active solids, which may have local access to sources and sinks of energy \citep{YavariGoriely2025Cauchy}. Thus, hyperelasticity should be regarded as an additional constitutive restriction and not as a consequence of elasticity alone.

There is a precise complementary version of Carroll's question: must a material for which strain is determined by stress possess a complementary energy? The answer is again negative. A particular class of complementary Cauchy elastic solids consists of those for which the strain response derives from a complementary strain energy function. We call such solids \emph{complementary Green elastic}, or \emph{complementary hyperelastic}. Therefore, complementary hyperelasticity is a particular case of complementary Cauchy elasticity, just as hyperelasticity is a particular case of Cauchy elasticity.

\subsection{Complementary energy in classical elasticity}

In linear elasticity, complementary energy leads naturally to a stress-based variational principle dual to the displacement-based minimum-energy principle. Extending this construction to finite elasticity is more difficult because the relation between deformation and strain is nonlinear, different stress measures have different work-conjugate strain measures, and convexity depends on the particular conjugate variables used. \citet{Hill1956}, \citet{Trifan1957}, and \citet{Pipes1962} examined the connections among complementary energy, convexity, uniqueness, stability, and constitutive monotonicity. \citet{Levinson1965} showed that a complementary-energy theorem in finite elasticity depends essentially on the choice of work-conjugate variables and formulated the theory in terms of the deformation gradient and the first Piola--Kirchhoff stress. \citet{Hill1967,Hill1968} subsequently emphasized the relation between loss of stability and bifurcation and the dependence of convexity on the chosen strain measure.

\citet{Sewell1969} placed complementary energy within the framework of convex duality and used the Legendre transformation to derive dual variational principles and two-sided bounds. \citet{Zubov1970} formulated a principle of stationary complementary work for finite elasticity in terms of the first Piola--Kirchhoff stress. \citet{Fraeijs1972} proposed a variational formulation in which the rotation appearing in the polar decomposition of the deformation gradient is retained as an independent field. \citet{Koiter1973} examined the conditions under which a minimum complementary-energy principle can be established, while \citet{Ogden1975} clarified the relation among the resulting variational formulations in nonlinear elastostatics. Unlike the classical linear theory, these nonlinear formulations generally lead to stationarity rather than a minimum principle; a minimum requires additional convexity and constitutive invertibility conditions on the branch under consideration.

\citet{Stumpf1978} derived dual extremum principles and error bounds for nonlinear elasticity under appropriate convexity assumptions. \citet{BerdichevskiiMisiura1979}, \citet{Washizu1980}, and \citet{Labisch1982} developed alternative dual and stationary formulations intended to accommodate the nonlinear kinematics and the local structure of the constitutive equations. 
\citet{DeCamposOden1985} established sufficient conditions for the existence of a complementary-energy functional in finite elastostatics and derived a stationary complementary-energy principle that avoids explicit inversion of the nonlinear equilibrium relation.
More recently, \citet{Gao2000} developed canonical dual transformations for nonconvex systems. \citet{KnopsTrimarcoWilliams2003} obtained uniqueness results using conditions formulated in stress space. These developments demonstrate that complementary energy in nonlinear elasticity does not provide a universal minimum principle. Whether a complementary functional exists, and whether its stationary point is a minimum, a maximum, or neither, depends on the chosen work-conjugate variables, the local invertibility of the stress-strain relation, and the convexity properties of the resulting functional.

\citet{ReynoldsBlume1993} considered constitutive equations in which the right Cauchy--Green strain is given as a function of the second Piola--Kirchhoff stress. The significance of this choice can be understood by comparing the objectivity requirements for stress and deformation responses. A constitutive equation $\mathbf{P}=\widehat{\mathbf{P}}(\mathbf{F},\mathbf{G},\mathbf{g})$ is objective provided that $\widehat{\mathbf{P}}(\mathbf{Q}\mathbf{F},\mathbf{G},\mathbf{g})=\mathbf{Q}\widehat{\mathbf{P}}(\mathbf{F},\mathbf{G},\mathbf{g})$ for every proper spatial rotation $\mathbf{Q}$ preserving $\mathbf{g}$. This permits its expression as a material response $\mathbf{S}=\widehat{\mathbf{S}}(\mathbf{C}^\flat,\mathbf{G})$, where $\mathbf{C}^\flat=\mathbf{F}^{*}\mathbf{g}$. Similarly, a constitutive equation $\mathbf{F}=\widehat{\mathbf{F}}(\mathbf{P},\mathbf{G},\mathbf{g})$ is compatible with objectivity provided that $\widehat{\mathbf{F}}(\mathbf{Q}\mathbf{P},\mathbf{G},\mathbf{g})=\mathbf{Q}\widehat{\mathbf{F}}(\mathbf{P},\mathbf{G},\mathbf{g})$. Such a response does not generally exclude rigid rotations, since both $\mathbf{F}$ and $\mathbf{P}$ rotate together. It cannot, however, include a stress-free state as a single-valued objective response. To see this, let $\mathbf{F}_0=\widehat{\mathbf{F}}(\mathbf{0},\mathbf{G},\mathbf{g})$. Objectivity requires $\widehat{\mathbf{F}}(\mathbf{Q}\mathbf{0},\mathbf{G},\mathbf{g})=\mathbf{Q}\mathbf{F}_0$, and hence $\mathbf{F}_0=\mathbf{Q}\mathbf{F}_0$ for every proper spatial rotation $\mathbf{Q}$ preserving $\mathbf{g}$. Since $\mathbf{F}_0$ is invertible, this would imply $\mathbf{Q}=\mathbf{I}$. Thus, a single-valued deformation response would assign a particular deformation gradient to the stress-free state and would not allow the body to undergo an arbitrary rigid rotation. The material strain response has no such difficulty, because $\mathbf{F}_0$ and $\mathbf{Q}\mathbf{F}_0$ determine the same $\mathbf{C}^\flat$. This reveals an asymmetry between the stress and deformation responses: an objective equation $\mathbf{P}=\widehat{\mathbf{P}}(\mathbf{F},\mathbf{G},\mathbf{g})$ can assign zero stress to every rotated deformation gradient $\mathbf{Q}\mathbf{F}_0$, whereas a single-valued objective equation $\mathbf{F}=\widehat{\mathbf{F}}(\mathbf{P},\mathbf{G},\mathbf{g})$ cannot assign all these deformation gradients to the same zero stress. Moreover, eliminating the spatial rotation yields an implicit relation between material strain and stress; expressing this relation as $\mathbf{C}^\flat=\widehat{\mathbf{C}}^\flat(\mathbf{S},\mathbf{G})$ requires an additional local solvability condition. Prescribing this material strain response directly therefore leaves the spatial rotation undetermined and avoids the obstruction associated with recovering the full deformation gradient at a stress-free state.

\citet{Bustamante2020LargeDeformations} considered the inflation and uniform axial extension of an incompressible circular annulus made of a complementary hyperelastic solid. The formulation uses the Hencky strain and Kirchhoff stress as work-conjugate variables. A complementary energy is expressed in terms of the Kirchhoff stress, and differentiation of this energy with respect to the Kirchhoff stress gives the Hencky strain. Incompressibility is imposed using the identity between the trace of the Hencky strain and the logarithm of the Jacobian. Requiring this trace to vanish restricts the admissible complementary energies and leads to a strain response that depends only on the deviatoric part of the stress. Consequently, the response is indifferent to the hydrostatic part of the stress. Because the response is derived from a complementary energy, its tangent compliance has major symmetry; hence, the conditions of incompressibility and pressure indifference coincide (see Example~\ref{ex:pressure-indifference}). Bustamante's treatment is tied specifically to the Hencky-strain--Kirchhoff-stress pair, and the identity used to impose incompressibility and the resulting restriction on the complementary energy depend on this choice of strain measure.

\subsection{Scope of complementary Cauchy elasticity}

The theory introduced in the present paper removes the assumption of complementary hyperelasticity from the outset. The fundamental constitutive object is the strain response function $\mathbf{C}^\flat=\mathbf{C}^\flat(\mathbf{S},\mathbf{G})$. No complementary energy is assumed to exist. This is the complementary analogue of Cauchy's original constitutive viewpoint: the response at a material point is determined by the present value of the independent mechanical variable, without any history dependence, but it is not required to derive from a scalar potential.

The distinction is already apparent in linear elasticity. For a linear stress-determined elastic material, infinitesimal strain is related to stress through the compliance tensor. This relation is generated by a quadratic complementary energy if and only if the compliance tensor has the major symmetry. A compliance tensor without major symmetry nevertheless defines an instantaneous and history-independent elastic response. Care is required, however, when comparing this description with stiffness-based elasticity, because the compliance tensor need not be invertible. Internal constraints provide an important example: the compliance tensor has a nontrivial null space containing the reactive stresses, and stresses that differ by a reactive stress produce the same strain \citep{Pipkin1976,ItskovAksel2002}. Consequently, no stiffness tensor exists as an ordinary inverse on the full stress space. One must instead work with appropriately reduced stress and strain spaces or introduce a suitable generalized inverse. Thus, odd elasticity can be described equivalently from the stiffness and compliance sides only when the relevant constitutive mapping, possibly a reduced mapping, is invertible.

The distinction between Cauchy elasticity and complementary Cauchy elasticity becomes more consequential at finite deformations. Even when $\mathbf{S}=\mathbf{S}(\mathbf{C}^\flat,\mathbf{G})$ is locally invertible at a given state, it need not have a global inverse. Several strains may correspond to the same stress, the inverse may have distinct branches, or it may become singular on subsets of the constitutive domain. Conversely, a well-defined strain response function need not determine stress uniquely. Complementary Cauchy elasticity is therefore not merely Cauchy elasticity with the constitutive equation algebraically inverted. It is a separate explicit representation of elastic response with its own constitutive domain and regularity requirements. 

One motivation for treating stress-determined relations as constitutive laws in their own right comes from small-displacement-gradient approximations and, in particular, from strain-limiting models for brittle materials. If only the strain measure is linearized while the nonlinear dependence on stress is retained, the infinitesimal strain may remain a nonlinear function of the Cauchy stress rather than reducing to the classical linear elastic constitutive equation $\boldsymbol{\sigma}=\mathbb{C}\!\!:\!\!\boldsymbol{\epsilon}$ \citep{Rajagopal2011LimitingStrain}. For the classical idealization of a sharp crack in linear elasticity, both the stress and strain fields generally have an inverse-square-root singularity at the crack tip. This singularity is a feature of the continuum idealization and does not mean that a physical material sustains infinite stress. It does, however, make the predicted strain unbounded as the crack tip is approached, contradicting the small-strain assumption underlying linearized elasticity. In a strain-limiting theory, the infinitesimal strain is instead prescribed as a nonlinear function of the Cauchy stress, $\boldsymbol{\epsilon}=\widehat{\boldsymbol{\epsilon}}(\boldsymbol{\sigma})$, and the constitutive response may be chosen so that the strain remains small, or approaches a prescribed finite bound, as the magnitude of the stress becomes arbitrarily large. For example, a bounded stress-to-strain response may be constructed in the following form
\begin{equation}
	\boldsymbol{\epsilon}=\epsilon_{\mathrm{L}}
	\frac{\boldsymbol{\sigma}}{\sqrt{\sigma_{0}^{2}+\lVert\boldsymbol{\sigma}\rVert^{2}}}\,.
\end{equation}
Here, $\epsilon_{\mathrm{L}}$ is a limiting strain and $\sigma_{0}$ is a stress scale. At stresses small compared with $\sigma_{0}$, the response is approximately linear, whereas the strain remains bounded as the stress magnitude increases. Such models provide an alternative framework for fracture problems traditionally treated within linear elasticity because they can accommodate the large stress concentrations predicted near an idealized crack tip without simultaneously predicting unbounded strains. Stress-determined strain-limiting relations have been applied to anti-plane and plane-strain crack problems \citep{RajagopalWalton2011,Gou2015}. These are particular constitutive models within complementary Cauchy elasticity; bounded strain is not a general property of complementary Cauchy elastic solids. If the stress-to-strain mapping is locally invertible, the same constitutive relation can also be expressed locally in Cauchy elastic form. The distinction becomes substantial when the inverse becomes singular at a limiting strain, fails to exist globally, or has multiple branches.

Another motivation comes from mechanical metamaterials with soft deformation modes. In linear elasticity, such modes are commonly associated with singular stiffness tensors. In complementary
Cauchy elasticity, they can instead be described in terms of stress constraints, which restrict the admissible stress states while leaving certain strain modes constitutively indeterminate. This stress-based formulation extends naturally to finite deformations and provides a framework for modeling metamaterials with soft modes, including pentamode and more general extremal composite materials.

Complementary Cauchy elasticity is particularly natural in stress-controlled experiments and in problems in which stress is the directly prescribed or measured mechanical variable. It may also be useful in modeling active materials whose strain response is rate independent but not generated by a complementary energy. Such materials may exchange energy with internal or external sources and should not be excluded merely because their constitutive response lacks a scalar potential. The experimental characterization of these materials requires loading protocols formulated directly in stress space.

In this paper, we formulate the nonlinear theory of complementary Cauchy elasticity. We first discuss its kinematics, constitutive structure, objectivity, and material symmetry. We distinguish
complementary Cauchy elasticity from complementary hyperelasticity and clarify its relation to conventional Cauchy elasticity and implicit elasticity. We then characterize isotropic and anisotropic
complementary Cauchy elastic solids. We study both kinematic and stress constraints, including their implications for internally constrained solids and extremal composite materials. We formulate
the balance laws and thermodynamics of complementary Cauchy elasticity and examine the role of covariance in deriving the balance of angular momentum. Finally, we linearize the constitutive
equations of implicit, Cauchy, and complementary Cauchy elasticity about finitely deformed and prestressed configurations and investigate the conditions under which the linearized response
is hyperelastic.

\paragraph{Contributions of this paper.} The contributions of this paper can be summarized as follows.
\begin{itemize}[topsep=0pt,noitemsep,leftmargin=10pt]

\item We formulate a nonlinear theory of complementary Cauchy elasticity in which the right Cauchy--Green strain is prescribed as a function of the second Piola--Kirchhoff stress, without assuming constitutive invertibility or the existence of a complementary strain-energy function. We define the complementary stress-work $1$-form and show that its exterior derivative is determined by the major-antisymmetric part of the tangent compliance tensor.

\item We place Cauchy elasticity and complementary Cauchy elasticity within implicit elasticity as the two explicit constitutive representations obtained by local solvability for stress and strain, respectively. On the stress-strain constitutive manifold, we show that the exterior derivatives of the stress-work and complementary stress-work $1$-forms are negatives of one another. Their simultaneous closure is characterized geometrically by the constitutive manifold being Lagrangian.

\item We formulate complementary versions of classical work postulates and use the Darboux classification of differential $1$-forms to obtain six local canonical forms for compressible anisotropic complementary Cauchy elastic solids. For isotropic solids, reduction to the three-dimensional invariant stress space yields the complementary Green, complementary Ericksen, and complementary Edelen classes.

\item We construct explicit isotropic non-hyperelastic responses belonging to the complementary Ericksen and complementary Edelen classes. In these examples, the failure of complementary hyperelasticity is detected both by the exterior derivative of the complementary stress-work $1$-form and by the failure of major symmetry of the tangent compliance tensor.

\item We study material symmetry for complementary Cauchy elastic solids and show that, on a constitutive branch where the Cauchy and complementary Cauchy response functions are locally inverse, the two response functions have the same material symmetry group.

\item We formulate stress constraints as restrictions on the admissible stress states and show that independent stress constraints are associated with constitutively indeterminate strain modes. This provides a stress-based description of soft modes in degenerate elastic materials and, in particular, a natural characterization of pentamode materials in terms of a one-dimensional admissible stress space and five constitutively indeterminate strain modes. We show that the elastic fluid provides a classical nonlinear-elastic realization of the isotropic pentamode idealization.

\item We study internal constraints in complementary Cauchy elasticity. For a kinematically constrained material, the constitutively relevant stress variable is an equivalence class of stresses modulo the reactive stress associated with the constraint, rather than an individual total stress. Satisfaction of the constraint requires the normal to the constraint manifold to lie in the left kernel of the tangent compliance, whereas the fact that a representative stress-to-strain map is constant on each reactive-stress equivalence class requires the same normal to lie in the right kernel. We derive these conditions for incompressibility and the corresponding pressure-equivalence classes, fiber inextensibility and the corresponding tension-equivalence classes, and a general scalar internal constraint. The left- and right-kernel conditions are distinct in general complementary Cauchy elasticity and coincide when the tangent compliance possesses major symmetry.

\item We show that kinematic internal constraints reveal a fundamental asymmetry between Cauchy and complementary Cauchy elasticity. Whereas a kinematic constraint restricts the domain of the strain-to-stress constitutive response in Cauchy elasticity, in complementary Cauchy elasticity the reactive part of the total stress carries no constitutive information about the strain. Consequently, the constrained complementary constitutive relation is naturally defined on equivalence classes of stresses modulo their reactive parts. Pressure-equivalence classes for incompressibility and tension-equivalence classes for fiber inextensibility are particular instances of this general quotient structure. In a representation using individual total stresses, the familiar notions of pressure indifference and tension indifference express the fact that all stresses belonging to the same equivalence class represent one and the same constitutive stress state.

\end{itemize}

\vskip 0.2in
This paper is organized as follows. In \S\ref{Sec:DifferentialGeometricStructure}, we introduce the complementary stress-work $1$-form, examine closed constitutive curves and implicit elastic
relations, formulate complementary versions of classical work postulates, discuss material symmetry, and obtain the Darboux classification of anisotropic and isotropic complementary Cauchy elastic solids. In \S\ref{Sec:InternalConstraints3DElasticity}, we study kinematic and
stress constraints in three-dimensional elasticity, including their implications for extremal composite materials. In \S\ref{Sec:InternalConstraints}, we examine kinematic constraints in implicit and complementary Cauchy elasticity and formulate the constitutive relations in terms of equivalence classes of stresses modulo reactive stresses. In \S\ref{Sec:BalanceLawsComplementaryCauchyElasticity}, we formulate the balance laws and thermodynamics of complementary Cauchy elasticity, with particular attention to the balance of angular momentum. In \S\ref{Sec:LinearizedElasticity}, we linearize the constitutive equations of implicit, Cauchy, and complementary Cauchy elasticity about finitely deformed and prestressed configurations and examine the hyperelasticity of the linearized response. Concluding remarks are given in \S\ref{Sec:Conclusions}.

\section{Geometric Structure of Complementary Cauchy Elasticity}\label{Sec:DifferentialGeometricStructure}

In this section, we develop the geometric and constitutive structure of complementary Cauchy elasticity. After reviewing the kinematics of finite deformations and the relevant stress measures, we introduce the complementary stress-work $1$-form on stress space and characterize complementary hyperelasticity through its exactness. We then place Cauchy and complementary Cauchy elasticity within the broader framework of implicit elasticity by introducing the stress-strain constitutive manifold and examining the stress-work and complementary stress-work $1$-forms intrinsically on this manifold. This formulation clarifies the relations among implicit, Cauchy, complementary Cauchy, hyperelastic, and complementary hyperelastic solids, and gives a geometric characterization of hyperelasticity in terms of the Lagrangian structure of the constitutive manifold. We next formulate complementary versions of classical work principles and study material symmetry for both isotropic and anisotropic responses. The Darboux classification of differential $1$-forms is then used to obtain local canonical representations of complementary Cauchy elastic solids, with separate treatments of compressible anisotropic and isotropic materials and explicit examples of non-hyperelastic isotropic responses.

\subsection{Kinematics of finite deformations}
\label{Sec:KinematicsFiniteDeformation}

Let $\mathcal{B}$ be a three-dimensional body manifold equipped with a material metric $\mathbf{G}$, and let $(\mathcal{S},\mathbf{g})$ be the ambient Riemannian manifold. For most applications, $\mathcal{S}=\mathbb{R}^3$ equipped with the Euclidean metric. A motion is a time-dependent family of embeddings $\varphi_t:\mathcal{B}\rightarrow\mathcal{S}$, $x=\varphi_t(X)=\varphi(X,t),~\forall X\in\mathcal{B}$. The image $\mathcal{C}_t=\varphi_t(\mathcal{B})$ is the current configuration at time $t$. 
The material velocity and material acceleration are spatial vector fields along the motion defined by
\begin{equation}
	\mathbf{V}(X,t)=\frac{\partial\varphi(X,t)}{\partial t}\in T_{\varphi_t(X)}\mathcal{S}\,,\qquad
	\mathbf{A}(X,t)=\frac{D\mathbf{V}(X,t)}{Dt}\in T_{\varphi_t(X)}\mathcal{S}\,,
\end{equation}
where $D/Dt$ denotes covariant differentiation along the material trajectory $t\mapsto\varphi(X,t)$.

The deformation gradient is the tangent map $\mathbf{F}=T\varphi_t:T_X\mathcal{B}\rightarrow T_{\varphi_t(X)}\mathcal{S}$. In local coordinates $\{X^A\}$ on $\mathcal{B}$ and $\{x^a\}$ on $\mathcal{S}$, it has the representation
\begin{equation}
	\mathbf{F}=F^a{}_A\frac{\partial}{\partial x^a}\otimes dX^A\,,\qquad
	F^a{}_A=\frac{\partial\varphi^a}{\partial X^A}\,.
\end{equation}
The dual map of $\mathbf{F}$ is $\mathbf{F}^{\star}:T_{\varphi_t(X)}^*\mathcal{S}\rightarrow T_X^*\mathcal{B}$, and is defined by
\begin{equation}
	\big\langle\mathbf{F}^{\star}\boldsymbol{\alpha},\mathbf{W}\big\rangle
	=\big\langle\boldsymbol{\alpha},\mathbf{F}\mathbf{W}\big\rangle\,,
	\qquad
	\boldsymbol{\alpha}\in T_{\varphi_t(X)}^*\mathcal{S}\,,\quad
	\mathbf{W}\in T_X\mathcal{B}\,,
\end{equation}
where $\langle .,.\rangle$ is the natural pairing of $1$-forms and vectors.
The right Cauchy--Green strain is the pullback of the spatial metric by the motion: $\mathbf{C}^{\flat}=\varphi_t^*\mathbf{g}=\mathbf{F}^{\star}\mathbf{g}\mathbf{F}$.
Thus, $\mathbf{C}^{\flat}(X)$ is a symmetric positive-definite material tensor. In components, $C_{AB}=g_{ab}(\varphi_t(X))F^a{}_AF^b{}_B$. The corresponding mixed material tensor is $\mathbf{C}=\mathbf{G}^{\sharp}\mathbf{C}^{\flat}$. The Jacobian (local volume ratio) is defined as
\begin{equation}
	J=\sqrt{\frac{\det\mathbf{g}}{\det\mathbf{G}}}\,\det\mathbf{F}
	=\sqrt{\det\!\big(\mathbf{G}^{\sharp}\mathbf{C}^{\flat}\big)}\,.
\end{equation}
In particular, a motion is locally isochoric when $J=1$.

For a spatial vector field $\mathbf{W}$ along $\varphi_t$, its covariant derivative in the direction of a material tangent vector $\mathbf{U}\in T_X\mathcal{B}$ is defined using the Levi--Civita connection of $\mathbf{g}$. We denote the resulting two-point tensor by $\nabla^{\varphi_t}\mathbf{W}$. Equivalently, $\nabla^{\varphi_t}$ is the connection induced on the pullback bundle $\varphi_t^*T\mathcal{S}$ by the Levi--Civita connection of $\mathbf{g}$. In components,
\begin{equation}
	W^a{}_{|A}=\frac{\partial W^a}{\partial X^A}
	+\gamma^a{}_{bc}(\varphi_t(X))\,F^b{}_A \,W^c\,,
\end{equation}
where $\gamma^a{}_{bc}$ are the connection coefficients of $\mathbf{g}$. The deformation gradient of an actual motion satisfies the following evolution identity
\begin{equation}
	\frac{D\mathbf{F}}{Dt}=\nabla^{\varphi_t}\mathbf{V}\,.
\end{equation}
Since the Levi--Civita connection is metric-compatible, the material time derivative of the right Cauchy--Green strain is written as
\begin{equation}
	\dot{C}_{AB}=g_{ab}\left[\left(\nabla^{\varphi_t}\mathbf{V}\right)^a{}_A\,F^b{}_B
	+F^a{}_A\left(\nabla^{\varphi_t}\mathbf{V}\right)^b{}_B\right]\,.
\end{equation}
When $\mathcal{S}=\mathbb{R}^3$ and Cartesian coordinates are used, these covariant derivatives reduce to ordinary derivatives. In this case, the deformation gradient may be regarded as an $\mathbb{R}^3$-valued differential $1$-form on $\mathcal{B}$: $\mathbf{F}=F^a{}_A\,dX^A\otimes\mathbf{e}_a$. The local compatibility condition is written as $d\mathbf{F}=\mathbf{0}$ \citep{Yavari2013}. In components, this condition reads
\begin{equation}
	\frac{\partial F^a{}_B}{\partial X^A}-\frac{\partial F^a{}_A}{\partial X^B}=0\,.
\end{equation}
By the Poincar\'e lemma, $d\mathbf{F}=\mathbf{0}$ implies locally that $\mathbf{F}=d\varphi_t=T\varphi_t$. If $\mathcal{B}$ is simply connected, this local condition is also sufficient for global compatibility. If $\mathcal{B}$ is not simply connected, however, the bulk compatibility equations $d\mathbf{F}=\mathbf{0}$ must be supplemented by global compatibility conditions requiring the periods of $\mathbf{F}$ to vanish. More precisely, for a set of closed curves $\{\gamma_\alpha\}$ representing generators of the first homology group $H_1(\mathcal{B})$, one must additionally require that \citep{Yavari2013,Yavari2020AlgebraicTopology}
\begin{equation}
	\int_{\gamma_\alpha}\mathbf{F}=\mathbf{0}\,,\qquad [\gamma_\alpha]\in H_1(\mathcal{B})\,.
\end{equation}
These conditions ensure that the closed $\mathbb{R}^3$-valued $1$-form $\mathbf{F}$ is globally exact and hence that $\mathbf{F}=T\varphi_t$ globally. The commonly used condition $\operatorname{Curl}\mathbf{F}=\mathbf{0}$ is the vector-calculus representation of the local condition $d\mathbf{F}=\mathbf{0}$.

In complementary Cauchy elasticity, $\mathbf{C}^{\flat}$ is prescribed constitutively as a function of the second Piola--Kirchhoff stress. For a given stress field $\mathbf{S}$, the constitutive equation therefore determines a symmetric positive-definite material tensor field $\mathbf{C}^{\flat}(\mathbf{S},\mathbf{G})$. This constitutively prescribed strain must, however, be kinematically realizable: it must coincide with the right Cauchy--Green strain induced by an actual deformation. Consequently, the complementary constitutive equation and the kinematics are coupled through the following relation
\begin{equation}
	\mathbf{F}^{\star}\mathbf{g}\mathbf{F}=\mathbf{C}^{\flat}(\mathbf{S},\mathbf{G})\,.
\end{equation}
This relation is pointwise and specifies the metric that the deformation must induce on the body. An arbitrary stress field $\mathbf{S}$, even when it lies pointwise in the constitutively admissible stress domain, need not produce through $\mathbf{C}^{\flat}(\mathbf{S},\mathbf{G})$ a strain field that is compatible with a deformation. Thus, the constitutive relation alone does not guarantee the existence of a deformation $\varphi_t$ whose deformation gradient $\mathbf{F}=T\varphi_t$ satisfies the above equation. The prescribed strain field must additionally satisfy the appropriate local compatibility conditions and, when the topology of $\mathcal{B}$ is nontrivial, the corresponding global compatibility conditions.

\subsection{Measures of stress}

We next recall the stress measures that will be used in the sequel. There are infinitely many possible stress measures in finite elasticity; here we restrict attention to the standard ones that arise naturally from the current and reference configurations.
Let $da$ denote an area element in the deformed configuration $\mathcal{C}$, and let $\mathbf{n}$ be its $\mathbf{g}$-unit normal, so that $\llangle \mathbf{n},\mathbf{n} \rrangle_{\mathbf{g}}=1$. The Cauchy stress $\boldsymbol{\sigma}$ determines the traction acting on this surface through $\mathbf{t}=\boldsymbol{\sigma}\mathbf{n}^\flat$, where $\mathbf{n}^\flat=\mathbf{g}\mathbf{n}$. The force associated with the area element is therefore $\mathbf{f}=\mathbf{t}\,da$. In components, $t^a=\sigma^{ab}n_b$, with $n_b=g_{bc}n^c$.
Now let us consider the corresponding area element $dA$ in the reference configuration $\mathcal{B}$, with $\mathbf{G}$-unit normal $\mathbf{N}$, i.e., $\llangle \mathbf{N},\mathbf{N} \rrangle_{\mathbf{G}}=1$. The first Piola--Kirchhoff stress $\mathbf{P}$ relates the same force to the reference area element and is defined by $\mathbf{t}\,da=\mathbf{P}\mathbf{N}^\flat\,dA$.
Nanson's formula relates the oriented area elements in the current and reference configurations according to $\mathbf{n}^\flat\,da=J\mathbf{F}^{-\star}\mathbf{N}^\flat\,dA$. Using $\mathbf{t}=\boldsymbol{\sigma}\mathbf{n}^\flat$, it follows that
\begin{equation}
\mathbf{P}=J\,\boldsymbol{\sigma}\mathbf{F}^{-\star}\,.
\end{equation}
Equivalently, in components, $P^{aA}=J\,\sigma^{ab}F^{-A}{}_{b}$.

A stress tensor entirely defined over the reference configuration is obtained by pulling the force back by $\mathbf{F}^{-1}$. The second Piola--Kirchhoff stress $\mathbf{S}$ is defined through $\mathbf{F}^{-1}\mathbf{t}\,da=\mathbf{S}\mathbf{N}^\flat\,dA$, and hence
\begin{equation}
\mathbf{S}=\mathbf{F}^{-1}\mathbf{P}=J\,\mathbf{F}^{-1}\boldsymbol{\sigma}\mathbf{F}^{-\star}\,.
\end{equation}
In components, $S^{AB}=F^{-A}{}_{a}\,P^{aB}=J\,F^{-A}{}_{a}\,\sigma^{ab}\,F^{-B}{}_{b}$. As a consequence of the balance of angular momentum, the second Piola--Kirchhoff stress is symmetric: $\mathbf{S}=\mathbf{S}^{\star}$, or equivalently $S^{AB}=S^{BA}$.
This symmetry will be derived from covariance of the energy balance in \S\ref{Sec:BalanceLawsComplementaryCauchyElasticity}.

\subsection{Complementary stress-work $1$-form}
\label{ComplementaryStressWorkForm}

The second Piola--Kirchhoff stress $\mathbf{S}$ is work-conjugate to the Green strain $\frac12(\mathbf{C}^\flat-\mathbf{G})$, or equivalently to $\frac12\mathbf{C}^\flat$, because the material metric $\mathbf{G}$ is fixed. Consequently, the local stress power per unit reference volume is $\frac12\mathbf{S}:\dot{\mathbf{C}}^\flat$.

In Cauchy elasticity, the strain is the independent constitutive variable and the stress response is written as $\mathbf{S}=\mathbf{S}(\mathbf{C}^\flat,\mathbf{G})$. The associated stress-work $1$-form is defined on strain space by \citep{Edelen1977,Cardin1991,Cardin1995,YavariGoriely2025Cauchy}
\begin{equation}
    \boldsymbol{\Omega}=\frac{1}{2}S^{AB}\,dC_{AB}=\frac{1}{2}\mathbf{S}\!:\!d\mathbf{C}^{\flat}\,.
\end{equation}
In complementary Cauchy elasticity, the constitutive dependence is reversed: strain is determined by stress, so that $\mathbf{C}^\flat=\mathbf{C}^\flat(\mathbf{S},\mathbf{G})$. It is therefore natural to regard stress space as the constitutive space and to define the complementary stress-work $1$-form by pairing the strain response with an infinitesimal increment of stress.

Throughout the constitutive developments of this section, the second Piola--Kirchhoff stress is taken to be symmetric, as required by the balance of angular momentum. This symmetry is not imposed as an additional constitutive assumption; it is derived from covariance of the energy balance in \S\ref{sec:AngularMomentum}.
At a material point $X\in\mathcal{B}$, let $\mathbb{S}_X$ denote the six-dimensional vector space of second Piola-Kirchhoff stress tensors. The material metric $\mathbf{G}$ is a symmetric positive-definite $2$-tensor on $\mathcal{B}$, and hence $\mathbf{G}(X)\in\operatorname{Sym}^{+}(T_X^*\mathcal{B})$. 
A complementary Cauchy elastic constitutive equation is defined by a smooth mapping
\begin{equation}
    \begin{aligned}
        \mathbf{C}^\flat:\mathbb{U}_X\times\operatorname{Sym}^{+}(T_X^*\mathcal{B})
        &\to\operatorname{Sym}^{+}(T_X^*\mathcal{B})\,,\\
        (\mathbf{S},\mathbf{G})
        &\mapsto\mathbf{C}^\flat(\mathbf{S},\mathbf{G})\,,
    \end{aligned}
\end{equation}
where $\mathbb{U}_X\subset\mathbb{S}_X$ is an open constitutive stress domain on which the strain response function is defined and takes values in $\operatorname{Sym}^{+}(T_X^*\mathcal{B})$. We call elements of $\mathbb{U}_X$ constitutively admissible stresses; no satisfaction of equilibrium or boundary conditions is implied.

\begin{definition} \label{Complementary-Stress-Work-Form}
The \emph{complementary stress-work $1$-form} is the differential $1$-form $\boldsymbol{\Lambda}\in\Lambda^1(\mathbb{U}_X)$ defined by
\begin{equation} \label{Complementary-Work-$1$-form}
	\boldsymbol{\Lambda}=\frac{1}{2}C_{AB}(\mathbf{S},\mathbf{G})\,dS^{AB}
	=\frac{1}{2}\mathbf{C}^{\flat}(\mathbf{S},\mathbf{G})\!:\!d\mathbf{S}\,.
\end{equation}
At each $\mathbf{S}\in\mathbb{U}_X$, let $\boldsymbol{\Lambda}_{\mathbf{S}}\in T_{\mathbf{S}}^*\mathbb{U}_X$ denote the value of $\boldsymbol{\Lambda}$ at $\mathbf{S}$.
Its pairing with a tangent vector $\mathbf{H}\in T_{\mathbf{S}}\mathbb{U}_X\simeq\mathbb{S}_X$
is given by
\begin{equation}
	\left\langle\boldsymbol{\Lambda}_{\mathbf{S}},\mathbf{H}\right\rangle
	=\frac{1}{2}C_{AB}(\mathbf{S},\mathbf{G})H^{AB}
	=\frac{1}{2}\mathbf{C}^{\flat}(\mathbf{S},\mathbf{G})\!:\!\mathbf{H}\,.
\end{equation}
\end{definition}

Thus, $\boldsymbol{\Lambda}_{\mathbf{S}}$ is a linear functional on the tangent space to stress space at $\mathbf{S}$. Its coefficients are the components of the strain response evaluated at that stress. The factor $\frac12$ appears because $\mathbf{S}$ is work-conjugate to $\frac12\mathbf{C}^\flat$.
For a fixed material point $X\in\mathcal{B}$, a smooth stress history defines a curve $\gamma_X:I\to\mathbb{U}_X$ in stress space by $\gamma_X(t)=\mathbf{S}(X,t)$. The pullback of $\boldsymbol{\Lambda}$ by this curve is the $1$-form on $I$ given by
\begin{equation}
    \gamma_X^*\boldsymbol{\Lambda}
    =\frac{1}{2}C_{AB}(\mathbf{S}(X,t),\mathbf{G}(X))\,\dot{S}^{AB}(X,t)\,dt
    =\frac{1}{2}\mathbf{C}^{\flat}(\mathbf{S}(X,t),\mathbf{G}(X))\!:\!\dot{\mathbf{S}}(X,t)\,dt\,.
\end{equation}
Accordingly, the complementary work along the stress-history curve $\gamma_X:[t_0,t_1]\to\mathbb{U}_X$ is defined by
\begin{equation}
	\mathscr{W}^c_{\gamma_X}=\int_{\gamma_X}\boldsymbol{\Lambda}
	=\frac{1}{2}\int_{t_0}^{t_1}C_{AB}(\mathbf{S}(X,t),\mathbf{G}(X))\,\dot{S}^{AB}(X,t)\,dt\,.
\end{equation}
The complementary work $\mathscr{W}^c_{\gamma_X}$ is not the mechanical work performed by stress during a deformation. The stress-work $1$-form $\boldsymbol{\Omega}$ pairs stress with an increment of strain and therefore represents mechanical stress-work, whereas the complementary stress-work $1$-form $\boldsymbol{\Lambda}$ pairs strain with an increment of stress. The complementary work $\mathscr{W}^c_{\gamma_X}$ is the integral of $\boldsymbol{\Lambda}$ along a stress path $\gamma_X$ and has no direct interpretation as mechanical work.

For a fixed material metric $\mathbf{G}$, the constitutive relation $\mathbf{C}^\flat=\mathbf{C}^\flat(\mathbf{S},\mathbf{G})$ allows the stress-work and complementary stress-work $1$-forms to be evaluated along the same path in stress space. Note that
\begin{equation}
	d\Big(\frac{1}{2}\mathbf{S}\!:\!\mathbf{C}^{\flat}\Big)
	=\frac{1}{2}\mathbf{S}\!:\!d\mathbf{C}^{\flat}+\frac{1}{2}\mathbf{C}^{\flat}\!:\!d\mathbf{S}
	=\boldsymbol{\Omega}+\boldsymbol{\Lambda}\,.
\end{equation}
Therefore,
\begin{equation}
    \boldsymbol{\Lambda}=d\Big(\frac{1}{2}\mathbf{S}\!:\!\mathbf{C}^{\flat}\Big)-\boldsymbol{\Omega}\,.
\end{equation}
This identity follows directly from the product rule and does not require the existence of either a strain energy or a complementary energy.

If the complementary stress-work $1$-form is exact, there exists locally a scalar function $\zeta$ such that $\boldsymbol{\Lambda}=d\zeta$. In components, this condition is equivalent to
\begin{equation}
    C_{AB}= 2\frac{\partial \zeta}{\partial S^{AB}}\,.
\end{equation}
The scalar function $\zeta$ is the complementary strain-energy function. A complementary Cauchy elastic solid for which such a function exists is called \textit{complementary hyperelastic}. In general, however, $\boldsymbol{\Lambda}$ need not be exact, and a complementary Cauchy elastic solid need not have a complementary strain-energy function. The local obstruction to the existence of a complementary strain-energy function is measured by the exterior derivative of $\boldsymbol{\Lambda}$:
\begin{equation}
    d\boldsymbol{\Lambda} = \frac{1}{2}\frac{\partial C_{AB}}{\partial S^{CD}}\,dS^{CD}\wedge dS^{AB}
    =\frac{1}{4}\left(\frac{\partial C_{AB}}{\partial S^{CD}} - \frac{\partial C_{CD}}{\partial S^{AB}}\right)
    dS^{CD}\wedge dS^{AB}\,.
\end{equation}
Hence, if $\boldsymbol{\Lambda}$ is closed, then
\begin{equation}
    \frac{\partial C_{AB}}{\partial S^{CD}}=\frac{\partial C_{CD}}{\partial S^{AB}}\,.
\end{equation}
This is the major-symmetry condition for the local tangent compliance tensor. On a simply connected stress domain, it is also sufficient for the existence of a complementary strain-energy function.

The complementary stress-work $1$-form is the fundamental constitutive object in complementary Cauchy elasticity. Its Darboux classification (see Appendix~\ref{App:DarbouxClassification}) will provide the local representation of general complementary Cauchy elastic constitutive equations in terms of the Darboux potentials.

\subsection{Closed curves on the implicit stress-strain constitutive manifold} \label{ClosedCurvesConstitutiveManifold}

The relation between curves in strain space and curves in stress space is most naturally understood within the framework of implicit elasticity. Let us fix a material point $X\in\mathcal{B}$ and a material metric $\mathbf{G}$. Let $\mathbb{C}_X=\operatorname{Sym}^{+}(T_X^*\mathcal{B})$ denote the space of symmetric positive-definite right Cauchy--Green strain tensors at $X$. Since $\operatorname{Sym}^{+}(T_X^*\mathcal{B})$ is an open subset of the six-dimensional vector space $\operatorname{Sym}(T_X^*\mathcal{B})$, it is a smooth six-dimensional manifold. Similarly, let $\mathbb{S}_X=\operatorname{Sym}(T_X\mathcal{B})$ denote the six-dimensional vector space, and hence smooth manifold, of symmetric second Piola--Kirchhoff stress tensors at $X$. The product $\mathbb{C}_X\times\mathbb{S}_X$ is therefore a smooth twelve-dimensional manifold whose points are stress-strain pairs $(\mathbf{C}^\flat,\mathbf{S})$.

An implicit elastic constitutive equation may be represented by a smooth map $\boldsymbol{\mathcal{F}}:\mathbb{C}_X\times\mathbb{S}_X\rightarrow\mathbb{V}_X$, where $\mathbb{V}_X$ is a six-dimensional vector space, such that
\begin{equation}
\boldsymbol{\mathcal{F}}(\mathbf{C}^\flat,\mathbf{S},\mathbf{G})=\mathbf{0}\,.
\end{equation}
For fixed $\mathbf{G}$, the set of constitutively admissible stress-strain pairs is the zero level set
\begin{equation} \label{Implicit-Constitutive-Manifold}
	\mathcal{K}_X=\boldsymbol{\mathcal{F}}_{\mathbf{G}}^{-1}(\mathbf{0})
	=\Big\{(\mathbf{C}^\flat,\mathbf{S})\in\mathbb{C}_X\times\mathbb{S}_X:
	\boldsymbol{\mathcal{F}}(\mathbf{C}^\flat,\mathbf{S},\mathbf{G})=\mathbf{0}\Big\}\,,
\end{equation}
where $\boldsymbol{\mathcal{F}}_{\mathbf{G}}(\mathbf{C}^\flat,\mathbf{S})=\boldsymbol{\mathcal{F}}(\mathbf{C}^\flat,\mathbf{S},\mathbf{G})$. Suppose that $\mathbf{0}$ is a regular value of $\boldsymbol{\mathcal{F}}_{\mathbf{G}}$, i.e., that for every $(\mathbf{C}^\flat,\mathbf{S})\in\mathcal{K}_X$ the derivative
\begin{equation}
D\boldsymbol{\mathcal{F}}_{\mathbf{G}}(\mathbf{C}^\flat,\mathbf{S}):T_{\mathbf{C}^\flat}\mathbb{C}_X\times T_{\mathbf{S}}\mathbb{S}_X\longrightarrow\mathbb{V}_X
\end{equation}
has rank six. By the regular level-set theorem, $\mathcal{K}_X$ is then an embedded smooth submanifold of $\mathbb{C}_X\times\mathbb{S}_X$ of codimension six and hence dimension six. Its tangent space at $(\mathbf{C}^\flat,\mathbf{S})\in\mathcal{K}_X$ is defined as
\begin{equation}
T_{(\mathbf{C}^\flat,\mathbf{S})}\mathcal{K}_X=\ker D\boldsymbol{\mathcal{F}}_{\mathbf{G}}(\mathbf{C}^\flat,\mathbf{S})\,.
\end{equation}
Thus, an implicit elastic constitutive equation is geometrically represented, at each material point, by a six-dimensional constitutive submanifold of the twelve-dimensional stress-strain space.
We refer to $\mathcal{K}_X$ as the \emph{implicit stress-strain constitutive manifold}. The natural projections onto the strain and stress spaces are
\begin{equation}
\begin{aligned}
\pi_{\mathcal{C}}:\mathcal{K}_X&\to\mathbb{C}_X\,, &\qquad\qquad \pi_{\mathcal{S}}:\mathcal{K}_X&\to\mathbb{S}_X\,,\\
(\mathbf{C}^\flat,\mathbf{S})&\mapsto\mathbf{C}^\flat\,, &\qquad\qquad (\mathbf{C}^\flat,\mathbf{S})&\mapsto\mathbf{S}\,.
\end{aligned}
\end{equation}

Let us consider first a Cauchy elastic constitutive equation $\mathbf{S}=\mathbf{S}(\mathbf{C}^\flat,\mathbf{G})$. In this case, $\mathcal{K}_X$ is the graph of the stress response function over $\mathbb{C}_X$. Let $\Gamma:[t_1,t_2]\to\mathbb{C}_X$ be a smooth curve in strain space, with $\Gamma(t)=\mathbf{C}^\flat(X,t)$. The curve $\Gamma$ has a unique lift to the constitutive manifold,
\begin{equation}
	\widehat{\Gamma}(t)
	=\big(\Gamma(t),\mathbf{S}(\Gamma(t),\mathbf{G})\big) \in\mathcal{K}_X\,.
\end{equation}
The corresponding stress curve $\gamma:[t_1,t_2]\to\mathbb{S}_X$ is the projection of $\widehat{\Gamma}$ onto stress space. Thus,
\begin{equation}
\gamma(t)=\pi_{\mathcal{S}}\big(\widehat{\Gamma}(t)\big)=\mathbf{S}\big(\Gamma(t),\mathbf{G}\big)\,.
\end{equation}
If $\Gamma$ is closed, then $\Gamma(t_1)=\Gamma(t_2)$. Since the stress response is single-valued,
\begin{equation}
\gamma(t_1)=\mathbf{S}\big(\Gamma(t_1),\mathbf{G}\big)=\mathbf{S}\big(\Gamma(t_2),\mathbf{G}\big)=\gamma(t_2)\,.
\end{equation}
Hence, every closed strain curve of a Cauchy elastic solid induces a closed stress curve.

The lifted curve $\widehat{\Gamma}$ should be distinguished from its
stress projection $\gamma$. Since $\pi_{\mathcal{C}}\circ\widehat{\Gamma}=\Gamma$, if $\Gamma$ is simple, then $\widehat{\Gamma}$ is also simple. Indeed, two distinct points
on the lifted curve cannot coincide unless their projections onto strain space also coincide.
The stress projection $\gamma=\pi_{\mathcal{S}}\circ\widehat{\Gamma}$, however, need not
be simple. If the stress response function is not injective on the image of $\Gamma$, distinct strain states along $\Gamma$ may correspond to the same stress state. Consequently, $\gamma$
may intersect itself, retrace part of its image, or collapse a portion of $\Gamma$ to a single stress state. None of these possibilities affects its closedness.

The converse does not hold in general. A non-closed strain curve may induce a closed stress curve if its distinct endpoint strains correspond to the same stress. Therefore, for a Cauchy elastic constitutive equation, closedness of $\Gamma$ implies closedness of $\gamma$, whereas closedness of $\gamma$ implies closedness of $\Gamma$ only when the stress response is injective on the relevant strain states.

Consider next a complementary Cauchy elastic constitutive equation $\mathbf{C}^\flat=\mathbf{C}^\flat(\mathbf{S},\mathbf{G})$. In this case, $\mathcal{K}_X$ is the graph of the strain response function over the admissible stress domain $\mathbb{U}_X\subset\mathbb{S}_X$. Let $\gamma:[t_1,t_2]\to\mathbb{U}_X$ be a smooth curve in stress space, with $\gamma(t)=\mathbf{S}(X,t)$. The curve $\gamma$ has a unique lift to the constitutive manifold,
\begin{equation}
	\widehat{\gamma}(t)
	=\big(\mathbf{C}^\flat(\gamma(t),\mathbf{G}),\gamma(t)\big) \in\mathcal{K}_X\,.
\end{equation}
The corresponding strain curve $\Gamma:[t_1,t_2]\to\mathbb{C}_X$ is the projection of $\widehat{\gamma}$ onto strain space. Thus,
\begin{equation}
	\Gamma(t)=\pi_{\mathcal{C}}\big(\widehat{\gamma}(t)\big)
	=\mathbf{C}^\flat\big(\gamma(t),\mathbf{G}\big)\,.
\end{equation}
If $\gamma$ is closed, then $\gamma(t_1)=\gamma(t_2)$. Since the strain response is single-valued,
\begin{equation}
	\Gamma(t_1)=\mathbf{C}^\flat\big(\gamma(t_1),\mathbf{G}\big)
	=\mathbf{C}^\flat\big(\gamma(t_2),\mathbf{G}\big)=\Gamma(t_2)\,.
\end{equation}
Hence, every closed stress curve of a complementary Cauchy elastic solid induces a closed strain curve.

If $\gamma$ is simple, then its lift $\widehat{\gamma}$ is also simple, since $\pi_{\mathcal{S}}\circ\widehat{\gamma}=\gamma$. Indeed, two distinct points on the lifted curve cannot coincide
unless their projections onto stress space also coincide. The strain projection $\Gamma=\pi_{\mathcal{C}}\circ\widehat{\gamma}$, however, need not be simple. If the strain response function is not injective on the image of $\gamma$, distinct stress states along $\gamma$
may correspond to the same strain state. Consequently, $\Gamma$ may intersect itself, retrace part of its image, or collapse a portion of $\gamma$ to a single strain state.

The converse does not hold in general. A non-closed stress curve may induce a closed strain curve if its distinct endpoint stresses correspond to the same strain. Therefore, for a complementary Cauchy elastic constitutive equation, closedness of $\gamma$ implies closedness of $\Gamma$, whereas closedness of $\Gamma$ implies closedness of $\gamma$ only when the strain response is injective on the relevant stress states.

The two explicit constitutive representations can be characterized in terms of the projection maps. If $D_{\mathbf{S}}\boldsymbol{\mathcal{F}}$ is invertible at a point of $\mathcal{K}_X$, the implicit-function theorem provides a local Cauchy elastic representation $\mathbf{S}=\mathbf{S}(\mathbf{C}^\flat,\mathbf{G})$. If $D_{\mathbf{C}^\flat}\boldsymbol{\mathcal{F}}$ is invertible, it provides a local complementary Cauchy elastic representation $\mathbf{C}^\flat=\mathbf{C}^\flat(\mathbf{S},\mathbf{G})$. In either case, a neighborhood in $\mathcal{K}_X$ is represented as the graph of the corresponding local response function; we call such a graph a constitutive branch. If both derivatives are invertible, the restrictions of $\pi_{\mathcal{C}}$ and $\pi_{\mathcal{S}}$ to the branch are local diffeomorphisms, and the two local response functions are inverses of one another. Consequently, within such a branch, a strain curve uniquely determines a stress curve and a stress curve uniquely determines a strain curve.

For a general implicit elastic constitutive equation, neither projection need be locally invertible. A curve in strain space or stress space may then admit multiple local lifts to $\mathcal{K}_X$\footnote{Multiple local lifts correspond to distinct constitutive branches over the same projected curve. For example, an implicit constitutive equation may associate two distinct stress states with the same strain state. A strain curve may then have two different lifts to the constitutive manifold, each following a different stress branch while having the same strain projection.}, or it may admit no lift through a prescribed constitutive state. Even when a local lift exists, its continuation along the entire projected curve is not guaranteed: the lift may encounter a point at which the relevant projection becomes singular. Moreover, a lift of a closed projected curve need not itself be closed; after continuation around the projected curve, it may terminate at a different point of the same fiber. Consequently, closedness of a curve in strain space or stress space does not, in general, imply closedness of a corresponding curve on the implicit constitutive manifold.

The stress-work and complementary stress-work $1$-forms have a particularly simple relation on $\mathcal{K}_X$. Let us define the stress-strain pairing by $\mathscr{P}(\mathbf{C}^\flat,\mathbf{S})=\frac{1}{2}\mathbf{S}\!:\!\mathbf{C}^\flat$. The product rule implies that
\begin{equation}
	d\mathscr{P}=\frac{1}{2}\mathbf{S}\!:\!d\mathbf{C}^\flat
	+\frac{1}{2}\mathbf{C}^\flat\!:\!d\mathbf{S}=\boldsymbol{\Omega}+\boldsymbol{\Lambda}\,.
\end{equation}
Let $\kappa:[t_1,t_2]\to\mathcal{K}_X$ be a closed curve on the constitutive manifold. Since the integral of an exact $1$-form along a closed curve vanishes, $\int_{\kappa}d\mathscr{P}=0$, and hence
\begin{equation}
\int_{\kappa}\boldsymbol{\Lambda}=-\int_{\kappa}\boldsymbol{\Omega}\,.
\end{equation}
For a Cauchy elastic constitutive equation, $\kappa=\widehat{\Gamma}$, and the two integrals may be evaluated along the strain curve $\Gamma$ and its induced stress curve $\gamma$, respectively. For a complementary Cauchy elastic constitutive equation, $\kappa=\widehat{\gamma}$, and the same identity relates the complementary stress-work along $\gamma$ to the stress-work along its induced strain curve $\Gamma$. Neither projected curve need be simple; the identity requires only that the lifted curve on $\mathcal{K}_X$ be closed.

The preceding discussion applies to general implicit elastic constitutive relations. For isotropic solids, the local invertibility conditions for the two projections can be made explicit.
At a fixed material point $X\in\mathcal{B}$, let us define the endomorphisms $\mathbf{T}=\mathbf{S}\mathbf{G}$ and $\mathbf{C}=\mathbf{G}^{\sharp}\mathbf{C}^{\flat}$ of $T_X\mathcal{B}$. Since $\mathbf{G}$ and $\mathbf{G}^{\sharp}$ are isomorphisms, specifying $\mathbf{T}$ is equivalent to specifying $\mathbf{S}$, and specifying $\mathbf{C}$ is equivalent to specifying $\mathbf{C}^{\flat}$. In terms of $\mathbf{T}$ and $\mathbf{C}$, the general isotropic implicit elastic constitutive relation may be written in the Rivlin--Ericksen form \citep{RivlinEricksen1955,Morgan1966} as
\begin{equation}\label{Implicit-Constitutive-Equation-Isotropic-Material}
\begin{aligned}
	\boldsymbol{\mathcal{F}}(\mathbf{C}^{\flat},\mathbf{S},\mathbf{G})
	={}&\alpha_0\mathbf{I}+\alpha_1\mathbf{T}+\alpha_2\mathbf{T}^2
	+\alpha_3\mathbf{C}+\alpha_4\mathbf{C}^{-1}
	+\alpha_5\left(\mathbf{T}\mathbf{C}+\mathbf{C}\mathbf{T}\right)
	+\alpha_6\left(\mathbf{T}^2\mathbf{C}+\mathbf{C}\mathbf{T}^2\right)\\
	&+\alpha_7\left(\mathbf{T}\mathbf{C}^{-1}+\mathbf{C}^{-1}\mathbf{T}\right)
	+\alpha_8\left(\mathbf{T}^2\mathbf{C}^{-1}+\mathbf{C}^{-1}\mathbf{T}^2\right)=\mathbf{0}\,,
\end{aligned}
\end{equation}
where $\mathbf{I}$ is the identity endomorphism of $T_X\mathcal{B}$ and $\alpha_i$, $i=0,\ldots,8$, are scalar response functions. A convenient set of ten joint invariants is
\begin{equation}\label{Implicit-Isotropic-Material-Invariants}
\begin{aligned}
	&I_1=\operatorname{tr}\mathbf{T}\,,\quad I_2=\operatorname{tr}\mathbf{T}^2\,,\quad 	
	I_3=\operatorname{tr}\mathbf{T}^3\,,\quad I_4=\operatorname{tr}\mathbf{C}\,,\quad 
	I_5=\frac{1}{2}\left[\left(\operatorname{tr}\mathbf{C}\right)^2
	-\operatorname{tr}\mathbf{C}^2\right]\,,\quad I_6=\det\mathbf{C}\,,\\
	&I_7=\operatorname{tr}\left(\mathbf{T}\mathbf{C}\right)\,,\quad 
	I_8=\operatorname{tr}\left(\mathbf{T}\mathbf{C}^2\right)\,,\quad 
	I_9=\operatorname{tr}\left(\mathbf{T}^2\mathbf{C}\right)\,,\quad 
	I_{10}=\operatorname{tr}\left(\mathbf{T}^2\mathbf{C}^2\right)\,.
\end{aligned}
\end{equation}
Thus, $\alpha_i=\alpha_i(I_1,\ldots,I_{10})$.

The local conditions for the two explicit constitutive representations can be evaluated at an arbitrary constitutively admissible state $(\mathbf{C}^{\flat},\mathbf{S})\in\mathcal{K}_X$. For convenience, let us introduce the following nine tensor-valued functions
\begin{equation}
\begin{aligned}
	&\mathbf{E}_0=\mathbf{I}\,,\qquad \mathbf{E}_1=\mathbf{T}\,,\qquad 
	\mathbf{E}_2=\mathbf{T}^2\,,\qquad \mathbf{E}_3=\mathbf{C}\,,\qquad 
	\mathbf{E}_4=\mathbf{C}^{-1}\,,\qquad 
	\mathbf{E}_5=\mathbf{T}\mathbf{C}+\mathbf{C}\mathbf{T}\,,\\
	& \mathbf{E}_6=\mathbf{T}^2\mathbf{C}+\mathbf{C}\mathbf{T}^2\,,\qquad 
	\mathbf{E}_7=\mathbf{T}\mathbf{C}^{-1}+\mathbf{C}^{-1}\mathbf{T}\,,\qquad 
	\mathbf{E}_8=\mathbf{T}^2\mathbf{C}^{-1}+\mathbf{C}^{-1}\mathbf{T}^2\,.
\end{aligned}
\end{equation}
Then \eqref{Implicit-Constitutive-Equation-Isotropic-Material} takes the compact form $\boldsymbol{\mathcal{F}}=\sum_{i=0}^8\alpha_i\mathbf{E}_i$. We write $\alpha_{i,a}=\partial\alpha_i/\partial I_a$.
To first order in the increment $\mathbf{H}$, the corresponding
increments of the invariants that depend on $\mathbf{T}$ are given by
$\delta I_1=\operatorname{tr}\mathbf{H}$,
$\delta I_2=2\operatorname{tr}(\mathbf{T}\mathbf{H})$,
$\delta I_3=3\operatorname{tr}(\mathbf{T}^2\mathbf{H})$,
$\delta I_7=\operatorname{tr}(\mathbf{C}\mathbf{H})$,
$\delta I_8=\operatorname{tr}(\mathbf{C}^2\mathbf{H})$,
$\delta I_9=\operatorname{tr}[(\mathbf{C}\mathbf{T}+\mathbf{T}\mathbf{C})\mathbf{H}]$,
and
$\delta I_{10}=\operatorname{tr}[(\mathbf{C}^2\mathbf{T}+\mathbf{T}\mathbf{C}^2)\mathbf{H}]$.\footnote{Here we have used the cyclic property of the trace,
\(\operatorname{tr}(\mathbf{A}\mathbf{B})=
\operatorname{tr}(\mathbf{B}\mathbf{A})\).} Accordingly, let us define
\begin{equation}\label{Stress-Variation-Alpha}
\begin{aligned}
	\vartheta_i^{\mathrm{S}}[\mathbf{H}]={}&\alpha_{i,1}\operatorname{tr}\mathbf{H}
	+2\alpha_{i,2}\operatorname{tr}(\mathbf{T}\mathbf{H})
	+3\alpha_{i,3}\operatorname{tr}(\mathbf{T}^2\mathbf{H})
	+\alpha_{i,7}\operatorname{tr}(\mathbf{C}\mathbf{H})
	+\alpha_{i,8}\operatorname{tr}(\mathbf{C}^2\mathbf{H})\\
	&+\alpha_{i,9}\operatorname{tr}\!\left[(\mathbf{C}\mathbf{T}
	+\mathbf{T}\mathbf{C})\mathbf{H}\right]
	+\alpha_{i,10}\operatorname{tr}\!\left[(\mathbf{C}^2\mathbf{T}
	+\mathbf{T}\mathbf{C}^2)\mathbf{H}\right]\,.
\end{aligned}
\end{equation}
Thus, $\vartheta_i^{\mathrm{S}}[\mathbf{H}]=D_{\mathbf{T}}\,\alpha_i[\mathbf{H}]$. Differentiating the tensorial part of \eqref{Implicit-Constitutive-Equation-Isotropic-Material} and collecting the terms gives usthe linear operator $\mathbb{A}_{(\mathbf{T},\mathbf{C})}$ defined by
\begin{equation}\label{General-Isotropic-Stress-Derivative}
\begin{aligned}
\mathbb{A}_{(\mathbf{T},\mathbf{C})}[\mathbf{H}]
\coloneq D_{\mathbf{T}}\boldsymbol{\mathcal{F}}[\mathbf{H}]
={}&\sum_{i=0}^{8}\vartheta_i^{\mathrm{S}}[\mathbf{H}]\,\mathbf{E}_i+\alpha_1\mathbf{H}+\alpha_2(\mathbf{T}\mathbf{H}+\mathbf{H}\mathbf{T})+\alpha_5(\mathbf{H}\mathbf{C}+\mathbf{C}\mathbf{H})\\
&+\alpha_6\!\left[(\mathbf{T}\mathbf{H}+\mathbf{H}\mathbf{T})\mathbf{C}+\mathbf{C}(\mathbf{T}\mathbf{H}+\mathbf{H}\mathbf{T})\right]
+\alpha_7(\mathbf{H}\mathbf{C}^{-1}+\mathbf{C}^{-1}\mathbf{H})\\
&+\alpha_8\!\left[(\mathbf{T}\mathbf{H}+\mathbf{H}\mathbf{T})\mathbf{C}^{-1}+\mathbf{C}^{-1}(\mathbf{T}\mathbf{H}+\mathbf{H}\mathbf{T})\right]\,.
\end{aligned}
\end{equation}
The first term in \eqref{General-Isotropic-Stress-Derivative} accounts for the dependence of the scalar response functions $\alpha_i$ on the invariants, whereas the remaining terms arise from the explicit dependence of the tensorial basis $\mathbf{E}_i$ on $\mathbf{T}$.

\begin{prop}\label{Prop:IsotropicImplicitCauchy}
Let $(\mathbf{C}^{\flat},\mathbf{S})\in\mathcal{K}_X$ be a regular constitutive state of the isotropic implicit relation \eqref{Implicit-Constitutive-Equation-Isotropic-Material}. The constitutive manifold admits a local Cauchy elastic representation $\mathbf{S}=\mathbf{S}(\mathbf{C}^{\flat},\mathbf{G})$ in a neighborhood of this state if and only if the linear operator $\mathbb{A}_{(\mathbf{T},\mathbf{C})}$ defined in \eqref{General-Isotropic-Stress-Derivative} is nonsingular. Equivalently,
\begin{equation}\label{General-Isotropic-Cauchy-Condition}
	\det\mathbb{A}_{(\mathbf{T},\mathbf{C})}\neq0\,.
\end{equation}
\end{prop}
\begin{proof}
The map $\mathbf{S}\mapsto\mathbf{T}=\mathbf{S}\mathbf{G}$ is a linear isomorphism. Therefore, $D_{\mathbf{S}}\boldsymbol{\mathcal{F}}$ is invertible if and only if $D_{\mathbf{T}}\boldsymbol{\mathcal{F}}=\mathbb{A}(\mathbf{T},\mathbf{C})$ is invertible.
Let $(\mathbf{C}^{\flat},\mathbf{S})\in\mathcal{K}_X$ be a regular constitutive state, with $\mathbf{G}$ fixed. A tangent vector to $\mathcal{K}_X$ at $(\mathbf{C}^{\flat},\mathbf{S})$ can be represented
by a smooth curve $t\mapsto(\mathbf{C}^{\flat}(t),\mathbf{S}(t))\in\mathcal{K}_X$ such that $\mathbf{C}^{\flat}(0)=\mathbf{C}^{\flat}$ and $\mathbf{S}(0)=\mathbf{S}$. Write $\mathbf{K}=\dot{\mathbf{C}}^{\flat}(0)$ and $\mathbf{H}=\dot{\mathbf{S}}(0)$. Differentiating $\boldsymbol{\mathcal{F}}(\mathbf{C}^{\flat}(t),\mathbf{S}(t),\mathbf{G})=\mathbf{0}$ at $t=0$ gives us
$D_{\mathbf{C}^{\flat}}\boldsymbol{\mathcal{F}}[\mathbf{K}] +D_{\mathbf{S}}\boldsymbol{\mathcal{F}}[\mathbf{H}]=\mathbf{0}$, where the derivatives are evaluated at $(\mathbf{C}^{\flat},\mathbf{S},\mathbf{G})$.

Along this curve, the strain projection is $\pi_{\mathcal{C}}(\mathbf{C}^{\flat}(t),\mathbf{S}(t))
=\mathbf{C}^{\flat}(t)$. Hence, $D\pi_{\mathcal{C}}(\mathbf{K},\mathbf{H})=\mathbf{K}$. A tangent
vector is therefore in the kernel of $D\pi_{\mathcal{C}}$ precisely when $\mathbf{K}=\mathbf{0}$. For such a tangent vector, the linearized constitutive equation becomes $D_{\mathbf{S}}\boldsymbol{\mathcal{F}}[\mathbf{H}]=\mathbf{0}$. Conversely, every $\mathbf{H}\in\ker D_{\mathbf{S}}\boldsymbol{\mathcal{F}}$ determines a tangent vector $(\mathbf{0},\mathbf{H})$ in
$\ker D\pi_{\mathcal{C}}$. Thus, $D\pi_{\mathcal{C}}$ has trivial kernel if and only if $D_{\mathbf{S}}\boldsymbol{\mathcal{F}}$ has trivial kernel.
Since the constitutive state is regular, $\mathcal{K}_X$ is six-dimensional near $(\mathbf{C}^{\flat},\mathbf{S})$, and $\mathbb{C}_X$ is also six-dimensional. Thus, $D\pi_{\mathcal{C}}:T_{(\mathbf{C}^{\flat},\mathbf{S})}\mathcal{K}_X \to\mathbb{C}_X$ is injective if and only if it is invertible.
Similarly, $D_{\mathbf{S}}\boldsymbol{\mathcal{F}}$ is a linear map between six-dimensional spaces, so it has trivial kernel if and only if it is invertible. Consequently, $D_{\mathbf{S}}\boldsymbol{\mathcal{F}}$ is invertible if and only if $D\pi_{\mathcal{C}}$ is invertible.
By the inverse-function theorem, $D\pi_{\mathcal{C}}$ is invertible if and only if $\pi_{\mathcal{C}}$ is a local diffeomorphism. In that case, the local inverse of $\pi_{\mathcal{C}}$ provides a unique local representation $\mathbf{S}=\mathbf{S}(\mathbf{C}^{\flat},\mathbf{G})$. Therefore, \eqref{General-Isotropic-Cauchy-Condition} is necessary and sufficient.
\end{proof}

We next consider increments of the strain while keeping the stress fixed. Let $\mathbf{K}$ be an arbitrary symmetric increment of $\mathbf{C}$. To first order in $\mathbf{K}$, the corresponding increments of the strain-dependent invariants are given by $\delta I_4=\operatorname{tr}\mathbf{K}$, $\delta I_5=(\operatorname{tr}\mathbf{C})(\operatorname{tr}\mathbf{K})-\operatorname{tr}(\mathbf{C}\mathbf{K})$, $\delta I_6=(\det\mathbf{C})\operatorname{tr}(\mathbf{C}^{-1}\mathbf{K})$, $\delta I_7=\operatorname{tr}(\mathbf{T}\mathbf{K})$, $\delta I_8=\operatorname{tr}[(\mathbf{T}\mathbf{C}+\mathbf{C}\mathbf{T})\mathbf{K}]$, $\delta I_9=\operatorname{tr}(\mathbf{T}^2\mathbf{K})$, and $\delta I_{10}=\operatorname{tr}[(\mathbf{T}^2\mathbf{C}
+\mathbf{C}\mathbf{T}^2)\mathbf{K}]$.

Accordingly, let us define
\begin{equation}\label{Strain-Variation-Alpha}
\begin{aligned}
	\vartheta_i^{\mathrm{C}}[\mathbf{K}]={}&\alpha_{i,4}\operatorname{tr}\mathbf{K}
	+\alpha_{i,5}\!\left[(\operatorname{tr}\mathbf{C})(\operatorname{tr}\mathbf{K})
	-\operatorname{tr}(\mathbf{C}\mathbf{K})\right] 
	+\alpha_{i,6}(\det\mathbf{C})\operatorname{tr}(\mathbf{C}^{-1}\mathbf{K})
	+\alpha_{i,7}\operatorname{tr}(\mathbf{T}\mathbf{K}) \\
	&+\alpha_{i,8}\operatorname{tr}\!\left[(\mathbf{T}\mathbf{C}
	+\mathbf{C}\mathbf{T})\mathbf{K}\right]
	+\alpha_{i,9}\operatorname{tr}(\mathbf{T}^2\mathbf{K}) 
	+\alpha_{i,10}\operatorname{tr}\!\left[(\mathbf{T}^2\mathbf{C}
	+\mathbf{C}\mathbf{T}^2)\mathbf{K}\right]\,.
\end{aligned}
\end{equation}
Thus,
$\vartheta_i^{\mathrm{C}}[\mathbf{K}]=D_{\mathbf{C}}\,\alpha_i[\mathbf{K}]$.
Using $D(\mathbf{C}^{-1})[\mathbf{K}]=-\mathbf{C}^{-1}\mathbf{K}\mathbf{C}^{-1}$,
differentiation of \eqref{Implicit-Constitutive-Equation-Isotropic-Material} with respect to $\mathbf{C}$ gives the linear operator
\begin{equation}\label{General-Isotropic-Strain-Derivative}
\begin{aligned}
	\mathbb{B}_{(\mathbf{T},\mathbf{C})}[\mathbf{K}]\coloneq{}
	&D_{\mathbf{C}}\boldsymbol{\mathcal{F}}[\mathbf{K}] \\
	={}&\sum_{i=0}^{8}\vartheta_i^{\mathrm{C}}[\mathbf{K}]\,\mathbf{E}_i
	+\alpha_3\mathbf{K}-\alpha_4\mathbf{C}^{-1}\mathbf{K}\mathbf{C}^{-1} 
	+\alpha_5(\mathbf{T}\mathbf{K}+\mathbf{K}\mathbf{T})
	+\alpha_6(\mathbf{T}^2\mathbf{K}+\mathbf{K}\mathbf{T}^2) \\
	&-\alpha_7\!\left(\mathbf{T}\mathbf{C}^{-1}\mathbf{K}\mathbf{C}^{-1}
	+\mathbf{C}^{-1}\mathbf{K}\mathbf{C}^{-1}\mathbf{T}\right) 
	-\alpha_8\!\left(\mathbf{T}^2\mathbf{C}^{-1}\mathbf{K}\mathbf{C}^{-1}
	+\mathbf{C}^{-1}\mathbf{K}\mathbf{C}^{-1}\mathbf{T}^2\right)\,.
\end{aligned}
\end{equation}
As in \eqref{General-Isotropic-Stress-Derivative}, the summation in \eqref{General-Isotropic-Strain-Derivative} contains the contribution associated with the invariant dependence of the response functions, while the remaining terms result from direct differentiation of the tensorial basis.

\begin{prop}\label{Prop:IsotropicImplicitComplementaryCauchy}
Let $(\mathbf{C}^{\flat},\mathbf{S})\in\mathcal{K}_X$ be a regular constitutive state of the isotropic implicit relation \eqref{Implicit-Constitutive-Equation-Isotropic-Material}. The constitutive manifold admits a local complementary Cauchy elastic representation $\mathbf{C}^{\flat}=\mathbf{C}^{\flat}(\mathbf{S},\mathbf{G})$ in a neighborhood of this state if and only if the linear operator $\mathbb{B}_{(\mathbf{T},\mathbf{C})}$ defined in \eqref{General-Isotropic-Strain-Derivative} is nonsingular. Equivalently,
\begin{equation}\label{General-Isotropic-Complementary-Cauchy-Condition}
	\det\mathbb{B}_{(\mathbf{T},\mathbf{C})}\neq0\,.
\end{equation}
\end{prop}
\begin{proof}
The map $\mathbf{C}^{\flat}\mapsto\mathbf{C}=\mathbf{G}^{\sharp}\mathbf{C}^{\flat}$ is a linear isomorphism. Therefore, $D_{\mathbf{C}^{\flat}}\boldsymbol{\mathcal{F}}$ is invertible if and
only if $D_{\mathbf{C}}\boldsymbol{\mathcal{F}}=\mathbb{B}(\mathbf{T},\mathbf{C})$
is invertible.  Let $(\mathbf{C}^{\flat},\mathbf{S})\in\mathcal{K}_X$ be a regular constitutive state, with $\mathbf{G}$ fixed. A tangent vector to $\mathcal{K}_X$ at $(\mathbf{C}^{\flat},\mathbf{S})$ can be represented by a smooth curve $t\mapsto(\mathbf{C}^{\flat}(t),\mathbf{S}(t))\in\mathcal{K}_X$ such that $\mathbf{C}^{\flat}(0)=\mathbf{C}^{\flat}$ and $\mathbf{S}(0)=\mathbf{S}$. Write
$\mathbf{K}=\dot{\mathbf{C}}^{\flat}(0)$ and $\mathbf{H}=\dot{\mathbf{S}}(0)$. Differentiating
$\boldsymbol{\mathcal{F}}(\mathbf{C}^{\flat}(t),\mathbf{S}(t),\mathbf{G})=\mathbf{0}$ at $t=0$ gives us $D_{\mathbf{C}^{\flat}}\boldsymbol{\mathcal{F}}[\mathbf{K}]+D_{\mathbf{S}}\boldsymbol{\mathcal{F}}[\mathbf{H}]=\mathbf{0}$, where the derivatives are evaluated at $(\mathbf{C}^{\flat},\mathbf{S},\mathbf{G})$.
Along this curve, the stress projection is $\pi_{\mathcal{S}}(\mathbf{C}^{\flat}(t),\mathbf{S}(t))
=\mathbf{S}(t)$. Consequently, differentiating this identity at $t=0$ yields $D\pi_{\mathcal{S}}(\mathbf{K},\mathbf{H})=\mathbf{H}$. Thus, a tangent vector belongs to the kernel of $D\pi_{\mathcal{S}}$ precisely when $\mathbf{H}=\mathbf{0}$. For such a tangent vector, the linearized constitutive equation reduces to $D_{\mathbf{C}^{\flat}}\boldsymbol{\mathcal{F}}[\mathbf{K}]=\mathbf{0}$. Conversely, every $\mathbf{K}\in\ker D_{\mathbf{C}^{\flat}}\boldsymbol{\mathcal{F}}$ determines a tangent vector $(\mathbf{K},\mathbf{0})$ in $\ker D\pi_{\mathcal{S}}$. Hence, $D\pi_{\mathcal{S}}$ has trivial kernel if and only if $D_{\mathbf{C}^{\flat}}\boldsymbol{\mathcal{F}}$ has trivial kernel.

Because the constitutive state is regular, $\mathcal{K}_X$ is a six-dimensional manifold near $(\mathbf{C}^{\flat},\mathbf{S})$. Stress space $\mathbb{S}_X$ is also six-dimensional. Therefore, the linear map $D\pi_{\mathcal{S}}:T_{(\mathbf{C}^{\flat},\mathbf{S})}\mathcal{K}_X \to\mathbb{S}_X$ is injective if and only if it is invertible. Moreover, $D_{\mathbf{C}^{\flat}}\boldsymbol{\mathcal{F}}$ is a linear map from the six-dimensional strain space to the six-dimensional target space, so it has trivial kernel if and only if it is invertible. It follows that $D_{\mathbf{C}^{\flat}}\boldsymbol{\mathcal{F}}$ is invertible if and only if $D\pi_{\mathcal{S}}$ is invertible.
The inverse-function theorem then implies that $\pi_{\mathcal{S}}$ is locally invertible if and only if $D_{\mathbf{C}^{\flat}}\boldsymbol{\mathcal{F}}$ is invertible. In that case, the inverse of $\pi_{\mathcal{S}}$ provides a unique local representation $\mathbf{C}^{\flat}=\mathbf{C}^{\flat}(\mathbf{S},\mathbf{G})$. Therefore, \eqref{General-Isotropic-Complementary-Cauchy-Condition} is
necessary and sufficient.
\end{proof}

\begin{remark}
Propositions~\ref{Prop:IsotropicImplicitCauchy} and \ref{Prop:IsotropicImplicitComplementaryCauchy} provide the local Cauchy and complementary Cauchy conditions at an arbitrary constitutively admissible state. Although the underlying constitutive relation is isotropic, the operators $\mathbb{A}(\mathbf{T},\mathbf{C})$ and $\mathbb{B}(\mathbf{T},\mathbf{C})$ need not possess the standard two-parameter isotropic representation at an arbitrary state, because the tensors $\mathbf{T}$ and $\mathbf{C}$ characterizing that state may introduce preferred directions. Consequently, there is no general algebraic implication between $\det\mathbb{A}_{(\mathbf{T},\mathbf{C})}\neq0$ and $\det\mathbb{B}_{(\mathbf{T},\mathbf{C})}\neq0$. Either condition may hold while the other fails, providing an explicit isotropic realization of the distinction between Cauchy and complementary Cauchy elasticity.

When both operators are nonsingular, the two local constitutive representations coexist and are inverses of one another. In particular, the local representations $\mathbf{S}=\mathbf{S}(\mathbf{C}^{\flat},\mathbf{G})$ and $\mathbf{C}^{\flat}=\mathbf{C}^{\flat}(\mathbf{S},\mathbf{G})$ are obtained by locally inverting the projections $\pi_{\mathcal{C}}$ and $\pi_{\mathcal{S}}$, respectively. With $\mathbf{G}$ fixed, these two representations describe the same constitutive manifold and are therefore local inverse maps. Since the two local constitutive representations are inverse maps, their differentials satisfy the chain rule $D_{\mathbf{S}}\mathbf{C}^{\flat}D_{\mathbf{C}^{\flat}}\mathbf{S}=\mathbb{I}$. Hence,
\begin{equation}
	D_{\mathbf{S}}\mathbf{C}^{\flat}=\left(D_{\mathbf{C}^{\flat}}\mathbf{S}\right)^{-1}\,.
\end{equation}
Thus, wherever both projections are locally invertible, the tangent stiffness $D_{\mathbf{C}^{\flat}}\mathbf{S}$ and tangent compliance $D_{\mathbf{S}}\mathbf{C}^{\flat}$ are inverse linear maps.

We now express these tangent operators directly in terms of the implicit constitutive equation. For a tangent vector $(\mathbf{K},\mathbf{H})$, with $\mathbf{K}=\dot{\mathbf{C}}^{\flat}(0)$ and $\mathbf{H}=\dot{\mathbf{S}}(0)$, the linearized constitutive relation established above is $D_{\mathbf{C}^{\flat}}\boldsymbol{\mathcal{F}}[\mathbf{K}]+D_{\mathbf{S}}\boldsymbol{\mathcal{F}}[\mathbf{H}]=\mathbf{0}$. If $D_{\mathbf{S}}\boldsymbol{\mathcal{F}}$ is invertible, this relation determines $\mathbf{H}$ uniquely in terms of $\mathbf{K}$:
\begin{equation}
	\mathbf{H}=-\left(D_{\mathbf{S}}\boldsymbol{\mathcal{F}}\right)^{-1}
	D_{\mathbf{C}^{\flat}}\boldsymbol{\mathcal{F}}[\mathbf{K}]\,.
\end{equation}
Since $\mathbf{S}=\mathbf{S}(\mathbf{C}^{\flat},\mathbf{G})$ locally, a curve $\mathbf{C}^{\flat}(t)$ on the constitutive manifold determines $\mathbf{S}(t)=\mathbf{S}(\mathbf{C}^{\flat}(t),\mathbf{G})$. Differentiating this local representation at $t=0$ gives $\mathbf{H}=D_{\mathbf{C}^{\flat}}\mathbf{S}[\mathbf{K}]$. Therefore,
\begin{equation}
	D_{\mathbf{C}^{\flat}}\mathbf{S}=-\left(D_{\mathbf{S}}\boldsymbol{\mathcal{F}}\right)^{-1}	
	D_{\mathbf{C}^{\flat}}\boldsymbol{\mathcal{F}}\,.
\end{equation}
Conversely, if $D_{\mathbf{C}^{\flat}}\boldsymbol{\mathcal{F}}$ is invertible, the same linearized relation determines $\mathbf{K}$ uniquely in terms of $\mathbf{H}$:
\begin{equation}
	\mathbf{K}=-\left(D_{\mathbf{C}^{\flat}}\boldsymbol{\mathcal{F}}\right)^{-1}
	D_{\mathbf{S}}\boldsymbol{\mathcal{F}}[\mathbf{H}]\,.
\end{equation}
Since $\mathbf{C}^{\flat}=\mathbf{C}^{\flat}(\mathbf{S},\mathbf{G})$ locally, a curve $\mathbf{S}(t)$ on the constitutive manifold determines $\mathbf{C}^{\flat}(t)=\mathbf{C}^{\flat}(\mathbf{S}(t),\mathbf{G})$. Differentiating this local representation at $t=0$ gives $\mathbf{K}=D_{\mathbf{S}}\mathbf{C}^{\flat}[\mathbf{H}]$. Therefore,
\begin{equation}
	D_{\mathbf{S}}\mathbf{C}^{\flat}
	=-\left(D_{\mathbf{C}^{\flat}}\boldsymbol{\mathcal{F}}\right)^{-1}
	D_{\mathbf{S}}\boldsymbol{\mathcal{F}}\,.
\end{equation}
These expressions are consistent with the inverse relation above: the tangent stiffness obtained by solving the linearized relation for $\mathbf{H}$ and the tangent compliance obtained by solving it for $\mathbf{K}$ are precisely the differentials of the two local inverse constitutive representations.
\end{remark}

\begin{example}[A Cauchy but not complementary Cauchy elastic solid]
Consider the isotropic subclass of \eqref{Implicit-Constitutive-Equation-Isotropic-Material} obtained by taking $\alpha_1=1$, $\alpha_3=-\mu$, and $\alpha_0=-\phi(I_4)$, where $\mu\neq0$ and $I_4=\operatorname{tr}\mathbf{C}$, with all the remaining response functions vanishing. The implicit constitutive equation reduces to
\begin{equation}\label{Example-Cauchy-Not-Complementary}
	\boldsymbol{\mathcal{F}}(\mathbf{C}^{\flat},\mathbf{S},\mathbf{G})
	=\mathbf{T}-\mu\mathbf{C}-\phi(I_4)\mathbf{I}=\mathbf{0}\,,
\end{equation}
or, equivalently, $\mathbf{T}=\mu\mathbf{C}+\phi(\operatorname{tr}\mathbf{C})\,\mathbf{I}$. Thus, the stress is an explicit single-valued function of the strain. If the reference configuration is required to be stress free, then $\mathbf{C}=\mathbf{I}$ and $\mathbf{T}=\mathbf{0}$ require $\phi(3)=-\mu$.

For an arbitrary symmetric increment $\mathbf{H}$ of $\mathbf{T}$, one has $D_{\mathbf{T}}\boldsymbol{\mathcal{F}}[\mathbf{H}]=\mathbf{H}$. Thus, $D_{\mathbf{T}}\boldsymbol{\mathcal{F}}$ is the identity map and is therefore invertible at every constitutive state, and the material is locally Cauchy elastic everywhere. Consider next an arbitrary symmetric increment $\mathbf{K}$ of $\mathbf{C}$. Since $DI_4[\mathbf{K}]=\operatorname{tr}\mathbf{K}$,
\begin{equation}
	D_{\mathbf{C}}\boldsymbol{\mathcal{F}}[\mathbf{K}]
	=-\mu\mathbf{K}-\phi'(I_4)\operatorname{tr}(\mathbf{K})\,\mathbf{I}
	=-\mu\operatorname{dev}\mathbf{K}-\frac{1}{3}\left[\mu+3\phi'(I_4)\right]
	\operatorname{tr}(\mathbf{K})\,\mathbf{I}\,.
\end{equation}
Here, $\operatorname{dev}\mathbf{K}=\mathbf{K}-\frac{1}{3}\operatorname{tr}(\mathbf{K})\mathbf{I}$ denotes the deviatoric part of $\mathbf{K}$. Therefore, $D_{\mathbf{C}}\boldsymbol{\mathcal{F}}$ has eigenvalue $-\mu$ on the five-dimensional deviatoric subspace and eigenvalue $-[\mu+3\phi'(I_4)]$ on the one-dimensional subspace. Since $\mu\neq0$, it follows that $D_{\mathbf{C}}\boldsymbol{\mathcal{F}}$ is invertible if and only if $\mu+3\phi'(I_4)\neq0$.
Consequently, at any constitutively admissible state for which $\mu+3\phi'(I_4)=0$, $D_{\mathbf{T}}\boldsymbol{\mathcal{F}}$ remains invertible, whereas $D_{\mathbf{C}}\boldsymbol{\mathcal{F}}$ becomes singular. The material is therefore locally Cauchy elastic but not locally complementary Cauchy elastic at such a state. Away from this singular locus, both local constitutive representations exist.
\end{example}

\begin{example}[A complementary Cauchy but not Cauchy elastic solid]
Consider the isotropic subclass of \eqref{Implicit-Constitutive-Equation-Isotropic-Material} obtained by taking $\alpha_3=1$, $\alpha_1=-\beta$, and $\alpha_0=-\psi(I_1)$, where $\beta\neq0$ and $I_1=\operatorname{tr}\mathbf{T}$, with all the remaining response functions vanishing. The implicit constitutive equation reduces to
\begin{equation}\label{Example-Complementary-Not-Cauchy}
	\boldsymbol{\mathcal{F}}(\mathbf{C}^{\flat},\mathbf{S},\mathbf{G})
	=\mathbf{C}-\beta\mathbf{T}-\psi(I_1)\,\mathbf{I}=\mathbf{0}\,,
\end{equation}
or, equivalently, $\mathbf{C}=\beta\mathbf{T}+\psi(\operatorname{tr}\mathbf{T})\,\mathbf{I}$. The admissible stress domain is restricted to those states for which $\beta\mathbf{T}+\psi(\operatorname{tr}\mathbf{T})\,\mathbf{I}$ is positive definite. Thus, the strain is an explicit single-valued function of the stress. If the reference configuration is required to be stress free, then $\mathbf{T}=\mathbf{0}$ and $\mathbf{C}=\mathbf{I}$ require $\psi(0)=1$.

For an arbitrary symmetric increment $\mathbf{K}$ of $\mathbf{C}$, one has $D_{\mathbf{C}}\boldsymbol{\mathcal{F}}[\mathbf{K}]=\mathbf{K}$. Thus, $D_{\mathbf{C}}\boldsymbol{\mathcal{F}}$ is the identity map and is therefore invertible at every constitutive state, and the material is locally complementary Cauchy elastic everywhere in its admissible domain. Consider next an arbitrary symmetric increment $\mathbf{H}$ of $\mathbf{T}$. Since $DI_1[\mathbf{H}]=\operatorname{tr}\mathbf{H}$,
\begin{equation}
	D_{\mathbf{T}}\boldsymbol{\mathcal{F}}[\mathbf{H}]
	=-\beta\mathbf{H}-\psi'(I_1)\operatorname{tr}(\mathbf{H})\mathbf{I}
	=-\beta\operatorname{dev}\mathbf{H}-\frac{1}{3}\left[\beta+3\psi'(I_1)\right]
	\operatorname{tr}(\mathbf{H})\,\mathbf{I}\,.
\end{equation}
Here, $\operatorname{dev}\mathbf{H}=\mathbf{H}-\frac{1}{3}\operatorname{tr}(\mathbf{H})\mathbf{I}$ denotes the deviatoric part of $\mathbf{H}$. Therefore, $D_{\mathbf{T}}\boldsymbol{\mathcal{F}}$ has eigenvalue $-\beta$ on the five-dimensional deviatoric subspace and eigenvalue $-[\beta+3\psi'(I_1)]$ on the one-dimensional subspace. Since $\beta\neq0$, it follows that $D_{\mathbf{T}}\boldsymbol{\mathcal{F}}$ is invertible if and only if $\beta+3\psi'(I_1)\neq0$.
Consequently, at any constitutively admissible state for which $\beta+3\psi'(I_1)=0$, $D_{\mathbf{C}}\boldsymbol{\mathcal{F}}$ remains invertible, whereas $D_{\mathbf{T}}\boldsymbol{\mathcal{F}}$ becomes singular. The material is therefore locally complementary Cauchy elastic but not locally Cauchy elastic at such a state. Away from this singular locus, both local constitutive representations exist.
\end{example}

\begin{example}[A non-Cauchy non-complementary Cauchy implicit elastic solid]
Consider the isotropic subclass of \eqref{Implicit-Constitutive-Equation-Isotropic-Material} obtained by taking $\alpha_1=\alpha_2=1$, $\alpha_3=-1$, and $\alpha_0=-\phi(I_4)$, where $I_4=\operatorname{tr}\mathbf{C}$, with all the remaining response functions vanishing. The implicit constitutive equation reduces to
\begin{equation}\label{Example-Neither-Cauchy-Nor-Complementary}
	\boldsymbol{\mathcal{F}}(\mathbf{C}^{\flat},\mathbf{S},\mathbf{G})
	=\mathbf{T}+\mathbf{T}^2-\mathbf{C}-\phi(I_4)\mathbf{I}=\mathbf{0}\,.
\end{equation}
Consider a constitutive state for which, in a common orthonormal basis, $\mathbf{T}=\operatorname{diag}(0,-1,1)$ and $\mathbf{C}=\operatorname{diag}(1,1,3)$, and suppose that $\phi(5)=-1$ and $\phi'(5)=-\frac{1}{3}$. Since $I_4=5$ and $\mathbf{T}+\mathbf{T}^2=\operatorname{diag}(0,0,2)=\mathbf{C}+\phi(5)\mathbf{I}$, this state satisfies the constitutive equation. For an arbitrary symmetric increment $\mathbf{H}$ in $\mathbf{T}$,
\begin{equation}
	D_{\mathbf{T}}\boldsymbol{\mathcal{F}}[\mathbf{H}]
	=\mathbf{H}+\mathbf{T}\mathbf{H}+\mathbf{H}\mathbf{T}\,.
\end{equation}
Writing $\mathbf{H}=(H_{ij})$, and using $\mathbf{T}=\operatorname{diag}(t_1,t_2,t_3)$ with $(t_1,t_2,t_3)=(0,-1,1)$, the diagonal components of $D_{\mathbf{T}}\boldsymbol{\mathcal{F}}[\mathbf{H}]$ are $H_{11}$, $-H_{22}$, and $3H_{33}$, while the off-diagonal components are $[D_{\mathbf{T}}\boldsymbol{\mathcal{F}}[\mathbf{H}]]_{12}=0$, $[D_{\mathbf{T}}\boldsymbol{\mathcal{F}}[\mathbf{H}]]_{13}=2H_{13}$, and $[D_{\mathbf{T}}\boldsymbol{\mathcal{F}}[\mathbf{H}]]_{23}=2H_{23}$. Hence, the only nonzero symmetric $\mathbf{H}$ for which $D_{\mathbf{T}}\boldsymbol{\mathcal{F}}[\mathbf{H}]=\mathbf{0}$ has $H_{12}=H_{21}$ as its only nonzero components. Thus,
\begin{equation}
	\ker D_{\mathbf{T}}\boldsymbol{\mathcal{F}}
	=\operatorname{span}\left\{\begin{pmatrix}0&1&0\\1&0&0\\0&0&0\end{pmatrix}\right\},\qquad 
	\operatorname{rank}D_{\mathbf{T}}\boldsymbol{\mathcal{F}}=5\,.
\end{equation}
Therefore, $D_{\mathbf{T}}\boldsymbol{\mathcal{F}}$ is singular and the material is not locally Cauchy elastic at this state. For an arbitrary symmetric increment $\mathbf{K}$ in $\mathbf{C}$,
\begin{equation}
	D_{\mathbf{C}}\boldsymbol{\mathcal{F}}[\mathbf{K}]
	=-\mathbf{K}-\phi'(5)\operatorname{tr}(\mathbf{K})\,\mathbf{I}
	=-\mathbf{K}+\frac{1}{3}\operatorname{tr}(\mathbf{K})\,\mathbf{I}=-\operatorname{dev}\mathbf{K}\,.
\end{equation}
Therefore,
\begin{equation}
	\ker D_{\mathbf{C}}\boldsymbol{\mathcal{F}}
	=\operatorname{span}\{\mathbf{I}\},\qquad 
	\operatorname{Im}D_{\mathbf{C}}\boldsymbol{\mathcal{F}}
	=\{\mathbf{A}\in\operatorname{Sym}(T_X\mathcal{B}):\operatorname{tr}\mathbf{A}=0\}\,,	
	\qquad \operatorname{rank}D_{\mathbf{C}}\boldsymbol{\mathcal{F}}=5\,.
\end{equation}
Thus, $D_{\mathbf{C}}\boldsymbol{\mathcal{F}}$ is singular and the material is not locally complementary Cauchy elastic at this state.
To examine the regularity of the full implicit constitutive relation, let $\mathbf{A}=D_{\mathbf{T}}\boldsymbol{\mathcal{F}}[\mathbf{H}]$ and $\mathbf{B}=D_{\mathbf{C}}\boldsymbol{\mathcal{F}}[\mathbf{K}]$. From the component form above, $\mathbf{A}$ has arbitrary diagonal components and arbitrary $A_{13}$ and $A_{23}$ components, while $A_{12}=0$. On the other hand, $\mathbf{B}=-\operatorname{dev}\mathbf{K}$ is an arbitrary deviatoric symmetric tensor and therefore can have an arbitrary $B_{12}$ component. Thus, the $12$ component that is missing from $\mathbf{A}$ can be supplied by $\mathbf{B}$, while the remaining components of $\mathbf{B}$ can be compensated by a suitable choice of $\mathbf{A}$. Hence, every symmetric tensor can be written as $\mathbf{A}+\mathbf{B}$, and
\begin{equation}
	\operatorname{Im}D_{\mathbf{T}}\boldsymbol{\mathcal{F}}
	+\operatorname{Im}D_{\mathbf{C}}\boldsymbol{\mathcal{F}}
	=\operatorname{Sym}(T_X\mathcal{B})\,.
\end{equation}
It follows that $\operatorname{rank}D\boldsymbol{\mathcal{F}}=6$. Therefore, the constitutive set is a regular six-dimensional implicit constitutive manifold in a neighborhood of this state, even though neither the Cauchy nor the complementary Cauchy local representation exists there.
\end{example}

\begin{remark}
In the first two examples, the analysis is carried out at a general constitutively admissible state, since the invertibility of the relevant partial derivative can be characterized directly in terms of the constitutive scalar functions. In the third example, a particular constitutive state is selected because the purpose is to exhibit explicitly that both partial derivatives are singular while their images together span $\operatorname{Sym}(T_X\mathcal B)$, so that the full implicit constitutive equation remains regular.
\end{remark}

\subsection{Stress-work and complementary stress-work $1$-forms on an implicit constitutive manifold}

The stress-work and complementary stress-work $1$-forms can be defined intrinsically on a general implicit constitutive manifold, without assuming that either stress or strain is an explicit function of the other. Let $\iota_X:\mathcal{K}_X\hookrightarrow\mathbb{C}_X\times\mathbb{S}_X$ denote the inclusion of the regular six-dimensional constitutive manifold into the strain-stress product space. On $\mathbb{C}_X\times\mathbb{S}_X$, let us define the $1$-forms $\boldsymbol{\Omega}=\frac{1}{2}\mathbf{S}\!:\!d\mathbf{C}^\flat$ and $\boldsymbol{\Lambda}=\frac{1}{2}\mathbf{C}^\flat\!:\!d\mathbf{S}$. Their restrictions to the constitutive manifold are
\begin{equation}
	\boldsymbol{\Omega}_{\mathcal{K}}=\iota_X^*\boldsymbol{\Omega}\,,\qquad 
	\boldsymbol{\Lambda}_{\mathcal{K}}=\iota_X^*\boldsymbol{\Lambda}\,.
\end{equation}
These restricted forms are defined regardless of whether either projection $\pi_{\mathcal{C}}:\mathcal{K}_X\rightarrow\mathbb{C}_X$ or $\pi_{\mathcal{S}}:\mathcal{K}_X\rightarrow\mathbb{S}_X$ is locally invertible.

Let us define the stress-strain pairing on the product space by $\mathcal{P}(\mathbf{C}^{\flat},\mathbf{S})=\frac{1}{2}\mathbf{S}\!:\!\mathbf{C}^{\flat}$. The product rule gives us $d\mathcal{P}=\boldsymbol{\Omega}+\boldsymbol{\Lambda}$, and hence, after restriction to $\mathcal{K}_X$,
\begin{equation}
	d\mathcal{P}_{\mathcal{K}}=\boldsymbol{\Omega}_{\mathcal{K}}
	+\boldsymbol{\Lambda}_{\mathcal{K}}\,,\qquad 
	\mathcal{P}_{\mathcal{K}}=\iota_X^*\mathcal{P}\,.
\end{equation}
Taking the exterior derivative yields $d\boldsymbol{\Omega}_{\mathcal{K}}=-d\boldsymbol{\Lambda}_{\mathcal{K}}$. Thus, the stress-work $1$-form is closed on $\mathcal{K}_X$ if and only if the complementary stress-work $1$-form is closed there. This conclusion does not require an explicit Cauchy or complementary Cauchy representation.

Let $\kappa(t)=\big(\mathbf{C}^{\flat}(t),\mathbf{S}(t)\big)\in\mathcal{K}_X$, $t\in[t_1,t_2]$, be a smooth constitutive curve. We call $\kappa$ a \emph{closed constitutive curve}, or equivalently a \emph{cyclic constitutive process}, if $\kappa(t_1)=\kappa(t_2)$, which is equivalent to requiring $\mathbf{C}^{\flat}(t_1)=\mathbf{C}^{\flat}(t_2)$ and $\mathbf{S}(t_1)=\mathbf{S}(t_2)$. For a general implicit elastic relation, this condition is stronger than the closedness of one projected curve alone. For a Cauchy elastic solid, the single-valuedness of the stress response implies that every closed strain curve lifts to a closed constitutive curve. Similarly, for a complementary Cauchy elastic solid, every closed stress curve lifts to a closed constitutive curve because the strain response is single-valued.
Since the integral of an exact $1$-form along a closed curve vanishes, one has
\begin{equation}
	\int_{\kappa}\boldsymbol{\Lambda}_{\mathcal{K}}
	=-\int_{\kappa}\boldsymbol{\Omega}_{\mathcal{K}}\,.
\end{equation}
Thus, along every cyclic constitutive process, the accumulated complementary work is the negative of the accumulated stress work. This identity is intrinsic to $\mathcal{K}_X$ and does not require either projection to define an explicit constitutive response.

Suppose that $\boldsymbol{\Omega}_{\mathcal{K}}$ is locally exact, so that there exists a scalar function $W_{\mathcal{K}}:\mathcal{K}_X\rightarrow\mathbb{R}$ such that $\boldsymbol{\Omega}_{\mathcal{K}}=dW_{\mathcal{K}}$. It then follows that $\boldsymbol{\Lambda}_{\mathcal{K}}$ is locally exact:
\begin{equation}
	\boldsymbol{\Lambda}_{\mathcal{K}}=d\big(\mathcal{P}_{\mathcal{K}}-W_{\mathcal{K}}\big)\,.
\end{equation}
Conversely, the local exactness of $\boldsymbol{\Lambda}_{\mathcal{K}}$ implies the local exactness of $\boldsymbol{\Omega}_{\mathcal{K}}$. Hence, local exactness is a property of the implicit constitutive manifold rather than of a particular explicit representation.

The function $W_{\mathcal{K}}$ is defined on the constitutive manifold and need not descend to a function of strain alone. If $\pi_{\mathcal{C}}$ is locally invertible, then $W=W_{\mathcal{K}}\circ(\pi_{\mathcal{C}}\big|_{\mathcal{K}_X})^{-1}$ is a local strain-energy function. Similarly, if $\pi_{\mathcal{S}}$ is locally invertible, then $\zeta=\big(\mathcal{P}_{\mathcal{K}}-W_{\mathcal{K}}\big)\circ(\pi_{\mathcal{S}}\big|_{\mathcal{K}_X})^{-1}$ is a local complementary strain-energy function. At a constitutive state where neither projection is locally invertible, the intrinsic potentials on $\mathcal{K}_X$ may still exist, although neither can be represented as a function on strain space or stress space alone.

The geometric obstruction to local exactness is obtained by introducing the symplectic $2$-form
\begin{equation}
	\boldsymbol{\omega}=\frac{1}{2}dS^{AB}\wedge dC_{AB}
	=d\boldsymbol{\Omega}=-d\boldsymbol{\Lambda}\,,
\end{equation}
on $\mathbb{C}_X\times\mathbb{S}_X$. Its restriction to the constitutive manifold is $\iota_X^*\boldsymbol{\omega}=d\boldsymbol{\Omega}_{\mathcal{K}}=-d\boldsymbol{\Lambda}_{\mathcal{K}}$. Because $\mathcal{K}_X$ has half the dimension of $\mathbb{C}_X\times\mathbb{S}_X$, the condition $\iota_X^*\boldsymbol{\omega}=0$ is equivalent to $\mathcal{K}_X$ being a Lagrangian submanifold\footnote{Let $(\mathcal{M},\boldsymbol{\omega})$ be a $2n$-dimensional symplectic manifold. An $n$-dimensional submanifold $\mathcal{L}\subset\mathcal{M}$ is called \emph{Lagrangian} if the restriction of $\boldsymbol{\omega}$ to $\mathcal{L}$ vanishes.} of the stress-strain product space. Thus, a regular implicit elastic constitutive relation admits local intrinsic work potentials if and only if its constitutive manifold is Lagrangian. The existence of global intrinsic work potentials requires, in addition, that the periods of the corresponding $1$-forms vanish over all closed curves in $\mathcal{K}_X$.

\subsection{Relations among implicit, Cauchy, complementary Cauchy, and hyperelastic solids}

At a material point $X\in\mathcal{B}$, an implicit elastic constitutive relation is represented by a six-dimensional constitutive manifold $\mathcal{K}_X \subset \operatorname{Sym}^{+}(T_X^*\mathcal{B})\times\mathbb{S}_X$, where $\operatorname{Sym}^{+}(T_X^*\mathcal{B})$ is the strain space and $\mathbb{S}_X$ is the stress space. A constitutive branch is Cauchy elastic when $\pi_{\mathcal{C}}$ restricts to a local diffeomorphism on that branch. The constitutive manifold can then be represented locally as the graph of a stress response $\mathbf{S}=\widehat{\mathbf{S}}(X,\mathbf{C}^{\flat},\mathbf{G})$. Similarly, a constitutive branch is complementary Cauchy elastic when $\pi_{\mathcal{S}}$ restricts to a local diffeomorphism, in which case it can be represented locally as the graph of a strain response $\mathbf{C}^{\flat}=\widehat{\mathbf{C}}^{\flat}(X,\mathbf{S},\mathbf{G})$. These two regions may overlap. On their intersection, both projections are locally invertible, and the two explicit response functions are local inverses of one another. Neither explicit representation need exist throughout the entire implicit constitutive manifold.
In this sense, Cauchy elasticity and complementary Cauchy elasticity are the two distinguished explicit subclasses of implicit elasticity, corresponding to local solvability of the constitutive relation for stress and strain, respectively.

To describe the corresponding classes of elastic solids, let $\mathfrak{E}_{\mathrm{I}}$ denote the class of implicit elastic solids, $\mathfrak{E}_{\mathrm{C}}$ the class of Cauchy elastic solids, and $\mathfrak{E}_{\mathrm{CC}}$ the class of complementary Cauchy elastic solids. Their relations are
\begin{equation}
	\mathfrak{E}_{\mathrm{C}}\subset\mathfrak{E}_{\mathrm{I}}\,,\qquad
	\mathfrak{E}_{\mathrm{CC}}\subset\mathfrak{E}_{\mathrm{I}}\,,\qquad
	\mathfrak{E}_{\mathrm{C}}\cap\mathfrak{E}_{\mathrm{CC}}\neq\varnothing\,.
\end{equation}
In general, neither $\mathfrak{E}_{\mathrm{C}}$ nor $\mathfrak{E}_{\mathrm{CC}}$ contains the other.

\begin{figure}[H]
\centering
\resizebox{0.8\textwidth}{!}{%
\begin{tikzpicture}[font=\small]
\draw[black,thick,rounded corners=4mm,fill=gray!6]
(-7,-3.8) rectangle (7,3.8);
\node[anchor=west] at (-6.85,4.15)
{$\mathfrak{E}_{\mathrm I}$: implicit elasticity};
\begin{scope}[yshift=-1mm]
\draw[blue!80!black,thick,fill=blue!15,fill opacity=0.55]
(-2.25,0) ellipse (4.2 and 2.8);
\draw[orange!85!black,thick,fill=orange!18,fill opacity=0.55]
(2.25,0) ellipse (4.2 and 2.8);
\node at (-4.15,3.15)
{$\mathfrak{E}_{\mathrm C}$: Cauchy elasticity};
\node at (3.45,3.15)
{$\mathfrak{E}_{\mathrm{CC}}$: complementary Cauchy elasticity};
\begin{scope}
\clip (-2.25,0) ellipse (4.2 and 2.8);
\fill[cyan!35,rounded corners=4mm]
(-5,-0.35) rectangle (5,1.25);
\draw[green!60!black,thick,rounded corners=4mm]
(-5,-0.35) rectangle (5,1.25);
\end{scope}
\begin{scope}
\clip (2.25,0) ellipse (4.2 and 2.8);
\fill[purple!30,fill opacity=0.70,rounded corners=4mm]
(-5,-0.35) rectangle (5,1.25);
\draw[purple!70!black,thick,rounded corners=4mm]
(-5,-0.35) rectangle (5,1.25);
\end{scope}
\begin{scope}
\clip (-2.25,0) ellipse (4.2 and 2.8);
\clip (2.25,0) ellipse (4.2 and 2.8);
\fill[orange!40,rounded corners=4mm]
(-5,-0.35) rectangle (5,1.25);
\end{scope}
\node at (-3.25,1.65)
{$\mathfrak{E}_{\mathrm H}$};
\node at (3.25,1.65)
{$\mathfrak{E}_{\mathrm{CH}}$};
\node[align=center,font=\footnotesize] at (0,0.45)
{$\mathfrak{E}_{\mathrm H}\cap\mathfrak{E}_{\mathrm{CC}}
=\mathfrak{E}_{\mathrm{CH}}\cap\mathfrak{E}_{\mathrm C}$};
\node at (-4,-1.65)
{$\mathfrak{E}_{\mathrm C}\setminus\mathfrak{E}_{\mathrm H}$};
\node at (4,-1.65)
{$\mathfrak{E}_{\mathrm{CC}}\setminus\mathfrak{E}_{\mathrm{CH}}$};
\end{scope}
\end{tikzpicture}%
}
\caption{Relations among the classes of elastic solids. The class $\mathfrak{E}_{\mathrm H}$ consists of the blue region together with the orange common region, whereas $\mathfrak{E}_{\mathrm{CH}}$ consists of the pink region together with the orange common region. Thus, within $\mathfrak{E}_{\mathrm C}\cap\mathfrak{E}_{\mathrm{CC}}$, the two hyperelastic classes coincide:
$\mathfrak{E}_{\mathrm H}\cap\mathfrak{E}_{\mathrm{CC}}
=\mathfrak{E}_{\mathrm{CH}}\cap\mathfrak{E}_{\mathrm C}$.}
\label{Fig:Elastic-Constitutive-Classes}
\end{figure}
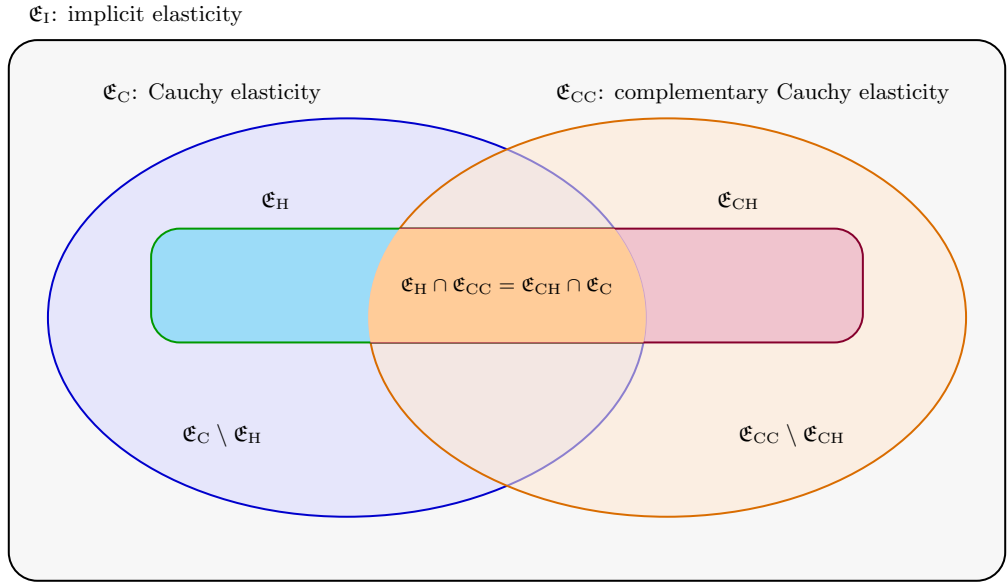

Let $\mathfrak{E}_{\mathrm{H}}\subset\mathfrak{E}_{\mathrm{C}}$ denote the class of hyperelastic solids. For such a solid, the stress-work $1$-form is locally exact, i.e., $\boldsymbol{\Omega}=dW$. Likewise, let $\mathfrak{E}_{\mathrm{CH}}\subset\mathfrak{E}_{\mathrm{CC}}$ denote the class of complementary hyperelastic solids, characterized locally by $\boldsymbol{\Lambda}=d\zeta$. Thus,
\begin{equation}
	\mathfrak{E}_{\mathrm{H}}\subset\mathfrak{E}_{\mathrm{C}}\,,\qquad
	\mathfrak{E}_{\mathrm{CH}}\subset\mathfrak{E}_{\mathrm{CC}}\,.
\end{equation}
A hyperelastic solid need not be complementary Cauchy elastic. Although its stress derives from a strain-energy function, the stress response may fail to be locally invertible, so that no single-valued complementary strain response exists. Similarly, a complementary hyperelastic solid need not admit a Cauchy elastic representation. Therefore,
\begin{equation}
	\mathfrak{E}_{\mathrm{H}}\not\subset\mathfrak{E}_{\mathrm{CH}}\,,\qquad
	\mathfrak{E}_{\mathrm{CH}}\not\subset\mathfrak{E}_{\mathrm{H}}\,.
\end{equation}
The two hyperelastic classes nevertheless have a nonempty intersection:
\begin{equation}
	\mathfrak{E}_{\mathrm{H}}\cap\mathfrak{E}_{\mathrm{CH}}\neq\varnothing\,,\qquad
	\mathfrak{E}_{\mathrm{H}}\cap\mathfrak{E}_{\mathrm{CH}}\subset
	\mathfrak{E}_{\mathrm{C}}\cap\mathfrak{E}_{\mathrm{CC}}\,.
\end{equation}
This intersection consists of hyperelastic solids whose stress response is locally invertible and whose local inverse defines a complementary hyperelastic response. For any solid belonging to $\mathfrak{E}_{\mathrm{C}}\cap\mathfrak{E}_{\mathrm{CC}}$, the Cauchy and complementary Cauchy constitutive equations are locally inverse representations of the same constitutive relation, and the stress-work and complementary stress-work $1$-forms satisfy the relation $d\boldsymbol{\Omega}=-d\boldsymbol{\Lambda}$. Consequently, $d\boldsymbol{\Omega}=0$ if and only if $d\boldsymbol{\Lambda}=0$ wherever both explicit representations exist. Thus, within $\mathfrak{E}_{\mathrm{C}}\cap\mathfrak{E}_{\mathrm{CC}}$, hyperelasticity and complementary hyperelasticity are equivalent. The relations among these classes are summarized schematically in Figure~\ref{Fig:Elastic-Constitutive-Classes}.

Every constitutive manifold considered here is six-dimensional. Cauchy elasticity and complementary Cauchy elasticity specify whether the constitutive relation can be represented explicitly in terms of strain or stress, respectively, while hyperelasticity and complementary hyperelasticity impose integrability conditions on the corresponding work $1$-forms. None of these conditions reduces the dimension of the constitutive manifold.

\vskip 0.2in

\subsection{Complementary work principles of elasticity}
\label{ComplementaryWorkTheorems}

Several constitutive restrictions in elasticity have been formulated in terms of the work of stress along paths in strain space \citep{TruesdellNoll2004}. Analogous restrictions may be considered in complementary Cauchy elasticity by replacing paths in strain space with paths in stress space. However, the resulting complementary work should not be confused with the mechanical work performed by stress.

At a material point $X\in\mathcal{B}$, let $\gamma_X:[t_1,t_2]\to\mathbb{U}_X$ be a smooth path in the admissible stress domain. The complementary stress-work along $\gamma_X$ is defined as
\begin{equation}
	\mathscr{W}^{c}_{\gamma_X}=\int_{\gamma_X}\boldsymbol{\Lambda}
	=\frac{1}{2}\int_{t_1}^{t_2}C_{AB}\big(\mathbf{S}(X,t),\mathbf{G}(X)\big)\,
	\dot{S}^{AB}(X,t)\,dt\,.
\end{equation}
The total complementary stress-work in a sub-body $\mathcal{U}\subset\mathcal{B}$ over the interval $[t_1,t_2]$ is
\begin{equation}
	\mathscr{W}^{c}(\mathcal{U},[t_1,t_2])=\int_{\mathcal{U}}\mathscr{W}^{c}_{\gamma_X}\,dV\,.
\end{equation}
Unlike stress work, complementary stress-work does not enter the balance of energy as mechanical work. In particular, there is no complementary analogue of the decomposition of total mechanical work into changes in kinetic energy and stress work.

A stress history in $\mathcal{U}$ is called \emph{cyclic} if $\mathbf{S}(X,t_1)=\mathbf{S}(X,t_2)$ for every $X\in\mathcal{U}$, so that $\gamma_X$ is a closed curve in stress space for every material point. Reversing its orientation, one concludes that
\begin{equation}
	\mathscr{W}^{c}_{-\gamma_X}=-\mathscr{W}^{c}_{\gamma_X}\,.
\end{equation}
This identity follows solely from the change of sign of a line integral when the orientation of the integration curve is reversed.

\subsubsection{A complementary version of Caprioli's work postulate}

\citet{Caprioli1955} imposed a cyclic-work restriction on elastic constitutive laws. In terms of the stress-work $1$-form, this restriction requires $\int_{\Gamma}\boldsymbol{\Omega}\geq 0$ for every closed strain path $\Gamma$. Since reversing the orientation changes the sign of the integral, application of the same inequality to $-\Gamma$ implies that $\int_{\Gamma}\boldsymbol{\Omega}=0$ for every closed strain path. Hence, $\boldsymbol{\Omega}$ is locally exact, and the solid is hyperelastic. An analogous cyclic-work criterion was introduced in plasticity by \citet{Ilyushin1961}, with positive net stress work indicating plastic deformation and zero net stress work indicating a purely elastic cycle. See \citep{TruesdellNoll2004,Hill1968} for further discussions.

The complementary analogue of Caprioli's postulate requires
\begin{equation}
	\int_{\gamma}\boldsymbol{\Lambda}\geq 0
\end{equation}
for every closed stress path $\gamma$. Applying the same inequality to $-\gamma$, and using $\int_{-\gamma}\boldsymbol{\Lambda}=-\int_{\gamma}\boldsymbol{\Lambda}$, gives
\begin{equation}
	\int_{\gamma}\boldsymbol{\Lambda}=0
\end{equation}
for every closed stress path. Hence, $\boldsymbol{\Lambda}$ is locally exact. On a simply connected stress domain, there exists a complementary strain-energy function $\zeta$ such that $\boldsymbol{\Lambda}=d\zeta$. Therefore, every complementary Cauchy elastic solid satisfying this postulate is complementary hyperelastic.

The complementary postulate has a different physical status from Caprioli's postulate. While $\int_{\Gamma}\boldsymbol{\Omega}$ represents mechanical stress work, $\int_{\gamma}\boldsymbol{\Lambda}$ does not represent mechanical work and does not enter the balance of energy. Consequently, the complementary version of Caprioli's postulate is an independent constitutive restriction on the strain response rather than a mechanical work restriction.

The relation between Caprioli's postulate and its complementary analogue can be formulated more generally for an implicit elastic constitutive equation. On the constitutive manifold $\mathcal{K}_X$,
the restricted stress-work and complementary stress-work $1$-forms we have $\boldsymbol{\Omega}_{\mathcal{K}}+\boldsymbol{\Lambda}_{\mathcal{K}} =d\mathcal{P}_{\mathcal{K}}$, where $\mathcal{P}_{\mathcal{K}}=\frac{1}{2}\mathbf{S}:\mathbf{C}^{\flat}$.
Therefore, for every closed curve $\widehat{\Gamma}\subset\mathcal{K}_X$,
\begin{equation}
    \int_{\widehat{\Gamma}}\boldsymbol{\Lambda}_{\mathcal{K}}
    =-\int_{\widehat{\Gamma}}\boldsymbol{\Omega}_{\mathcal{K}}\,.
\end{equation}
A generalized cyclic-work postulate may be formulated by requiring
either
\begin{equation}
    \int_{\widehat{\Gamma}}\boldsymbol{\Omega}_{\mathcal{K}}\geq 0
    \qquad\text{or}\qquad
    \int_{\widehat{\Gamma}}\boldsymbol{\Lambda}_{\mathcal{K}}\geq 0
\end{equation}
for every closed curve $\widehat{\Gamma}$ on $\mathcal{K}_X$.
Since reversing the orientation of a closed curve changes the
sign of either integral, each of these conditions implies that
the corresponding integral vanishes. The preceding identity
then implies that
\begin{equation}
    \int_{\widehat{\Gamma}}\boldsymbol{\Omega}_{\mathcal{K}}
    =\int_{\widehat{\Gamma}}\boldsymbol{\Lambda}_{\mathcal{K}} =0\,,
\end{equation}
for every closed curve $\widehat{\Gamma}$ on $\mathcal{K}_X$. Consequently, both restricted $1$-forms are globally exact, and there exist intrinsic work and complementary work potentials
on $\mathcal{K}_X$. In particular, $\iota_X^*\boldsymbol{\omega}=0$, so that the constitutive
manifold is Lagrangian. Thus, either cyclic-work postulate implies the existence of both intrinsic potentials for a general implicit elastic constitutive equation, without requiring either constitutive projection to be invertible.

\subsubsection{A complementary version of Ericksen's work theorem}
\label{ComplementaryEricksenWorkTheorem}

Motivated by Carath\'eodory's approach to thermodynamics, \citet{Ericksen1956} considered stress-work-free motions, namely, deformation paths along which $\boldsymbol{\Omega}=0$. He postulated that every deformation has arbitrarily close neighboring deformations that cannot be reached by such a path. 
Carath\'eodory's theorem then implies the local representation $\boldsymbol{\Omega}=\phi\,d\psi$, for some scalar functions $\phi$ and $\psi$ on strain space. Thus, the stress response is locally determined by two scalar functions. This representation is hyperelastic precisely when $d\phi\wedge d\psi=0$, in which case $\phi$ depends locally on strain through $\psi$ alone. Following \citet{YavariGoriely2025Cauchy}, Cauchy elastic solids satisfying Ericksen's inaccessibility postulate are called \emph{Ericksen materials}. The postulate is an additional constitutive assumption rather than a consequence of the balance laws or the second law of thermodynamics.

The complementary version is formulated on stress space. Consider the Pfaffian equation $
\boldsymbol{\Lambda}=\frac{1}{2}\mathbf{C}^\flat\!:\!d\mathbf{S}=0$. Its integral curves are stress paths $\gamma$ for which $\boldsymbol{\Lambda}(\dot{\gamma})=0$. The complementary stress-work therefore vanishes along every segment of such a path. Suppose that every stress state has arbitrarily close neighboring stress states that cannot be reached from it by an integral curve of this equation. We call this the local inaccessibility postulate for $\boldsymbol{\Lambda}$.
By Carath\'eodory's theorem, the local inaccessibility postulate implies that there exist scalar functions $\kappa$ and $\zeta$ such that $\boldsymbol{\Lambda}=\kappa\,d\zeta$. Comparing this representation with the definition of $\boldsymbol{\Lambda}$, one concludes that
\begin{equation}
	C_{AB}=2\kappa\frac{\partial\zeta}{\partial S^{AB}}\,.
\end{equation}
Thus, the strain response is locally determined by two scalar functions on stress space. Moreover, $d\boldsymbol{\Lambda}=d\kappa\wedge d\zeta$. Hence, $\boldsymbol{\Lambda}$ is closed if and only if $d\kappa\wedge d\zeta=0$. Wherever $d\zeta\neq0$, this condition implies that $\kappa$ is locally a function of $\zeta$. In that case,
\begin{equation}
	\boldsymbol{\Lambda}=\kappa(\zeta)\,d\zeta=d\left(\int_{\zeta_0}^{\zeta}\kappa(s)\,ds\right)\,.
\end{equation}
and the material is complementary hyperelastic.

A complementary Cauchy elastic solid satisfying the local inaccessibility postulate is called a \emph{complementary Ericksen material}. Such a material admits a local two-function representation, but it is complementary hyperelastic only when $\kappa$ depends locally on stress through $\zeta$ alone. Complementary Ericksen materials therefore form the simplest class arising from the inaccessibility postulate that need not have a complementary strain-energy function.

\subsection{Material symmetry in complementary Cauchy elasticity}
\label{MaterialSymmetryComplementaryCauchyElasticity}

A complementary constitutive equation $\mathbf{C}^{\flat}=\widehat{\mathbf{C}}^{\flat}(X,\mathbf{S},\mathbf{G})$ is expressed entirely in terms of material tensors and is therefore objective. Material symmetry concerns a different invariance: the response of the solid under transformations of the material tangent space that preserve the material metric. The orthogonal group associated with $\mathbf{G}$ is defined as
\begin{equation}
	\mathrm{Orth}(\mathbf{G})=\left\{\mathbf{K}:T_X\mathcal{B}\rightarrow 
	T_X\mathcal{B}\mid\mathbf{K}^{\star}\mathbf{G}\mathbf{K}=\mathbf{G}\right\}\,.
\end{equation}
The natural actions of $\mathbf{K}\in\mathrm{Orth}(\mathbf{G})$ on $\mathbf{C}^{\flat}$ and $\mathbf{S}$ are defined as
\begin{equation}
	\mathbf{K}^*\mathbf{C}^{\flat}=\mathbf{K}^{\star}\mathbf{C}^{\flat}\mathbf{K}\,,\qquad 
	\mathbf{K}^*\mathbf{S}=\mathbf{K}^{-1}\mathbf{S}\mathbf{K}^{-\star}\,.
\end{equation}
In components, $(\mathbf{K}^*\mathbf{C}^{\flat})_{AB}=K^M{}_A\,G_{MN}\,K^N{}_B$ and $(\mathbf{K}^*\mathbf{S})^{AB}=K^{-A}{}_M\,S^{MN}\,K^{-B}{}_N$.
The material symmetry group of a complementary Cauchy elastic solid at $X$ is the subgroup $\mathcal{G}^c_X\leq\mathrm{Orth}(\mathbf{G})$ consisting of those transformations that preserve the admissible stress domain and satisfy the condition
\begin{equation}
	\widehat{\mathbf{C}}^{\flat}(X,\mathbf{K}^*\mathbf{S},\mathbf{G})
	=\mathbf{K}^*\widehat{\mathbf{C}}^{\flat}(X,\mathbf{S},\mathbf{G})\,,\qquad \forall 
	\mathbf{S}\in\mathbb{U}_X\,, \forall \mathbf{K}\in\mathcal{G}^c_X
\end{equation}
Thus, applying a material symmetry to the stress and then evaluating the strain response gives the same result as first evaluating the response and then applying the symmetry to the resulting strain.

\begin{prop}
Suppose that, on a constitutive branch, the complementary Cauchy elastic response $\widehat{\mathbf{C}}^{\flat}$ is the inverse of the Cauchy elastic response $\widehat{\mathbf{S}}$, and suppose that the branch is invariant under the action of all $\mathbf{K}\in\mathrm{Orth}(\mathbf{G})$. Then the Cauchy and complementary Cauchy elastic responses have the same material symmetry group on that branch.
\end{prop}

\begin{proof}
Suppose that, on a constitutive branch, the complementary response $\widehat{\mathbf{C}}^{\flat}$ is the inverse of the Cauchy elastic response $\widehat{\mathbf{S}}$, and suppose that the branch is invariant under the action of $\mathbf{K}$. The inverse relations are
\begin{equation} \label{C-Inverse}
	\widehat{\mathbf{C}}^{\flat}\bigl(X,\widehat{\mathbf{S}}(X,\mathbf{C}^{\flat},\mathbf{G}),
	\mathbf{G}\bigr)=\mathbf{C}^{\flat}\,,
\end{equation}
and
\begin{equation} \label{S-Inverse}
	\widehat{\mathbf{S}}\bigl(X,\widehat{\mathbf{C}}^{\flat}(X,\mathbf{S},\mathbf{G}),
	\mathbf{G}\bigr)=\mathbf{S}\,.
\end{equation}
Assume first that $\mathbf{K}\in \mathcal{G}$ is a material symmetry of the Cauchy elastic response, i.e.,
\begin{equation}
	\widehat{\mathbf{S}}(X,\mathbf{K}^*\mathbf{C}^{\flat},\mathbf{G})
	=\mathbf{K}^*\widehat{\mathbf{S}}(X,\mathbf{C}^{\flat},\mathbf{G})\,.
\end{equation}
Let $\mathbf{S}$ be any stress in the stress domain of the branch and define its corresponding strain by $\mathbf{C}^{\flat}=\widehat{\mathbf{C}}^{\flat}(X,\mathbf{S},\mathbf{G})$. The inverse relation \eqref{S-Inverse} implies that $\mathbf{S}=\widehat{\mathbf{S}}(X,\mathbf{C}^{\flat},\mathbf{G})$. Therefore, the symmetry relation implies that
\begin{equation}
	\mathbf{K}^*\mathbf{S}
	=\mathbf{K}^*\widehat{\mathbf{S}}(X,\mathbf{C}^{\flat},\mathbf{G})
	=\widehat{\mathbf{S}}(X,\mathbf{K}^*\mathbf{C}^{\flat},\mathbf{G})\,.
\end{equation}
Evaluating the complementary response $\widehat{\mathbf{C}}^{\flat}(X,\,\cdot\,,\mathbf{G})$ at the two equal stress tensors in this identity and using the inverse relation \eqref{C-Inverse} yields
\begin{equation}
\begin{aligned}
	\widehat{\mathbf{C}}^{\flat}(X,\mathbf{K}^*\mathbf{S},\mathbf{G})
	=\widehat{\mathbf{C}}^{\flat}
	\bigl(X,\widehat{\mathbf{S}}(X,\mathbf{K}^*\mathbf{C}^{\flat},\mathbf{G}),\mathbf{G}\bigr)
	=\mathbf{K}^*\mathbf{C}^{\flat}\,.
\end{aligned}
\end{equation}
Recalling that $\mathbf{C}^{\flat}=\widehat{\mathbf{C}}^{\flat}(X,\mathbf{S},\mathbf{G})$, this identity becomes
\begin{equation}
	\widehat{\mathbf{C}}^{\flat}(X,\mathbf{K}^*\mathbf{S},\mathbf{G})
	=\mathbf{K}^*\mathbf{C}^{\flat}
	=\mathbf{K}^*\widehat{\mathbf{C}}^{\flat}(X,\mathbf{S},\mathbf{G})\,.
\end{equation}
Hence, every material symmetry of $\widehat{\mathbf{S}}$ is also a material symmetry of $\widehat{\mathbf{C}}^{\flat}$, i.e., $\mathbf{K}\in\mathcal{G}^c$.

Conversely, suppose that $\mathbf{K}$ is a material symmetry of the complementary response, i.e.,
\begin{equation}
	\widehat{\mathbf{C}}^{\flat}(X,\mathbf{K}^*\mathbf{S},\mathbf{G})
	=\mathbf{K}^*\widehat{\mathbf{C}}^{\flat}(X,\mathbf{S},\mathbf{G})\,.
\end{equation}
Let $\mathbf{C}^{\flat}$ be any strain in the strain domain of the branch and define $\mathbf{S}=\widehat{\mathbf{S}}(X,\mathbf{C}^{\flat},\mathbf{G})$. The inverse relation \eqref{C-Inverse} implies that $\mathbf{C}^{\flat}=\widehat{\mathbf{C}}^{\flat}(X,\mathbf{S},\mathbf{G})$, and hence
\begin{equation}
\begin{aligned}
	\widehat{\mathbf{C}}^{\flat}(X,\mathbf{K}^*\mathbf{S},\mathbf{G})
	=\mathbf{K}^*\widehat{\mathbf{C}}^{\flat}(X,\mathbf{S},\mathbf{G})
	=\mathbf{K}^*\mathbf{C}^{\flat}\,.
\end{aligned}
\end{equation}
Evaluating the Cauchy elastic response $\widehat{\mathbf{S}}(X,\,\cdot\,,\mathbf{G})$ at the two equal strain tensors in this identity and using the second inverse relation gives
\begin{equation}
\begin{aligned}
	\widehat{\mathbf{S}}(X,\mathbf{K}^*\mathbf{C}^{\flat},\mathbf{G})
	=\widehat{\mathbf{S}}
	\bigl(X,\widehat{\mathbf{C}}^{\flat}(X,\mathbf{K}^*\mathbf{S},\mathbf{G}),\mathbf{G}\bigr)
	=\mathbf{K}^*\mathbf{S}
	=\mathbf{K}^*\widehat{\mathbf{S}}(X,\mathbf{C}^{\flat},\mathbf{G})\,.
\end{aligned}
\end{equation}
Thus, every material symmetry of $\widehat{\mathbf{C}}^{\flat}$ is also a material symmetry of $\widehat{\mathbf{S}}$, i.e., $\mathbf{K}\in\mathcal{G}$, and the two material symmetry groups coincide on the constitutive branch. Without local invertibility, the inverse identities used above are unavailable, and the symmetry group of one explicit representation does not, in general, determine that of the other.
\end{proof}

\subsubsection{Isotropic complementary Cauchy elasticity}

A complementary Cauchy elastic solid is isotropic at $X$ when $\mathcal{G}^c_X=\mathrm{Orth}(\mathbf{G})$. The representation theorem for isotropic tensor functions then tells us that
\begin{equation} \label{Isotropic-CC-Elasticity}
	\mathbf{C}^{\flat}=a_0\mathbf{G}+a_1\mathbf{S}^{\flat}+a_2\mathbf{S}^{2\flat}\,,
\end{equation}
where $\mathbf{S}^{\flat}=\mathbf{G}\mathbf{S}\mathbf{G}$ and $\mathbf{S}^{2\flat}=\mathbf{G}\mathbf{S}\mathbf{G}\mathbf{S}\mathbf{G}$. The scalar response functions $a_i=a_i(X,J_1,J_2,J_3)$ depend on stress through the principal invariants of the material stress endomorphism $\mathbf{S}\mathbf{G}$:\footnote{One can equivalently write these as 
\begin{equation} \label{Invariants-Stress}
	J_1=\operatorname{tr}_{\mathbf{G}}\mathbf{S}\,,\qquad 
	J_2=\frac{1}{2}\left[J_1^2-\operatorname{tr}_{\mathbf{G}}\mathbf{S}^{2\sharp}\right]\,,\qquad 
	J_3=\det(\mathbf{S}\mathbf{G})\,.
\end{equation}
}
\begin{equation} \label{Invariants-Stress}
	J_1=\operatorname{tr}(\mathbf{S}\mathbf{G})\,,\qquad 
	J_2=\frac{1}{2}\left[J_1^2-\operatorname{tr}((\mathbf{S}\mathbf{G})^2)\right]\,,\qquad 
	J_3=\det(\mathbf{S}\mathbf{G})\,.
\end{equation}
This representation supplies the constitutive basis used below in the classification of compressible isotropic complementary Cauchy elastic solids.

\subsubsection{Anisotropic complementary Cauchy elasticity}
\label{Sec:Anisotropic-complementary-Cauchy-elasticity}

The anisotropy of a complementary Cauchy elastic solid may be described by a finite collection of structural tensors $\boldsymbol{\zeta}_i$, $i=1,\ldots,N$, characterizing its material symmetry group. The complementary constitutive response has the following form
\begin{equation}
	\mathbf{C}^{\flat}=\mathbf{C}^{\flat}(X,\mathbf{S},\mathbf{G},\boldsymbol{\zeta}_1,\ldots,
	\boldsymbol{\zeta}_N)\,.
\end{equation}
When the structural tensors are included among the constitutive arguments, the response may be treated as an isotropic tensor function of the augmented collection of tensors \citep{liu1982,boehler1987,zheng1993,zheng1994theory,Lu2000,MazzucatoRachele2006}. It may therefore be represented using a finite integrity basis of scalar invariants and a corresponding finite tensor basis.

As an example, consider a transversely isotropic complementary Cauchy elastic solid. Let $\mathbf{N}$ be a unit vector specifying the material preferred direction. In addition to the three principal stress invariants $J_1$, $J_2$, and $J_3$, we define the following two extra invariants
\begin{equation}
	J_4=N^AN^B S_{AB}\,,\qquad J_5=N^AN^B S_{AM}S^M{}_B\,.
\end{equation}
The strain response admits the component representation
\begin{equation}
\begin{aligned}
	C_{AB}={}&a_0G_{AB}+a_1N_AN_B+a_2S_{AB}+a_3S_{AM}S^M{}_B 
	+a_4\left(N_AS_{BM}N^M+N_BS_{AM}N^M\right) \\
	&+a_5\left(N_AS_{BK}S^K{}_MN^M+N_BS_{AK}S^K{}_MN^M\right)\,,
\end{aligned}
\end{equation}
where $N_A=G_{AB}N^B$ and
\begin{equation}
	a_i=a_i(X,J_1,J_2,J_3,J_4,J_5)\,,\qquad i=0,\ldots,5\,.
\end{equation}
The response functions are restricted so that $\mathbf{C}^{\flat}$ is positive definite throughout the constitutively admissible stress domain.

Other material symmetry classes can be treated in the same manner by introducing structural tensors appropriate to the symmetry group under consideration. The resulting complementary strain response can then be expressed in terms of the corresponding finite integrity basis of scalar invariants and tensor generators.

\subsection{Darboux classification of complementary Cauchy elastic solids}
\label{Complementary-Edelen-Darboux-Potentials}

We next apply the Darboux classification to the complementary stress-work $1$-form. We first consider compressible anisotropic solids, whose constitutive variables span the full six-dimensional stress space, and then specialize the classification to compressible isotropic solids, whose constitutive response can be expressed in terms of the three principal stress invariants.

\subsubsection{Compressible anisotropic complementary Cauchy elasticity}

Let us fix a material point $X\in\mathcal{B}$ and a material metric $\mathbf{G}$. The complementary stress-work $1$-form is given in \eqref{Complementary-Work-$1$-form}. This is a differential $1$-form on the six-dimensional admissible stress domain $\mathbb{U}_X\subset\mathbb{S}_X$.\footnote{Here, we assume the balance of angular momentum and,
consequently, the symmetry of the second Piola--Kirchhoff stress. A detailed discussion of the balance of angular momentum and its derivation from covariance of the energy balance is given in
\S\ref{Sec:BalanceLawsComplementaryCauchyElasticity}.} Darboux's theorem implies that $\boldsymbol{\Lambda}$ can locally assume one of the following six canonical forms (see Appendix \ref{App:DarbouxClassification}):
\begin{empheq}[left={\empheqlbrace }]{align}
    \label{Complementary-Cauchy-1}
    \boldsymbol{\Lambda}&= d\zeta_1\,,  \\
    \label{Complementary-Cauchy-2}
    \boldsymbol{\Lambda}  &= \kappa_1d\zeta_1\,,    \\
    \label{Complementary-Cauchy-3}
    \boldsymbol{\Lambda} &= \kappa_1d\zeta_1+d\zeta_2\,,    \\
    \label{Complementary-Cauchy-4}
    \boldsymbol{\Lambda} &= \kappa_1d\zeta_1+\kappa_2d\zeta_2\,,    \\
    \label{Complementary-Cauchy-5}
    \boldsymbol{\Lambda} &= \kappa_1d\zeta_1+\kappa_2d\zeta_2+d\zeta_3\,,    \\
    \label{Complementary-Cauchy-6}
    \boldsymbol{\Lambda} &= \kappa_1d\zeta_1+\kappa_2d\zeta_2+\kappa_3d\zeta_3\,.
\end{empheq}
Here, $\zeta_i=\zeta_i(X,\mathbf{S},\mathbf{G})$ and $\kappa_i=\kappa_i(X,\mathbf{S},\mathbf{G})$, $i=1,2,3$. We call the functions $\zeta_i$ the \emph{generalized complementary energy functions}, or the \emph{complementary Edelen--Darboux potentials}, and the functions $\kappa_i$ their scalar multipliers.

The first canonical form \eqref{Complementary-Cauchy-1} characterizes complementary hyperelasticity. In this case, $\zeta_1$ is the complementary strain-energy function and $d\boldsymbol{\Lambda}=d\circ d\zeta_1=0$. The complementary stress-work is then path independent. More specifically,
\begin{equation}
	W_{\mathrm{c}}(\mathcal{U},[t_1,t_2])=\int_{\mathcal{U}}\int_{\gamma_X}d\zeta_1\,dV
	=\int_{\mathcal{U}}\left[\zeta_1(X,\mathbf{S}(X,t_2),\mathbf{G}(X))
	-\zeta_1(X,\mathbf{S}(X,t_1),\mathbf{G}(X))\right]dV\,.
\end{equation}
Consequently, the complementary stress-work vanishes for a cyclic stress history for which $\mathbf{S}(X,t_1)=\mathbf{S}(X,t_2)$ for every $X\in\mathcal{U}$.

We call complementary Cauchy elastic solids described by the canonical forms \eqref{Complementary-Cauchy-1}--\eqref{Complementary-Cauchy-6} \emph{complementary Edelen elastic solids} of types I, II, III, IV, V, and VI, respectively. Type I corresponds to complementary Green elastic, or complementary hyperelastic, solids. Type II corresponds to complementary Ericksen elastic solids for which $d\kappa_1\wedge d\zeta_1\neq0$. If $d\kappa_1\wedge d\zeta_1=0$, then $\kappa_1$ depends locally on stress through $\zeta_1$ alone, and the type II representation reduces to type I.

\begin{remark}
It should be emphasized that the complementary Edelen--Darboux potentials are not unique. For example, the type VI form \eqref{Complementary-Cauchy-6} is invariant under the following transformations
\begin{equation}
	\widetilde{\zeta}_i=g_i(\zeta_i)\,,\qquad \widetilde{\kappa}_i=\frac{\kappa_i}{g_i'(\zeta_i)}\,,
	\qquad i=1,2,3\,,
\end{equation}
where $g_i$ are arbitrary differentiable functions with non-vanishing derivatives. Thus, the constitutive content is carried by the complementary stress-work $1$-form $\boldsymbol{\Lambda}$ and not by a particular choice of its generalized complementary energy functions.
\end{remark}

\paragraph{Generalized complementary Doyle--Ericksen formulas.}

The canonical representations of the complementary stress-work $1$-form give corresponding representations of the strain response. For any scalar function $\zeta=\zeta(X,\mathbf{S},\mathbf{G})$ on stress space,
\begin{equation}
	d\zeta
	=\frac{\partial\zeta}{\partial\mathbf{S}}\!:\!d\mathbf{S}\,.
\end{equation}
Comparing each of the six canonical forms with $\boldsymbol{\Lambda}=\frac{1}{2}\mathbf{C}^{\flat}\!:\!d\mathbf{S}$ and using the arbitrariness of $d\mathbf{S}$ gives the following constitutive representations:
\begin{equation} \label{Darboux-CC}
\begin{dcases}
	\mathbf{C}^{\flat}
	=2\displaystyle\frac{\partial\zeta_1}{\partial\mathbf{S}}\,,
	&\text{type I}\,,\\[1mm]
	\mathbf{C}^{\flat}
	=2\kappa_1\displaystyle\frac{\partial\zeta_1}{\partial\mathbf{S}}\,,
	&\text{type II}\,,\\[1mm]
	\mathbf{C}^{\flat}
	=2\left(
	\kappa_1\displaystyle\frac{\partial\zeta_1}{\partial\mathbf{S}}
	+\displaystyle\frac{\partial\zeta_2}{\partial\mathbf{S}}\right)\,,
	&\text{type III}\,,\\[1mm]
	\mathbf{C}^{\flat}
	=2\left(
	\kappa_1\displaystyle\frac{\partial\zeta_1}{\partial\mathbf{S}}
	+\kappa_2\displaystyle\frac{\partial\zeta_2}{\partial\mathbf{S}}\right)\,,
	&\text{type IV}\,,\\[1mm]
	\mathbf{C}^{\flat}
	=2\left(
	\kappa_1\displaystyle\frac{\partial\zeta_1}{\partial\mathbf{S}}
	+\kappa_2\displaystyle\frac{\partial\zeta_2}{\partial\mathbf{S}}
	+\displaystyle\frac{\partial\zeta_3}{\partial\mathbf{S}}\right)\,,
	&\text{type V}\,,\\[1mm]
	\mathbf{C}^{\flat}
	=2\left(
	\kappa_1\displaystyle\frac{\partial\zeta_1}{\partial\mathbf{S}}
	+\kappa_2\displaystyle\frac{\partial\zeta_2}{\partial\mathbf{S}}
	+\kappa_3\displaystyle\frac{\partial\zeta_3}{\partial\mathbf{S}}\right)\,,
	&\text{type VI}\,.
\end{dcases}
\end{equation}
We call these relations the \emph{generalized complementary Doyle--Ericksen formulas}. Type I gives the complementary hyperelastic relation \eqref{Darboux-CC}$_1$, which is the complementary counterpart of the classical Doyle--Ericksen formula \citep{doyle1956nonlineari,MarsdenHughes1983}
\begin{equation}
	\mathbf{S}
	=2\frac{\partial W}{\partial\mathbf{C}^{\flat}}\,.
\end{equation}
The generalized complementary formulas do not require the complementary constitutive equation to be invertible. They follow directly from the Darboux representation of $\boldsymbol{\Lambda}$ and therefore remain meaningful even when no explicit Cauchy elastic response $\mathbf{S}=\mathbf{S}(\mathbf{C}^{\flat},\mathbf{G})$ exists.

\subsubsection{Compressible isotropic complementary Cauchy elasticity}
\label{Compressible-Isotropic-Complementary-Cauchy}

Consider a compressible isotropic complementary Cauchy elastic solid with strain response function $\mathbf{C}^\flat=\mathbf{C}^\flat(\mathbf{S},\mathbf{G})$. We define the material stress endomorphism as $\mathbf{S}\mathbf{G}$ (with components $S^A{}_B$) and denote its principal invariants given in \eqref{Invariants-Stress}. Let us introduce the invariant map
\begin{equation}
	\mathbf{J}:\mathbb{U}_X\rightarrow\mathbb{R}^3\,,\qquad 
	\mathbf{J}(\mathbf{S})=(J_1,J_2,J_3)\,.
\end{equation}
The representation theorem for isotropic tensor functions implies that the scalar response functions depend on stress through the invariants $J_1$, $J_2$, and $J_3$. Consequently, the complementary stress-work $1$-form on $\mathbb{U}_X$ is the pullback by $\mathbf{J}$ of a $1$-form on the three-dimensional space of stress invariants. Denoting this reduced $1$-form by $\boldsymbol{\Lambda}_{\mathcal{J}}$, we have
\begin{equation}
	\boldsymbol{\Lambda}=\mathbf{J}^*\boldsymbol{\Lambda}_{\mathcal{J}}\,,\qquad 
	\boldsymbol{\Lambda}_{\mathcal{J}}	
	=f_1(X,J_1,J_2,J_3)\,dJ_1+f_2(X,J_1,J_2,J_3)\,dJ_2+f_3(X,J_1,J_2,J_3)\,dJ_3\,.
\end{equation}
On the open subset of stress space where the invariant map $\boldsymbol{\mathcal{J}}=(J_1,J_2,J_3)$ has maximal rank, the differentials $dJ_1$, $dJ_2$, and $dJ_3$ are linearly independent, and the invariants provide local coordinates on a three-dimensional invariant space. Darboux's theorem then implies that $\boldsymbol{\Lambda}_{\mathcal{J}}$ has one of the following three canonical forms:\footnote{On a smooth subset where $\operatorname{rank}D\boldsymbol{\mathcal{J}}=2$, only the first two canonical forms are possible. On a smooth subset where $\operatorname{rank}D\boldsymbol{\mathcal{J}}=1$, the induced $1$-form is locally exact, and only the first canonical form is possible. These lower-dimensional subsets must be treated separately.}
\begin{empheq}[left={\empheqlbrace }]{align}
	\label{Isotropic-Complementary-Cauchy-1}
	\boldsymbol{\Lambda}_{\mathcal{J}}&=d\zeta_1\,,\\
	\label{Isotropic-Complementary-Cauchy-2}
	\boldsymbol{\Lambda}_{\mathcal{J}}&=\kappa_1d\zeta_1\,,\\
	\label{Isotropic-Complementary-Cauchy-3}
	\boldsymbol{\Lambda}_{\mathcal{J}}&=\kappa_1d\zeta_1+d\zeta_2\,.
\end{empheq}
Here, $\zeta_i=\zeta_i(X,J_1,J_2,J_3)$, $i=1,2$, and $\kappa_1=\kappa_1(X,J_1,J_2,J_3)$. The first canonical form characterizes complementary Green elastic, or complementary hyperelastic, solids. The second characterizes complementary Ericksen elastic solids when $d\kappa_1\wedge d\zeta_1\neq0$; when
$d\kappa_1\wedge d\zeta_1=0$, it reduces locally to the first canonical form. A solid described by the third canonical form, with $\boldsymbol{\Lambda}_{\mathcal{J}}\wedge d\boldsymbol{\Lambda}_{\mathcal{J}}\neq0$, is called an \emph{isotropic complementary Edelen elastic solid}.

On a region where $d\zeta_1\neq0$ and, for a complementary Ericksen solid, $\kappa_1\neq0$, the maximal integral manifolds of $\boldsymbol{\Lambda}_{\mathcal{J}}=0$ are the level surfaces $\zeta_1(J_1,J_2,J_3)=\text{constant}$. Complementary Green and complementary Ericksen elastic solids therefore have the local inaccessibility property in stress-invariant space. In contrast, an isotropic complementary Edelen elastic solid for which $\boldsymbol{\Lambda}_{\mathcal{J}}\wedge d\boldsymbol{\Lambda}_{\mathcal{J}}\neq0$ possesses the local accessibility property in stress-invariant space.

\paragraph{Constitutive equations in terms of the complementary Edelen--Darboux potentials.} Let us introduce the notation
\begin{equation}
	\zeta_{ij}=\frac{\partial\zeta_i}{\partial J_j}\,,\qquad i=1,2\,,\qquad j=1,2,3\,.
\end{equation}
The two indices label the potential and the stress invariant, respectively; they do not denote a second derivative and carry no symmetry. Pulling the three canonical forms back to stress space and comparing them with $\boldsymbol{\Lambda}=\frac{1}{2}\mathbf{C}^{\flat}\!:\!d\mathbf{S}$ gives us the constitutive equations for complementary Green, complementary Ericksen, and complementary Edelen isotropic elastic solids, respectively:
\begin{equation}
\begin{dcases}
	\mathbf{C}^{\flat}=2\displaystyle\sum_{j=1}^{3}\zeta_{1j}\frac{\partial J_j}{\partial\mathbf{S}}\,,
	\\[0.5mm]
	\mathbf{C}^{\flat}=2\kappa_1\displaystyle\sum_{j=1}^{3}
	\zeta_{1j}\frac{\partial J_j}{\partial\mathbf{S}}\,,
	\\[0.5mm]
	\mathbf{C}^{\flat}=2\displaystyle\sum_{j=1}^{3}(\kappa_1\zeta_{1j}
	+\zeta_{2j})\frac{\partial J_j}{\partial\mathbf{S}}\,.
\end{dcases}
\end{equation}
The derivatives of the stress invariants are written as\footnote{Let $\delta\mathbf{S}=\mathbf{H}$ be arbitrary. The product rule implies that $\delta(\mathbf{S}^m)=\sum_{k=0}^{m-1}\mathbf{S}^k\mathbf{H}\mathbf{S}^{m-1-k}$. By the cyclic property of the trace, all the terms in this sum have the same trace, and hence $\delta\operatorname{tr}(\mathbf{S}^m)=m\operatorname{tr}(\mathbf{S}^{m-1}\mathbf{H})=m\mathbf{S}^{(m-1)\flat}\!:\!\mathbf{H}$. For example, $\delta(\mathbf{S}^2)=\mathbf{S}\mathbf{H}+\mathbf{H}\mathbf{S}$, and therefore $\delta\operatorname{tr}(\mathbf{S}^2)=2\operatorname{tr}(\mathbf{S}\mathbf{H})$. The Cayley--Hamilton identity $\mathbf{S}^3-J_1\mathbf{S}^2+J_2\mathbf{S}-J_3\mathbf{I}=\mathbf{0}$
implies that $J_3=\frac{1}{6}\bigl[J_1^3-3J_1\operatorname{tr}(\mathbf{S}^2)+2\operatorname{tr}(\mathbf{S}^3)\bigr]$. Using this identity together with $J_1=\operatorname{tr}\mathbf{S}$ and $J_2=\frac{1}{2}\bigl[J_1^2-\operatorname{tr}(\mathbf{S}^2)\bigr]$ gives us $\delta J_1=\mathbf{G}\!:\!\mathbf{H}$, $\delta J_2=(J_1\mathbf{G}-\mathbf{S}^{\flat})\!:\!\mathbf{H}$, and $\delta J_3=(J_2\mathbf{G}-J_1\mathbf{S}^{\flat}+\mathbf{S}^{2\flat})\!:\!\mathbf{H}$. Since $\mathbf{H}$ is arbitrary, the stated derivatives follow.}
\begin{equation} \label{Stress-Derivatives}
	\frac{\partial J_1}{\partial\mathbf{S}}=\mathbf{G}\,,\qquad
	\frac{\partial J_2}{\partial\mathbf{S}}=J_1\mathbf{G}-\mathbf{S}^{\flat}\,,\qquad
	\frac{\partial J_3}{\partial\mathbf{S}}=J_2\mathbf{G}-J_1\mathbf{S}^{\flat}+\mathbf{S}^{2\flat}\,,
\end{equation}
where $\mathbf{S}^{\flat}=\mathbf{G}\mathbf{S}\mathbf{G}$ and $\mathbf{S}^{2\flat}=\mathbf{G}\mathbf{S}\mathbf{G}\mathbf{S}\mathbf{G}$. Substitution of the derivatives of the stress invariants gives a common representation for the three material classes. Let us define
\begin{equation}
	q_j=\begin{cases}
	\zeta_{1j}\,, & \text{for complementary Green elasticity}\,,\\
	\kappa_1\zeta_{1j}\,, & \text{for complementary Ericksen elasticity}\,,\\
	\kappa_1\zeta_{1j}+\zeta_{2j}\,, & \text{for complementary Edelen elasticity}\,,
	\end{cases}\qquad j=1,2,3\,.
\end{equation}
The corresponding constitutive equations are written as
\begin{equation}
	\mathbf{C}^{\flat}=2\alpha\mathbf{G}-2\beta\mathbf{S}^{\flat}+2q_3\mathbf{S}^{2\flat}\,,
\end{equation}
where $\alpha=q_1+J_1q_2+J_2q_3$ and $\beta=q_2+J_1q_3$.
The positive-definiteness of $\mathbf{C}^{\flat}$ imposes additional constitutive restrictions on the complementary Edelen--Darboux potentials and their scalar multipliers. These restrictions are independent of the local Darboux classification.

\paragraph{Cyclic stress histories and paths of zero complementary stress-work.} The admissible stress-invariant domain is defined as
\begin{equation}
	\mathcal{J}=\left\{(J_1,J_2,J_3)\in\mathbb{R}^3:(J_1,J_2,J_3)
	\text{ are the invariants of an admissible stress state}\right\}\,.
\end{equation}
Unlike the principal invariants of the right Cauchy--Green strain, the stress invariants are not required to be positive. The domain $\mathcal{J}$ is determined by the constitutively admissible stress states.
A cyclic stress history is a closed path $\gamma:[t_1,t_2]\to\mathbb{U}_X$ such that $\gamma(t_1)=\gamma(t_2)$. Its projection under the invariant map $\mathbf{J}:\mathbb{U}_X\to\mathcal{J}$ is the closed curve $\gamma_{\mathcal{J}}=\mathbf{J}\circ\gamma$ in invariant space. The converse does not hold in general: a closed curve in $\mathcal{J}$ need not lift to a closed curve in $\mathbb{U}_X$, since distinct stress tensors may have the same principal invariants.
Since $\boldsymbol{\Lambda}=\mathbf{J}^*\boldsymbol{\Lambda}_{\mathcal{J}}$, the complementary stress-work along $\gamma$ can be evaluated along its invariant projection:
\begin{equation}
	\int_{\gamma}\boldsymbol{\Lambda}=\int_{\gamma_{\mathcal{J}}}\boldsymbol{\Lambda}_{\mathcal{J}}\,.
\end{equation}
For a complementary Green elastic solid, $\boldsymbol{\Lambda}_{\mathcal{J}}=d\zeta_1$, and hence
\begin{equation}
	\int_{\gamma_{\mathcal{J}}}\boldsymbol{\Lambda}_{\mathcal{J}}
	=\zeta_1(\gamma_{\mathcal{J}}(t_2))-\zeta_1(\gamma_{\mathcal{J}}(t_1))\,.
\end{equation}
Thus, the complementary stress-work vanishes for every cyclic stress history.

A path $c$ in $\mathcal{J}$ is an integral curve of the Pfaffian equation $\boldsymbol{\Lambda}_{\mathcal{J}}=0$ if $\boldsymbol{\Lambda}_{\mathcal{J}}(\dot{c})=0$ at every point along the path. This pointwise condition implies that the complementary stress-work vanishes along every segment of $c$; it is stronger than requiring only that the total integral along the entire path vanish. For both complementary Green and complementary Ericksen elastic solids, the maximal integral manifolds are locally the surfaces
\begin{equation}
	\zeta_1(J_1,J_2,J_3)=\mathrm{constant}\,.
\end{equation}
These two classes therefore have the Carath\'eodory local inaccessibility property in invariant stress space.

\paragraph{Paths of zero complementary stress-work for an isotropic complementary Edelen solid.}Apart from replacing the strain invariants and the stress-work $1$-form by the stress invariants and the complementary stress-work $1$-form, the following construction is identical to the corresponding construction for isotropic Cauchy elasticity \citep{YavariGoriely2025Cauchy}. We repeat the argument to make the present discussion self-contained and to formulate the local accessibility property in invariant stress space. Let us consider an isotropic complementary Edelen elastic solid for which
\begin{equation}
	\boldsymbol{\Lambda}_{\mathcal{J}}=\kappa_1d\zeta_1+d\zeta_2\,,\qquad 
	\boldsymbol{\Lambda}_{\mathcal{J}}\wedge d\boldsymbol{\Lambda}_{\mathcal{J}}
	=d\zeta_2\wedge d\kappa_1\wedge d\zeta_1\neq0\,.
\end{equation}
The non-vanishing of this $3$-form implies that $(\kappa_1,\zeta_1,\zeta_2)$ form a local coordinate system on $\mathcal{J}$. By translating and redefining these coordinates while preserving the canonical form of $\boldsymbol{\Lambda}_{\mathcal{J}}$, we may assign coordinates $(0,0,0)$ to a given state $x$ and $(\mathring{\kappa}_1,\mathring{\zeta}_1,\mathring{\zeta}_2)$ to a neighboring state $y$. Choose a curve $c:[0,1]\rightarrow\mathbb{R}^2$ in the $(\kappa_1,\zeta_1)$-plane such that $c(0)=(0,0)$ and $c(1)=(\mathring{\kappa}_1,\mathring{\zeta}_1)$. Let us define the third coordinate along the curve by
\begin{equation}
	\zeta_2(t)=-\int_0^t\kappa_1(\tau)\,\dot{\zeta}_1(\tau)\,d\tau\,.
\end{equation}
The endpoint value $\zeta_2(1)$ is the negative of the line integral of $\kappa_1\,d\zeta_1$ along $c$. Inserting a sufficiently small loop into $c$ changes this integral by the signed area enclosed by the loop while leaving the endpoints of $c$ unchanged. Since $y$ is sufficiently close to $x$, the loop may be chosen so that $\zeta_2(1)=\mathring{\zeta}_2$. The resulting curve
\begin{equation}
	\gamma_{\mathcal{J}}(t)=(\kappa_1(t),\zeta_1(t),\zeta_2(t))\,,\qquad t\in[0,1]\,,
\end{equation}
connects $x$ to $y$ and satisfies the condition $\boldsymbol{\Lambda}_{\mathcal{J}}(\dot{\gamma}_{\mathcal{J}})=\kappa_1\dot{\zeta}_1+\dot{\zeta}_2=0$. Therefore, any two sufficiently close points of $\mathcal{J}$ can be connected by an integral curve of $\boldsymbol{\Lambda}_{\mathcal{J}}=0$. Isotropic complementary Edelen elastic solids satisfying $\boldsymbol{\Lambda}_{\mathcal{J}}\wedge d\boldsymbol{\Lambda}_{\mathcal{J}}\neq0$ consequently have the local accessibility property in invariant stress space.

\subsection{Examples of isotropic complementary Cauchy elastic solids}
\label{Sec:Examples}

We begin by expressing the complementary stress-work $1$-form of a general compressible isotropic complementary Cauchy elastic solid in terms of the principal stress invariants. We then use this representation to construct explicit responses that are not complementary hyperelastic and to identify the failure of major symmetry directly from the response functions. Recall that the constitutive equation has the isotropic representation \eqref{Isotropic-CC-Elasticity}. The associated complementary stress-work $1$-form is written as
\begin{equation}
	\boldsymbol{\Lambda}=\frac{1}{2}\mathbf{C}^{\flat}\!:\!d\mathbf{S}
	=\frac{a_0}{2}\mathbf{G}\!:\!d\mathbf{S}+\frac{a_1}{2}\mathbf{S}^{\flat}\!:\!d\mathbf{S}
	+\frac{a_2}{2}\mathbf{S}^{2\flat}\!:\!d\mathbf{S}\,.
\end{equation}
The first contraction is simplified to read $\mathbf{G}\!:\!d\mathbf{S}=dJ_1$. Differentiating the definition of $J_2$ gives us $dJ_2=J_1\,dJ_1-\mathbf{S}^{\flat}\!:\!d\mathbf{S}$, and hence $\mathbf{S}^{\flat}\!:\!d\mathbf{S}=J_1\,dJ_1-dJ_2$. The Cayley--Hamilton identity implies that\footnote{The Cayley--Hamilton identity for $\mathbf{S}\mathbf{G}$ reads $(\mathbf{S}\mathbf{G})^3-J_1(\mathbf{S}\mathbf{G})^2+J_2(\mathbf{S}\mathbf{G})-J_3\mathbf{I}=\mathbf{0}$.
Taking its trace and using $\operatorname{tr}\mathbf{I}=3$ gives us $\operatorname{tr}(\mathbf{S}^3)-J_1\operatorname{tr}(\mathbf{S}^2)+J_2\operatorname{tr}\mathbf{S}-3J_3=0$,  where the powers and traces are understood using the material metric $\mathbf{G}$. Substitution of $\operatorname{tr}\mathbf{S}=J_1$ and $\operatorname{tr}(\mathbf{S}^2)=J_1^2-2J_2$ yields the stated identity.}
\begin{equation}
	\operatorname{tr}(\mathbf{S}^3)=J_1^3-3J_1J_2+3J_3\,.
\end{equation}
On the other hand, $d\operatorname{tr}(\mathbf{S}^3)=3\mathbf{S}^{2\flat}\!:\!d\mathbf{S}$. Differentiating the preceding invariant identity and dividing by $3$ therefore implies that $\mathbf{S}^{2\flat}\!:\!d\mathbf{S}
=(J_1^2-J_2)\,dJ_1-J_1\,dJ_2+dJ_3$. Substitution of these three contractions into the complementary stress-work $1$-form yields
\begin{equation}
	\boldsymbol{\Lambda}=\frac{1}{2}a_0\,dJ_1+\frac{1}{2}a_1\left(J_1\,dJ_1-dJ_2\right)
	+\frac{1}{2}a_2\left[(J_1^2-J_2)\,dJ_1-J_1\,dJ_2+dJ_3\right]\,.
\end{equation}
Equivalently,
\begin{equation} \label{General-Isotropic-Complementary-Work-Form}
	\boldsymbol{\Lambda}=\frac{1}{2}\left(P\,dJ_1+Q\,dJ_2+R\,dJ_3\right)\,,
\end{equation}
where
\begin{equation}
	P=a_0+J_1a_1+(J_1^2-J_2)a_2\,,\qquad Q=-(a_1+J_1a_2)\,,\qquad R=a_2\,.
\end{equation}

Taking the exterior derivative of $\boldsymbol{\Lambda}$, one obtains
\begin{equation}\label{Exterior-Derivative-General-Isotropic-Example}
	d\boldsymbol{\Lambda}
	=\frac{1}{2}\left(\frac{\partial Q}{\partial J_1}
	-\frac{\partial P}{\partial J_2}\right)dJ_1\wedge dJ_2
	+\frac{1}{2}\left(\frac{\partial R}{\partial J_1}
	-\frac{\partial P}{\partial J_3}\right)dJ_1\wedge dJ_3
	+\frac{1}{2}\left(\frac{\partial R}{\partial J_2}
	-\frac{\partial Q}{\partial J_3}\right)dJ_2\wedge dJ_3\,.
\end{equation}
This identity holds throughout the constitutive stress domain. The invariant map $\boldsymbol{\mathcal{J}}=(J_1,J_2,J_3)$ may have rank one, two, or three. Rank zero is excluded because $dJ_1=\mathbf{G}\!:\!d\mathbf{S}$ is nonzero.\footnote{As a matter of fact, for the stress variation $\mathbf{H}=\mathbf{G}^{\sharp}$, $dJ_1[\mathbf{H}]=\mathbf{G}\!:\!\mathbf{G}^{\sharp}=3$, and hence $dJ_1$ is nowhere zero on stress space.} At a state of stress with three distinct principal values, $\boldsymbol{\mathcal{J}}$ has rank three; at a stress with exactly two distinct principal values, it has rank two; and at a hydrostatic stress, it has rank one.

On the open subset where $\boldsymbol{\mathcal{J}}$ has rank three, the $1$-forms $dJ_1$, $dJ_2$, and $dJ_3$ are linearly independent. Consequently, the three $2$-forms $dJ_1\wedge dJ_2$, $dJ_1\wedge dJ_3$, and $dJ_2\wedge dJ_3$ are linearly independent, and $\boldsymbol{\Lambda}$ is closed if and only if
\begin{equation}\label{General-Isotropic-Closure-Conditions}
	\frac{\partial P}{\partial J_2}=\frac{\partial Q}{\partial J_1}\,,\qquad
	\frac{\partial P}{\partial J_3}=\frac{\partial R}{\partial J_1}\,,\qquad
	\frac{\partial Q}{\partial J_3}=\frac{\partial R}{\partial J_2}\,.
\end{equation}
In terms of the original response functions, these conditions become
\begin{equation}\label{General-Isotropic-Closure-Conditions-ai}
\begin{aligned}
	&\frac{\partial a_1}{\partial J_1}+J_1\frac{\partial a_2}{\partial J_1}
	+\frac{\partial a_0}{\partial J_2}+J_1\frac{\partial a_1}{\partial J_2}
	+(J_1^2-J_2)\frac{\partial a_2}{\partial J_2}=0\,,\\
	&\frac{\partial a_0}{\partial J_3}+J_1\frac{\partial a_1}{\partial J_3}
	+(J_1^2-J_2)\frac{\partial a_2}{\partial J_3}-\frac{\partial a_2}{\partial J_1}=0\,,\\
	&\frac{\partial a_1}{\partial J_3}+J_1\frac{\partial a_2}{\partial J_3}
	+\frac{\partial a_2}{\partial J_2}=0\,.
\end{aligned}
\end{equation}
Thus, violation of any one of \eqref{General-Isotropic-Closure-Conditions-ai} at a rank-three stress implies that the solid is not complementary hyperelastic in a neighborhood of that stress.

Because the constitutively admissible stress domain $\mathbb{U}_X$ is open in the six-dimensional stress space $\mathbb{S}_X$, the stresses in $\mathbb{U}_X$ with three distinct principal values form an open dense subset of $\mathbb{U}_X$.\footnote{The set is open because the eigenvalues of a symmetric endomorphism depend continuously on the endomorphism and distinct eigenvalues remain distinct under sufficiently small perturbations. It is dense because any repeated eigenvalue can be split by an arbitrarily small symmetric perturbation; see, for example, \citep{Kato1995,Bhatia1997}.} Consequently, every stress in the domain with repeated principal values is the limit of a sequence of stresses with three distinct principal values. Since the response functions $a_i$ are smooth, the coefficients in \eqref{Exterior-Derivative-General-Isotropic-Example} are continuous. Therefore, if the closure conditions hold at every stress with three distinct principal values, they also hold at stresses with repeated principal values by continuity.

\subsubsection{Response functions depending only on the first stress invariant}

Consider the subclass $a_i=a_i(J_1)$, $i=0,1,2$. In this case, \eqref{Exterior-Derivative-General-Isotropic-Example} reduces to
\begin{equation}\label{Exterior-Derivative-J1-Dependent}
	d\boldsymbol{\Lambda}=-\frac{1}{2}\left(a_1'+J_1a_2'\right)dJ_1\wedge dJ_2
	+\frac{1}{2}a_2'\,dJ_1\wedge dJ_3\,,
\end{equation}
where a prime denotes differentiation with respect to $J_1$. On a rank-three region, the $2$-forms $dJ_1\wedge dJ_2$ and $dJ_1\wedge dJ_3$ are linearly independent. Therefore, $\boldsymbol{\Lambda}$ is closed if and only if
\begin{equation}
	a_2'=0\,,\qquad a_1'+J_1a_2'=0\,.
\end{equation}
The first condition implies that $a_2$ is constant. The second condition then reduces to $a_1'=0$, and hence $a_1$ is also constant. No restriction is imposed on $a_0(J_1)$. By the density and continuity argument given above, the same conclusion holds throughout the constitutively admissible stress domain.

For constant $a_1$ and $a_2$, the three terms in the complementary stress-work $1$-form can be recognized as exact differentials:
\begin{equation}
\begin{aligned}
	a_0(J_1)\,dJ_1&=d\left(\int_{\mathring{J}_1}^{J_1}a_0(s)\,ds\right)\,,\\
	J_1\,dJ_1-dJ_2 &=d\left(\frac{1}{2}J_1^2-J_2\right)\,,\\
	(J_1^2-J_2)\,dJ_1-J_1\,dJ_2+dJ_3&=d\left(\frac{1}{3}J_1^3-J_1J_2+J_3\right)\,.
\end{aligned}
\end{equation}
Substitution of these identities into the expression for $\boldsymbol{\Lambda}$ gives us
\begin{equation}
\begin{aligned}
	\boldsymbol{\Lambda}=d\Bigg[
	\frac{1}{2}\int_{\mathring{J}_1}^{J_1}a_0(s)\,ds+\frac{a_1}{2}\left(\frac{1}{2}J_1^2-J_2\right)
	+\frac{a_2}{2}\left(\frac{1}{3}J_1^3-J_1J_2+J_3\right)\Bigg]\,.
\end{aligned}
\end{equation}
Thus, a local complementary strain-energy function is given as
\begin{equation}
	\zeta=\frac{1}{2}\int_{\mathring{J}_1}^{J_1}a_0(s)\,ds
	+\frac{a_1}{2}\left(\frac{1}{2}J_1^2-J_2\right)
	+\frac{a_2}{2}\left(\frac{1}{3}J_1^3-J_1J_2+J_3\right)\,,
\end{equation}
and $\boldsymbol{\Lambda}=d\zeta$. The choice of the constant $\mathring{J}_1$ changes $\zeta$ only by an additive constant.

\paragraph{An explicit non-hyperelastic response.}
Let $k_1>0$ and $k_2\neq0$ be material constants, and consider $a_0=k_1$, $a_1=k_2J_1$, and $a_2=0$. The corresponding constitutive equation reads
\begin{equation}\label{Explicit-Isotropic-Complementary-Ericksen-Response}
	\mathbf{C}^{\flat}=k_1\mathbf{G}+k_2J_1\mathbf{S}^{\flat}\,.
\end{equation}
This response is isotropic and compressible. Its constitutively admissible stress domain may be chosen to be any connected open set on which the strain response is positive-definite, i.e., $k_1\mathbf{G}+k_2J_1\mathbf{S}^{\flat}\in\operatorname{Sym}^{+}(T_X^*\mathcal{B})$. In a $\mathbf{G}$-orthonormal principal basis of $\mathbf{S}$, the principal values of $\mathbf{C}^{\flat}$ are $k_1+k_2J_1s_a$, $a=1,2,3$, where $s_a$ are the principal values of $\mathbf{S}$. Therefore, positive definiteness is equivalent to
\begin{equation}
	k_1+k_2J_1s_a>0\,,\qquad a=1,2,3\,.
\end{equation}
Since $J_1=s_1+s_2+s_3$, where $s_a$ are the principal stresses, the triangle inequality gives us
\begin{equation}
	|J_1| \leq |s_1|+|s_2|+|s_3| \leq3\lVert\mathbf{S}\rVert_{\mathrm{op}}\,,
\end{equation}
where $\lVert\mathbf{S}\rVert_{\mathrm{op}}=\max_a|s_a|$ is the operator norm of $\mathbf{S}\mathbf{G}$. Consequently, $|k_2J_1s_a| \leq3|k_2|\,\lVert\mathbf{S}\rVert_{\mathrm{op}}^2$, $a=1,2,3$. Therefore, the condition $3|k_2|\,\lVert\mathbf{S}\rVert_{\mathrm{op}}^2<k_1$ is sufficient for positive definiteness. Thus, one possible constitutively admissible stress domain is the open ball
\begin{equation}
	\mathbb{U}_X=\left\{\mathbf{S}\in\mathbb{S}_X:\lVert\mathbf{S}\rVert_{\mathrm{op}}
	<\sqrt{\frac{k_1}{3|k_2|}}\,\right\}\,.
\end{equation}
For the response \eqref{Explicit-Isotropic-Complementary-Ericksen-Response}, the complementary stress-work $1$-form is written as
\begin{equation} \label{Explicit-Isotropic-Complementary-Ericksen-Form}
	\boldsymbol{\Lambda}=\frac{1}{2}\left(k_1+k_2J_1^2\right)dJ_1-\frac{k_2}{2}J_1\,dJ_2\,.
\end{equation}
Its exterior derivative is
\begin{equation}\label{Explicit-Isotropic-Complementary-Ericksen-Exterior-Derivative}
	d\boldsymbol{\Lambda}=-\frac{k_2}{2}\,dJ_1\wedge dJ_2\,.
\end{equation}
Hence, on every rank-three region, $d\boldsymbol{\Lambda}\neq0$. The response therefore does not admit a complementary strain-energy function and is not complementary hyperelastic. Nevertheless, $\boldsymbol{\Lambda}\wedge d\boldsymbol{\Lambda}=0$. Both terms in \eqref{Explicit-Isotropic-Complementary-Ericksen-Form} lie in the span of $dJ_1$ and $dJ_2$, whereas \eqref{Explicit-Isotropic-Complementary-Ericksen-Exterior-Derivative} is proportional to $dJ_1\wedge dJ_2$. Therefore, the response is an isotropic complementary Ericksen elastic solid on every region where $\boldsymbol{\Lambda}\neq0$ and $d\boldsymbol{\Lambda}\neq0$.

The tangent compliance tensor associated with \eqref{Explicit-Isotropic-Complementary-Ericksen-Response} is written as
\begin{equation}
	\mathbb{A}_{ABCD}=k_2S_{AB}G_{CD}+\frac{k_2J_1}{2}
	\left(G_{AC}G_{BD}+G_{AD}G_{BC}\right)\,.
\end{equation}
Its major-antisymmetric part is $\mathbb{A}_{ABCD}-\mathbb{A}_{CDAB}=k_2\left(S_{AB}G_{CD}-S_{CD}G_{AB}\right)$. This tensor does not vanish at a generic non-hydrostatic stress. Thus, the failure of complementary hyperelasticity may be seen equivalently from the non-vanishing of $d\boldsymbol{\Lambda}$ or from the failure of major symmetry of the tangent compliance tensor.

\subsubsection{Response functions depending only on the third stress invariant}
\label{Sec:IsotropicComplementaryCauchyJ3Example}

Consider the following isotropic complementary Cauchy elastic response
\begin{equation} \label{Isotropic-Complementary-Cauchy-J3-General}
	\mathbf{C}^{\flat}=a_0(J_3)\mathbf{G}+a_1(J_3)\,\mathbf{S}^{\flat}
	+a_2(J_3)\,\mathbf{S}^{2\flat}\,.
\end{equation}
Although the scalar response functions depend only on $J_3$, the strain response itself does not. The tensors $\mathbf{S}^{\flat}$ and $\mathbf{S}^{2\flat}$ retain information about the principal directions and the individual principal values of the stress. 
Substitution into \eqref{General-Isotropic-Complementary-Work-Form}, the complementary stress-work $1$-form is written as
\begin{equation}\label{Complementary-Work-Form-J3-General}
	\boldsymbol{\Lambda}=\frac{1}{2}\left[a_0+J_1a_1+(J_1^2-J_2)a_2\right]dJ_1
	-\frac{1}{2}(a_1+J_1a_2)dJ_2+\frac{1}{2}a_2\,dJ_3\,,
\end{equation}
where the response functions are evaluated at $J_3$. In what follows, a prime denotes differentiation with respect to $J_3$. Taking the exterior derivative gives us
\begin{equation} \label{Exterior-Derivative-J3-General}
	d\boldsymbol{\Lambda}=-\frac{1}{2}
	\left[a_0'+J_1a_1'+(J_1^2-J_2)a_2'\right]dJ_1\wedge dJ_3
	+\frac{1}{2}\left(a_1'+J_1a_2'\right)dJ_2\wedge dJ_3\,.
\end{equation}
The coefficient of $dJ_1\wedge dJ_2$ vanishes because the contributions arising from the explicit dependence of the coefficients in \eqref{Complementary-Work-Form-J3-General} on $J_1$ and $J_2$ cancel.

On an open region where the invariant map has rank three, the $2$-forms $dJ_1\wedge dJ_3$ and $dJ_2\wedge dJ_3$ are linearly independent. Hence, $\boldsymbol{\Lambda}$ is closed if and only if
\begin{equation} \label{Closure-Conditions-J3-General}
	a_0'+J_1a_1'+(J_1^2-J_2)a_2'=0\,, \qquad a_1'+J_1a_2'=0\,.
\end{equation}
On a region where the invariant map has rank three, $J_1$, $J_2$, and $J_3$ are locally independent and provide local coordinates on invariant space. Therefore, $J_2$ may be varied locally while $J_1$ and $J_3$ are held fixed. Since the functions $a_i'$ depend only on $J_3$, the first identity in \eqref{Closure-Conditions-J3-General} is affine in $J_2$ and vanishes on an open interval of $J_2$. Its coefficient of $J_2$ must consequently vanish, and hence $a_2'=0$. The second identity then reduces to $a_1'=0$, and substitution into the first gives $a_0'=0$. Therefore, the complementary stress-work $1$-form is closed if and only if all three response functions $a_0$, $a_1$, and $a_2$ are constant. By the density and continuity argument given earlier, this conclusion extends from the subset of stresses with three distinct principal values to the entire constitutively admissible stress domain.

Having established that $a_0$, $a_1$, and $a_2$ are constants, we now construct the corresponding complementary strain-energy function. The complementary stress-work $1$-form \eqref{Complementary-Work-Form-J3-General} may first be rearranged as
\begin{equation} \label{C-SW}
	\boldsymbol{\Lambda}=\frac{a_0}{2}\,dJ_1+\frac{a_1}{2}\left(J_1\,dJ_1-dJ_2\right)
	+\frac{a_2}{2}\left[(J_1^2-J_2)\,dJ_1-J_1\,dJ_2+dJ_3\right]\,.
\end{equation}
Since the coefficients $a_0$, $a_1$, and $a_2$ are constant, each term in this expression is an exact differential:
\begin{equation}
\begin{aligned}
	dJ_1&=d(J_1)\,,\\
	J_1\,dJ_1-dJ_2&=d\left(\frac{1}{2}J_1^2-J_2\right)\,,\\
	(J_1^2-J_2)\,dJ_1-J_1\,dJ_2+dJ_3&=d\left(\frac{1}{3}J_1^3-J_1J_2+J_3\right)\,.
\end{aligned}
\end{equation}
It follows that
\begin{equation}
	\boldsymbol{\Lambda}=d\left[\frac{a_0}{2}J_1+\frac{a_1}{2}\left(\frac{1}{2}J_1^2-J_2\right)
	+\frac{a_2}{2}\left(\frac{1}{3}J_1^3-J_1J_2+J_3\right)\right]\,.
\end{equation}
Therefore, one may choose the following local complementary strain-energy function:
\begin{equation}
	\zeta=\frac{a_0}{2}J_1+\frac{a_1}{2}\left(\frac{1}{2}J_1^2-J_2\right)
	+\frac{a_2}{2}\left(\frac{1}{3}J_1^3-J_1J_2+J_3\right)\,.
\end{equation}
Conversely, if any of the response functions varies with $J_3$, the complementary stress-work $1$-form fails to be closed on some region where the invariant map has rank three, and the response is not complementary hyperelastic there.

\paragraph{An explicit complementary Edelen elastic solid.}
Let $k_1>0$ and $k_2\neq0$ be material constants, and choose the following response functions
\begin{equation} \label{Explicit-J3-Coefficients}
	a_0=k_1\,, \qquad a_1=k_2J_3\,, \qquad a_2=0\,.
\end{equation}
The resulting constitutive equation reads
\begin{equation} \label{Explicit-J3-Response}
	\mathbf{C}^{\flat}=k_1\mathbf{G}+k_2J_3\mathbf{S}^{\flat}\,.
\end{equation}
This response is isotropic and compressible. In a $\mathbf{G}$-orthonormal principal basis of $\mathbf{S}$, the principal values of $\mathbf{C}^{\flat}$ are $k_1+k_2J_3s_a$, $a=1,2,3$, where $s_a$ are the corresponding principal stresses. The strain response is positive-definite precisely when $k_1+k_2J_3s_a>0$, $a=1,2,3$. Since $J_3=s_1s_2s_3$ and $\lVert\mathbf{S}\rVert_{\mathrm{op}}=\max_a|s_a|$, one has
\begin{equation}
	|J_3| \leq\lVert\mathbf{S}\rVert_{\mathrm{op}}^3\,, \qquad
	|J_3s_a| \leq\lVert\mathbf{S}\rVert_{\mathrm{op}}^4\,, \qquad a=1,2,3\,.
\end{equation}
Therefore, the condition $|k_2|\,\lVert\mathbf{S}\rVert_{\mathrm{op}}^4<k_1$ is sufficient for positive definiteness of $\mathbf{C}^\flat$. Thus, one possible connected constitutively admissible stress domain is the open ball
\begin{equation}
	\mathbb{U}_X =\left\{ \mathbf{S}\in\mathbb{S}_X: \lVert\mathbf{S}\rVert_{\mathrm{op}}
	<\left(\frac{k_1}{|k_2|}\right)^{\frac{1}{4}}\right\}\,.
\end{equation}
Substitution of \eqref{Explicit-J3-Coefficients} into \eqref{Complementary-Work-Form-J3-General} gives us
\begin{equation} \label{Explicit-J3-Work-Form}
	\boldsymbol{\Lambda}=\frac{1}{2}\left(k_1+k_2J_1J_3\right)dJ_1
	-\frac{k_2}{2}J_3\,dJ_2\,.
\end{equation}
Taking the exterior derivative, one obtains
\begin{equation} \label{Explicit-J3-Exterior-Derivative}
	d\boldsymbol{\Lambda}=-\frac{k_2}{2}J_1\,dJ_1\wedge dJ_3
	+\frac{k_2}{2}\,dJ_2\wedge dJ_3\,.
\end{equation}
On a region where the invariant map has rank three, the $2$-forms $dJ_1\wedge dJ_3$ and $dJ_2\wedge dJ_3$ are linearly independent. Since $k_2\neq0$, the second coefficient in \eqref{Explicit-J3-Exterior-Derivative} does not vanish, and hence $d\boldsymbol{\Lambda}\neq0$. The response therefore does not admit a complementary strain-energy function and is not complementary hyperelastic. Moreover, a direct calculation gives us
\begin{equation} \label{Explicit-J3-Contact-Form}
	\boldsymbol{\Lambda}\wedge d\boldsymbol{\Lambda} 
	=\frac{k_1k_2}{4}\,dJ_1\wedge dJ_2\wedge dJ_3\,.
\end{equation}
Since $k_1>0$ and $k_2\neq0$, this $3$-form does not vanish on any region where the invariant map has rank three. Therefore, the Pfaffian equation $\boldsymbol{\Lambda}=0$ possesses the local accessibility property in invariant stress space. 
By the Darboux classification of isotropic complementary Cauchy elastic solids, \eqref{Explicit-J3-Response} defines an isotropic complementary Edelen elastic solid rather than a complementary Ericksen elastic solid.

To examine the failure of major symmetry directly, let us define $\mathbf{M}=\frac{\partial J_3}{\partial\mathbf{S}}=J_2\mathbf{G}-J_1\mathbf{S}^{\flat}+\mathbf{S}^{2\flat}$. Differentiation of \eqref{Explicit-J3-Response} with respect to stress gives us
\begin{equation}
	\mathbb{A}_{ABCD}=k_2S_{AB}M_{CD}
	+\frac{k_2J_3}{2}\left(G_{AC}G_{BD}+G_{AD}G_{BC}\right)\,.
\end{equation}
Consequently, the major-antisymmetric part of the tangent compliance tensor becomes $\mathbb{A}_{ABCD}-\mathbb{A}_{CDAB}=k_2\left(S_{AB}M_{CD}-S_{CD}M_{AB}\right)$. This tensor does not vanish at a generic stress. Thus, the non-vanishing of $d\boldsymbol{\Lambda}$ in \eqref{Explicit-J3-Exterior-Derivative} and the failure of major symmetry of the local tangent compliance tensor provide equivalent tests of the failure of complementary hyperelasticity.

\subsection{Isotropic Cauchy elasticity revisited}
\label{Sec:IsotropicCauchyRevisited}

Examples of isotropic Cauchy elastic solids that are not hyperelastic are known. However, the conditions for hyperelasticity of a general isotropic Cauchy elastic response do not appear to have been systematically expressed in terms of the principal strain invariants. For completeness, and to make the parallel with the complementary Cauchy elasticity theory developed above explicit, we carry out this analysis here. In particular, we express the stress-work $1$-form in invariant strain space and derive the differential conditions on the isotropic response functions that characterize hyperelasticity. Their violation provides a systematic means of identifying isotropic Cauchy elastic responses that are not hyperelastic.

Recall that a general compressible isotropic Cauchy elastic solid has the following representation
\begin{equation}
	\mathbf{S}=b_0\mathbf{G}^{\sharp}+b_1\mathbf{C}^{\sharp}+b_2\mathbf{C}^{2\sharp}\,,
\end{equation}
where $b_i=b_i(I_1,I_2,I_3)$, $i=0,1,2$, and $I_1$, $I_2$, and $I_3$ are the principal invariants of $\mathbf{C}=\mathbf{G}^{\sharp}\mathbf{C}^{\flat}$. The associated stress-work $1$-form is written as
\begin{equation}
	\boldsymbol{\Omega}=\frac{1}{2}\mathbf{S}\!:\!d\mathbf{C}^{\flat}
	=\frac{b_0}{2}\mathbf{G}^{\sharp}\!:\!d\mathbf{C}^{\flat}
	+\frac{b_1}{2}\mathbf{C}^{\sharp}\!:\!d\mathbf{C}^{\flat}
	+\frac{b_2}{2}\mathbf{C}^{2\sharp}\!:\!d\mathbf{C}^{\flat}\,.
\end{equation}
Using the definitions of the principal invariants, one has $\mathbf{G}^{\sharp}\!:\!d\mathbf{C}^{\flat}=dI_1$ and $\mathbf{C}^{\sharp}\!:\!d\mathbf{C}^{\flat}=I_1\,dI_1-dI_2$. 
The Cayley--Hamilton identity similarly gives us $\mathbf{C}^{2\sharp}:d\mathbf{C}^{\flat}=(I_1^2-I_2)\,dI_1-I_1\,dI_2+dI_3$. Hence,
\begin{equation}
	\boldsymbol{\Omega}=\frac{1}{2}b_0\,dI_1+\frac{1}{2}b_1\big(I_1\,dI_1-dI_2\big)
	+\frac{1}{2}b_2\left[\big(I_1^2-I_2\big)\,dI_1-I_1\,dI_2+dI_3\right]\,.
\end{equation}
Equivalently,
\begin{equation}\label{General-Isotropic-Cauchy-Work-Form}
\boldsymbol{\Omega}=\frac{1}{2}\big(P\,dI_1+Q\,dI_2+R\,dI_3\big)\,,
\end{equation}
where
\begin{equation}
P=b_0+I_1b_1+\big(I_1^2-I_2\big)b_2\,,\qquad Q=-\big(b_1+I_1b_2\big)\,,\qquad R=b_2\,.
\end{equation}

Taking the exterior derivative, one obtains
\begin{equation}\label{Exterior-Derivative-General-Isotropic-Cauchy}
	d\boldsymbol{\Omega}=\frac{1}{2}\bigg(\frac{\partial Q}{\partial I_1}-\frac{\partial P}{\partial 	I_2}\bigg)dI_1\wedge dI_2+\frac{1}{2}\bigg(\frac{\partial R}{\partial I_1}-\frac{\partial P}{\partial 	I_3}\bigg)dI_1\wedge dI_3+\frac{1}{2}\bigg(\frac{\partial R}{\partial I_2}-\frac{\partial Q}{\partial 	I_3}\bigg)dI_2\wedge dI_3\,.
\end{equation}
On the open subset of strain space where the invariant map $\boldsymbol{\mathcal I}=(I_1,I_2,I_3)$ has rank three, the $1$-forms $dI_1$, $dI_2$, and $dI_3$ are linearly independent. Consequently, the three $2$-forms $dI_1\wedge dI_2$, $dI_1\wedge dI_3$, and $dI_2\wedge dI_3$ are linearly independent, and $\boldsymbol{\Omega}$ is closed if and only if
\begin{equation}\label{General-Isotropic-Cauchy-Closure-Conditions}
	\frac{\partial P}{\partial I_2}=\frac{\partial Q}{\partial I_1}\,,\qquad 
	\frac{\partial P}{\partial I_3}=\frac{\partial R}{\partial I_1}\,,\qquad 
	\frac{\partial Q}{\partial I_3}=\frac{\partial R}{\partial I_2}\,.
\end{equation}
In terms of the original response functions, these conditions become
\begin{equation}\label{General-Isotropic-Cauchy-Closure-Conditions-bi}
\begin{aligned}
	&\frac{\partial b_1}{\partial I_1}+I_1\frac{\partial b_2}{\partial I_1}
	+\frac{\partial b_0}{\partial I_2}+I_1\frac{\partial b_1}{\partial I_2}
	+(I_1^2-I_2)\frac{\partial b_2}{\partial I_2}=0\,,\\
	&\frac{\partial b_0}{\partial I_3}+I_1\frac{\partial b_1}{\partial I_3}
	+(I_1^2-I_2)\frac{\partial b_2}{\partial I_3}-\frac{\partial b_2}{\partial I_1}=0\,,\\
	&\frac{\partial b_1}{\partial I_3}+I_1\frac{\partial b_2}{\partial I_3}
	+\frac{\partial b_2}{\partial I_2}=0\,.
\end{aligned}
\end{equation}
Thus, violation of any one of \eqref{General-Isotropic-Cauchy-Closure-Conditions-bi} at a strain where the invariant map has rank three implies that the solid is not hyperelastic in a neighborhood of that strain.

Because the constitutively admissible strain domain is open in the six-dimensional strain space, the strains with three distinct principal values form an open dense subset of the admissible domain. Consequently, every strain with repeated principal values is the limit of a sequence of strains with three distinct principal values. Since the response functions $b_i$ are smooth, the coefficients in \eqref{Exterior-Derivative-General-Isotropic-Cauchy} are continuous. Therefore, if the closure conditions hold at every strain with three distinct principal values, they also hold at strains with repeated principal values by continuity.

\subsubsection{Response functions depending only on the first strain invariant}

Consider the subclass $b_i=b_i(I_1)$, $i=0,1,2$. In this case, \eqref{Exterior-Derivative-General-Isotropic-Cauchy} reduces to
\begin{equation}\label{Exterior-Derivative-I1-Dependent}
	d\boldsymbol{\Omega}=-\frac{1}{2}\big(b_1'+I_1b_2'\big)dI_1\wedge dI_2
	+\frac{1}{2}b_2'\,dI_1\wedge dI_3\,,
\end{equation}
where a prime denotes differentiation with respect to $I_1$. On a region where the invariant map has rank three, the $2$-forms $dI_1\wedge dI_2$ and $dI_1\wedge dI_3$ are linearly independent. Therefore, $\boldsymbol{\Omega}$ is closed if and only if $b_2'= b_1'+I_1b_2'=0$. The first condition implies that $b_2$ is constant. The second condition then reduces to $b_1'=0$, and hence $b_1$ is also constant. No restriction is imposed on $b_0(I_1)$. By density and continuity, the same conclusion extends throughout the constitutively admissible strain domain.

For constant $b_1$ and $b_2$, the stress-work $1$-form is exact:
\begin{equation}
	\boldsymbol{\Omega}=d\Bigg[\frac{1}{2}\int_{\mathring{I}_1}^{I_1}b_0(s)\,ds
	+\frac{b_1}{2}\bigg(\frac{1}{2}I_1^2-I_2\bigg)
	+\frac{b_2}{2}\bigg(\frac{1}{3}I_1^3-I_1I_2+I_3\bigg)\Bigg]\,.
\end{equation}
Thus, a local strain-energy function is given as
\begin{equation}
	\psi=\frac{1}{2}\int_{\mathring{I}_1}^{I_1}b_0(s)\,ds
	+\frac{b_1}{2}\bigg(\frac{1}{2}I_1^2-I_2\bigg)
	+\frac{b_2}{2}\bigg(\frac{1}{3}I_1^3-I_1I_2+I_3\bigg)\,.
\end{equation}

\paragraph{An explicit non-hyperelastic response.}
Let $k_2\neq0$ be a material constant and consider the subclass $b_0=k_1$, $b_1=k_2I_1$, $b_2=0$. The corresponding constitutive equation is
\begin{equation}\label{Explicit-Isotropic-Cauchy-Ericksen-Response}
	\mathbf{S}=k_1\mathbf{G}^{\sharp}+k_2I_1\mathbf{C}^{\sharp}\,.
\end{equation}
The corresponding stress-work $1$-form and its exterior derivative are 
\begin{equation}
	\boldsymbol{\Omega}=\frac{1}{2}\big(k_1+k_2I_1^2\big)dI_1-\frac{k_2}{2}I_1\,dI_2\,,\qquad
	d\boldsymbol{\Omega}=-\frac{k_2}{2}\,dI_1\wedge dI_2\,.
\end{equation}
Thus, $d\boldsymbol{\Omega}\neq0$ on every region where the invariant map has rank three, and the response is not hyperelastic. On the other hand, $\boldsymbol{\Omega}\wedge d\boldsymbol{\Omega}=0$. Therefore, on every region where $\boldsymbol{\Omega}\neq0$ and $d\boldsymbol{\Omega}\neq0$, the response defines an isotropic Ericksen elastic solid.
The tangent stiffness tensor is written as
\begin{equation}
	\mathbb{C}^{ABCD}=k_2C^{AB}G^{CD}
	+\frac{k_2I_1}{2}\big(G^{AC}G^{BD}+G^{AD}G^{BC}\big)\,.
\end{equation}
Its major-antisymmetric part is $\mathbb{C}^{ABCD}-\mathbb{C}^{CDAB}=k_2\big(C^{AB}G^{CD}-C^{CD}G^{AB}\big)$, which is nonzero at a generic non-dilatational strain. Thus, the failure of hyperelasticity is detected equivalently by $d\boldsymbol{\Omega}\neq0$ and by the failure of major symmetry of the tangent stiffness tensor.

\subsubsection{Response functions depending only on the third strain invariant}

Consider the isotropic Cauchy elastic response $\mathbf{S}=b_0(I_3)\mathbf{G}^{\sharp}+b_1(I_3)\mathbf{C}^{\sharp}+b_2(I_3)\mathbf{C}^{2\sharp}$. Although the scalar response functions depend only on $I_3$, the stress response itself does not. The tensors $\mathbf{C}^{\sharp}$ and $\mathbf{C}^{2\sharp}$ retain information about the principal directions and the individual principal values of the strain.
Substitution into \eqref{General-Isotropic-Cauchy-Work-Form} gives us
\begin{equation}
	\boldsymbol{\Omega}=\frac{1}{2}\big[b_0+I_1b_1+(I_1^2-I_2)b_2\big]dI_1
	-\frac{1}{2}(b_1+I_1b_2)dI_2+\frac{1}{2}b_2\,dI_3\,,
\end{equation}
where the response functions are evaluated at $I_3$. A prime will denote differentiation with respect to $I_3$. Taking the exterior derivative, one obtains
\begin{equation}
	d\boldsymbol{\Omega}=-\frac{1}{2}\big[b_0'+I_1b_1'+(I_1^2-I_2)b_2'\big]dI_1
	\wedge dI_3+\frac{1}{2}\big(b_1'+I_1b_2'\big)dI_2\wedge dI_3\,.
\end{equation}
On an open region where the invariant map has rank three, $\boldsymbol{\Omega}$ is closed if and only if
\begin{equation}\label{Closure-Conditions-I3-General}
	b_0'+I_1b_1'+(I_1^2-I_2)b_2'=0\,,\qquad b_1'+I_1b_2'=0\,.
\end{equation}
Since $I_1$, $I_2$, and $I_3$ are locally independent on such a region, $I_2$ may be varied while $I_1$ and $I_3$ are held fixed. The first identity is affine in $I_2$, and therefore its coefficient of $I_2$ must vanish. Hence $b_2'=0$. The second identity then implies $b_1'=0$, and substitution into the first gives $b_0'=0$. Thus, the stress-work $1$-form is closed if and only if all three response functions are constant.
For constant $b_0$, $b_1$, and $b_2$, one may choose the following local strain-energy function
\begin{equation}
	\psi=\frac{b_0}{2}I_1+\frac{b_1}{2}\bigg(\frac{1}{2}I_1^2-I_2\bigg)
	+\frac{b_2}{2}\bigg(\frac{1}{3}I_1^3-I_1I_2+I_3\bigg)\,.
\end{equation}
Conversely, if any of the response functions varies with $I_3$, the stress-work $1$-form fails to be closed on some region where the invariant map has rank three, and the response is not hyperelastic there.

\paragraph{An explicit Edelen elastic solid.}
Let $k_2\neq0$ and choose the following subclass: $b_0=k_1$, $b_1=k_2I_3$, $b_2=0$. The resulting constitutive equation is
\begin{equation}\label{Explicit-I3-Cauchy-Response}
	\mathbf{S}=k_1\mathbf{G}^{\sharp}+k_2I_3\mathbf{C}^{\sharp}\,.
\end{equation}
The corresponding stress-work $1$-form and its exterior derivative are
\begin{equation}
	\boldsymbol{\Omega}=\frac{1}{2}\big(k_1+k_2I_1I_3\big)dI_1-\frac{k_2}{2}I_3\,dI_2\,,\qquad
	d\boldsymbol{\Omega}=-\frac{k_2}{2}I_1\,dI_1\wedge dI_3+\frac{k_2}{2}\,dI_2\wedge dI_3\,.
\end{equation}
On a region where the invariant map has rank three, $d\boldsymbol{\Omega}\neq0$. Moreover,
\begin{equation}\label{Explicit-I3-Cauchy-Contact-Form}
	\boldsymbol{\Omega}\wedge d\boldsymbol{\Omega}
	=\frac{k_1k_2}{4}\,dI_1\wedge dI_2\wedge dI_3\,.
\end{equation}
Hence, when $k_1k_2\neq0$, this $3$-form does not vanish on any region where the invariant map has rank three. Since $\boldsymbol{\Omega}\wedge d\boldsymbol{\Omega}\neq0$, the Pfaffian equation $\boldsymbol{\Omega}=0$ does not satisfy the Frobenius integrability condition. Thus, by the Darboux classification of isotropic Cauchy elastic solids, \eqref{Explicit-I3-Cauchy-Response} defines an isotropic Edelen elastic solid rather than an Ericksen elastic solid.

\section{Internal Constraints in Three-Dimensional Elasticity}
\label{Sec:InternalConstraints3DElasticity}

In this section, we study two distinct types of constitutive constraints: kinematic constraints, which restrict the admissible strain states, and stress constraints, which restrict the admissible stress states. We first discuss kinematic constraints and their associated reactive stresses, and then turn to stress constraints and the restrictions they impose on the admissible stress space. As will be seen, kinematic constraints are more easily incorporated in Cauchy elasticity, whereas stress constraints are more easily treated within complementary Cauchy elasticity.

\subsection{Kinematic constraints}

\paragraph{A brief history of kinematic constraints in elasticity.}
The study of kinematic constraints in elasticity has a long history, dating back to the nineteenth century.
In the context of linearized elasticity, \citet{Poincare1892} considered internal constraints expressed as functions of the displacement gradients and derived the corresponding equilibrium equations using the principle of virtual work and Lagrange multipliers.
In his foundational work on finite elasticity, \citet{Rivlin1948} developed a constitutive framework for isotropic hyperelastic solids, including incompressible materials. For incompressible solids, the stress contains an arbitrary hydrostatic pressure that is not determined by the strain-energy function.
\citet{EricksenRivlin1954} extended the theory of finite elasticity to materials subject to internal kinematic constraints more general than incompressibility. They formulated the corresponding constitutive equations and showed how the constraints give rise to indeterminate reaction stresses represented by Lagrange multipliers.
\citet{Spencer1972} developed a systematic theory of the finite deformation of fiber-reinforced elastic materials, emphasizing the mechanical consequences of fiber inextensibility. 
\citet{CohenWang1987} formulated the constitutive response of internally constrained elastic materials in terms of affine spaces of admissible stresses. They investigated material frame-indifference and material symmetry, establishing conditions under which a determinate-stress representative can satisfy both requirements.

Let $\mathbf{C}=\mathbf{F}^{\mathsf T}\mathbf{F}\in\operatorname{Sym}^{+}(3)$. At each material point, $\mathbf{C}$ has six independent components. Internal constraints may be associated intrinsically with material line (one-dimensional), surface (two-dimensional), and volume (three-dimensional) elements.

\subsubsection{Fiber inextensibility constraint}

Fiber inextensibility is a classical internal constraint in nonlinear elasticity and has been studied extensively in the mechanics of fiber-reinforced solids \citep{Adkins1955,RivlinEricksen1955,Wang1969,Boehler1977,TruesdellNoll2004,Saccomandi2002}. The constraint requires preservation of length along the prescribed fiber direction and gives rise to an associated reactive stress, represented through a Lagrange multiplier that is commonly interpreted as a fiber tension.

For a material fiber with unit tangent $\mathbf{N}$, the only intrinsic geometric quantity is length. Prescribed length is expressed as $\llangle \mathbf{N},\mathbf{C}\mathbf{N}\rrangle_{\mathbf{G}}=\lambda_0^2$, where $(\mathbf{C}\mathbf{N})^A=C^A{}_B N^B$. Inextensibility corresponds to $\lambda_0=1$. Let us define the constraint function $\phi_{\mathsf{F}}(\mathbf{C})=\llangle \mathbf{N},\mathbf{C}\mathbf{N}\rrangle_{\mathbf{G}}-\lambda_0^2$. Its derivative with respect to $\mathbf{C}^{\flat}$ is
\begin{equation}
	\phi_{\mathsf{F},\mathbf{C}^{\flat}}=\frac{\partial\phi_{\mathsf{F}}}{\partial\mathbf{C}^{\flat}}
	=\mathbf{N}\otimes\mathbf{N}\,.
\end{equation}
Thus, the reaction stress space associated with the constraint is given by $\mathcal{N}_{\mathsf{F}}(\mathbf{N})=\operatorname{span}\{\mathbf{N}\otimes\mathbf{N}\}$.
Physically, this space characterizes the admissible reaction stresses associated with the fiber constraint. Since admissible changes of $\mathbf{C}$ satisfy the constraint, the reactive stress does no work on such changes. For an inextensible fiber, the reaction stress is consequently proportional to $\mathbf{N}\otimes\mathbf{N}$ and represents the indeterminate fiber tension or compression required to enforce the prescribed fiber length. Thus, the reactive stress associated with the fiber constraint has the form $\mathbf{S}_{r}=T_0\,\mathbf{N}\otimes\mathbf{N}$, where $T_0=T_0(X)$ is an indeterminate tension field that is determined as part of the solution of the boundary-value problem.

Hence, one fiber family imposes one scalar constraint. Two fiber families with unit tangents $\mathbf{N}_1$ and $\mathbf{N}_2$ are pointwise linearly independent if $\mathbf{N}_1$ and $\mathbf{N}_2$ are not parallel. More generally, $p$ fiber families with unit tangents $\mathbf{N}_1,\ldots,\mathbf{N}_p$ are pointwise linearly independent if the vectors $\mathbf{N}_1,\ldots,\mathbf{N}_p$ are linearly independent at each material point, i.e., $\sum_{\alpha=1}^{p}a_\alpha\mathbf{N}_\alpha=\mathbf{0}$ implies $a_\alpha=0$ for every $\alpha$. Consequently, in three dimensions there can be at most three pointwise linearly independent fiber families.
For two linearly independent fiber families, the reaction stress has the form
\begin{equation}
    \mathbf{S}_{\mathrm{R}}
    =T_1\,\mathbf{N}_1\otimes\mathbf{N}_1
    +T_2\,\mathbf{N}_2\otimes\mathbf{N}_2\,,
\end{equation}
where $T_1=T_1(X)$ and $T_2=T_2(X)$ are independent, indeterminate fiber tension fields. Similarly, for three linearly independent fiber families, the reaction stress is
\begin{equation}
    \mathbf{S}_{\mathrm{R}}
    =T_1\,\mathbf{N}_1\otimes\mathbf{N}_1
    +T_2\,\mathbf{N}_2\otimes\mathbf{N}_2
    +T_3\,\mathbf{N}_3\otimes\mathbf{N}_3\,,
\end{equation}
where $T_1=T_1(X)$, $T_2=T_2(X)$, and $T_3=T_3(X)$ are independent, indeterminate fiber tension fields. Each tension field may take positive or negative values, corresponding to tensile or compressive reaction stresses.

\subsubsection{Surface-area preservation constraint}

Let $\mathbf{N}$ denote the reference unit normal to a material surface, so that $\llangle\mathbf{N},\mathbf{N}\rrangle_{\mathbf{G}}=1$, and let $\mathbf{N}^{\flat}=\mathbf{G}\mathbf{N}$ be its metric dual, with components $N_A=G_{AB}N^B$. The projection tensor
\begin{equation}
	\boldsymbol{\mathsf{P}}=\mathbf{G}^{\sharp}-\mathbf{N}\otimes\mathbf{N}\,,\qquad 
	\mathsf{P}^{AB}=G^{AB}-N^A N^B\,,
\end{equation}
represents the restriction of the inverse material metric to the tangent plane of the surface. The intrinsic geometric quantity associated with a material surface is its area. We first derive the corresponding surface area stretch.

Let $dA$ and $da$ denote the reference and current area elements, respectively. Nanson's formula reads $\mathbf{n}^{\flat}\,da=J\,\mathbf{F}^{-\star}\mathbf{N}^{\flat}\,dA$, where $\mathbf{F}^{-\star}:T_X^*\mathcal{B}\rightarrow T_x^*\mathcal{S}$ denotes the inverse dual of $\mathbf{F}$ and $\mathbf{n}^{\flat}$ is the spatial unit conormal. Since $\|\mathbf{n}^{\flat}\|_{\mathbf{g}^{\sharp}}=1$, taking the norm induced by $\mathbf{g}^{\sharp}$ gives us
\begin{equation}
	J_s\coloneq\frac{da}{dA}=J\,\|\mathbf{F}^{-\star}\mathbf{N}^{\flat}\|_{\mathbf{g}^{\sharp}}\,.
\end{equation}
The square of the norm appearing in this expression is given as
\begin{equation}
	\|\mathbf{F}^{-\star}\mathbf{N}^{\flat}\|_{\mathbf{g}^{\sharp}}^2
	=\llangle \mathbf{F}^{-\star}\mathbf{N}^{\flat},
	\mathbf{F}^{-\star}\mathbf{N}^{\flat}\rrangle_{\mathbf{g}^{\sharp}}
	=\llangle \mathbf{N}^{\flat},\mathbf{N}^{\flat}\rrangle_{\mathbf{C}^{-\sharp}}\,.
\end{equation}
Therefore, the surface area stretch is written as
\begin{equation}
	J_s=J\sqrt{\llangle\mathbf{N}^{\flat},\mathbf{N}^{\flat}\rrangle_{\mathbf{C}^{-\sharp}}}
	=J\sqrt{C^{-AB}N_A N_B}\,.
\end{equation}
Prescribed area stretch is expressed as $J_s=J_{s0}$, and surface incompressibility corresponds to $J_s=1$. Let us define the constraint function
\begin{equation}
	\phi_{\mathsf{A}}(\mathbf{C})
	=J\sqrt{\llangle\mathbf{N}^{\flat},\mathbf{N}^{\flat}\rrangle_{\mathbf{C}^{-\sharp}}}-J_{s0}\,.
\end{equation}
To characterize the reaction stress space associated with the constraint, let $q=C^{-AB}N_A N_B$, so that $J_s=J\sqrt{q}$. Differentiation with respect to $C_{AB}$ gives
\begin{equation}
	\frac{\partial J}{\partial C_{AB}}=\frac{J}{2}C^{-AB}\,,\qquad
	\frac{\partial q}{\partial C_{AB}}=-C^{-AC}N_C\,C^{-BD}N_D\,.
\end{equation}
It follows that
\begin{equation}
	\left(\phi_{\mathsf{A},\mathbf{C}^{\flat}}\right)^{AB}
	=\frac{\partial\phi_{\mathsf{A}}}{\partial C_{AB}}
	=\frac{J_s}{2}\left[C^{-AB}-\frac{C^{-AC}N_C\,C^{-BD}N_D}{C^{-CD}N_C N_D}\right]\,.
\end{equation}
Equivalently, the derivative can be written intrinsically as
\begin{equation}
	\phi_{\mathsf{A},\mathbf{C}^{\flat}}
	=\frac{J_s}{2}\left[\mathbf{C}^{-\sharp}-\frac{\left(\mathbf{C}^{-\sharp}\mathbf{N}^{\flat}\right)
	\otimes\left(\mathbf{C}^{-\sharp}\mathbf{N}^{\flat}\right)}
	{\llangle\mathbf{N}^{\flat},\mathbf{N}^{\flat}\rrangle_{\mathbf{C}^{-\sharp}}}\right]\,.
\end{equation}
Thus, the reaction stress space associated with the constraint is given by
$\mathcal{N}_{\mathsf{A}}(\mathbf{N};\mathbf{C})=\operatorname{span}\{\phi_{\mathsf{A},\mathbf{C}^{\flat}}\}$. Physically, this one-dimensional normal space specifies the direction of the reactive stress required to enforce the prescribed surface area. The reactive stress therefore has the form $\mathbf{S}_r=T_s\,\phi_{\mathsf{A},\mathbf{C}^{\flat}}$, where $T_s=T_s(X)$ is an indeterminate scalar field determined as part of the solution of the boundary-value problem. Hence, a surface-area constraint imposes one scalar constraint.

\subsubsection{Surface-conformality constraint}

Preservation of all intrinsic angles in a material surface requires its induced metric to remain conformal to the reference surface metric. To see this, let us consider two arbitrary nonzero tangent vectors $\mathbf{U}$ and $\mathbf{V}$ to the material surface. In the reference configuration, the angle $\Theta$ between them is given as
\begin{equation}
	\cos\Theta
	=\frac{\llangle\mathbf{U},\mathbf{V}\rrangle_{\mathbf{G}}}
	{\sqrt{\llangle\mathbf{U},\mathbf{U}\rrangle_{\mathbf{G}}\,
	\llangle\mathbf{V},\mathbf{V}\rrangle_{\mathbf{G}}}}\,.
\end{equation}
After deformation, their images are $\mathbf{F}\mathbf{U}$ and $\mathbf{F}\mathbf{V}$, and hence the angle $\theta$ between them is given as
\begin{equation}
	\cos\theta
	=\frac{\llangle\mathbf{U},\mathbf{C}\mathbf{V}\rrangle_{\mathbf{G}}}
	{\sqrt{\llangle\mathbf{U},\mathbf{C}\mathbf{U}\rrangle_{\mathbf{G}}\,
	\llangle\mathbf{V},\mathbf{C}\mathbf{V}\rrangle_{\mathbf{G}}}}\,.
\end{equation}
A standard characterization of conformal linear maps between inner-product spaces states that preservation of all angles is equivalent to preservation of the metric up to a positive scalar factor. In the present two-dimensional setting, this result can be seen directly. Let $\{\mathbf{E}_1,\mathbf{E}_2\}$ be an orthonormal basis of the reference tangent plane, so that $\llangle\mathbf{E}_\alpha,\mathbf{E}_\beta\rrangle_{\mathbf{G}}=\delta_{\alpha\beta}$, $\alpha,\beta=1,2$. Since $\mathbf{E}_1$ and $\mathbf{E}_2$ are orthogonal, angle preservation requires their images to remain orthogonal, and hence
\begin{equation}
	\llangle \mathbf{F}\mathbf{E}_1,\mathbf{F}\mathbf{E}_2\rrangle_{\mathbf{g}}
	=\llangle\mathbf{E}_1,\mathbf{C}\mathbf{E}_2\rrangle_{\mathbf{G}}=0\,.
\end{equation}
Moreover, $\mathbf{E}_1+\mathbf{E}_2$ and $\mathbf{E}_1-\mathbf{E}_2$ are orthogonal in the reference configuration. Their images must therefore also be orthogonal, and hence
\begin{equation}
	\llangle\mathbf{E}_1+\mathbf{E}_2,\mathbf{C}(\mathbf{E}_1
	-\mathbf{E}_2)\rrangle_{\mathbf{G}}=0\,.
\end{equation}
Using the symmetry of $\mathbf{C}^{\flat}$ and the preceding orthogonality condition, it follows that
\begin{equation}
	\llangle\mathbf{E}_1,\mathbf{C}\mathbf{E}_1\rrangle_{\mathbf{G}}
	=\llangle\mathbf{E}_2,\mathbf{C}\mathbf{E}_2\rrangle_{\mathbf{G}}\,.
\end{equation}
Thus, the two orthonormal tangent directions remain orthogonal and undergo the same local stretch. Let us denote this stretch by $\lambda_s>0$. Thus,
\begin{equation}
	\llangle\mathbf{E}_\alpha,\mathbf{C}\mathbf{E}_\beta\rrangle_{\mathbf{G}}
	=\lambda_s^2\,\delta_{\alpha\beta}\,,\qquad \alpha,\beta=1,2\,.
\end{equation}
By bilinearity, it follows for arbitrary tangent vectors $\mathbf{U}$ and $\mathbf{V}$ that
\begin{equation}
	\llangle\mathbf{U},\mathbf{C}\mathbf{V}\rrangle_{\mathbf{G}}
	=\lambda_s^2\llangle\mathbf{U},\mathbf{V}\rrangle_{\mathbf{G}}\,.
\end{equation}
Therefore, preservation of all intrinsic angles is equivalent to the metric induced by the deformation on the material surface being an isotropic scaling of the reference surface metric. This is precisely the surface-conformality condition. The local surface stretch $\lambda_s=\lambda_s(X)$ is not prescribed by this constraint and remains indeterminate.

The surface-conformality constraint can therefore be represented by the two independent constraint functions
\begin{equation}
	\phi_{\mathsf{C}}^1(\mathbf{C})
	=\llangle\mathbf{E}_1,\mathbf{C}\mathbf{E}_1\rrangle_{\mathbf{G}}
	-\llangle\mathbf{E}_2,\mathbf{C}\mathbf{E}_2\rrangle_{\mathbf{G}}\,,\qquad
	\phi_{\mathsf{C}}^2(\mathbf{C})
	=\llangle\mathbf{E}_1,\mathbf{C}\mathbf{E}_2\rrangle_{\mathbf{G}}\,.
\end{equation}
Their derivatives with respect to $\mathbf{C}^{\flat}$ are
\begin{equation}
	\phi_{\mathsf{C},\mathbf{C}^{\flat}}^1
	=\mathbf{E}_1\otimes\mathbf{E}_1-\mathbf{E}_2\otimes\mathbf{E}_2\,,\qquad
	\phi_{\mathsf{C},\mathbf{C}^{\flat}}^2
	=\frac{1}{2}\left(\mathbf{E}_1\otimes\mathbf{E}_2+\mathbf{E}_2\otimes\mathbf{E}_1\right)\,.
\end{equation}
Thus, $\phi_{\mathsf{C},\mathbf{C}^{\flat}}^1$ and $\phi_{\mathsf{C},\mathbf{C}^{\flat}}^2$ are independent, and the reaction stress space is defined as
\begin{equation}
	\mathcal{N}_{\mathsf{C}}(\mathbf{N})
	=\operatorname{span}\Big\{
	\mathbf{E}_1\otimes\mathbf{E}_1-\mathbf{E}_2\otimes\mathbf{E}_2,\,
	\mathbf{E}_1\otimes\mathbf{E}_2+\mathbf{E}_2\otimes\mathbf{E}_1\Big\}\,.
\end{equation}
Physically, the first normal direction generates a reactive stress that opposes unequal stretching of the two tangent directions, whereas the second generates a reactive shear stress that prevents a change in the angle between them. Neither reaction opposes an equal isotropic stretching of both tangent directions, consistently with the fact that the local surface stretch $\lambda_s$ remains unrestricted by the conformality constraint. The reactive stress can therefore be written as
\begin{equation}
\mathbf{S}_r
=T_1\left(\mathbf{E}_1\otimes\mathbf{E}_1-\mathbf{E}_2\otimes\mathbf{E}_2\right)
+T_2\left(\mathbf{E}_1\otimes\mathbf{E}_2+\mathbf{E}_2\otimes\mathbf{E}_1\right)\,,
\end{equation}
where $T_1=T_1(X)$ and $T_2=T_2(X)$ are indeterminate scalar fields determined as part of the solution of the boundary-value problem. Hence, surface conformality imposes two independent scalar constraints while leaving one intrinsic surface-deformation mode, namely isotropic surface dilation, unrestricted.

\subsubsection{Surface inextensibility constraint}

An intrinsically inextensible material surface preserves the lengths of all tangent vectors and, consequently, its complete intrinsic metric. Thus, for every tangent vector $\mathbf{U}$, $\llangle\mathbf{U},\mathbf{C}\mathbf{U}\rrangle_{\mathbf{G}}=\llangle\mathbf{U},\mathbf{U}\rrangle_{\mathbf{G}}$. By polarization,\footnote{The inner product of two tangent vectors is determined entirely by their lengths. Using the symmetry and bilinearity of the inner product, one has
\begin{equation}
	\llangle\mathbf{U}+\mathbf{V},\mathbf{U}+\mathbf{V}\rrangle_{\mathbf{G}}
	=\llangle\mathbf{U},\mathbf{U}\rrangle_{\mathbf{G}}
	+2\llangle\mathbf{U},\mathbf{V}\rrangle_{\mathbf{G}}
	+\llangle\mathbf{V},\mathbf{V}\rrangle_{\mathbf{G}}\,.
\end{equation}
Solving this relation for the inner product of $\mathbf{U}$ and $\mathbf{V}$ gives us
\begin{equation}
	\llangle\mathbf{U},\mathbf{V}\rrangle_{\mathbf{G}}
	=\frac{1}{2}\Big[\llangle\mathbf{U}+\mathbf{V},\mathbf{U}+\mathbf{V}\rrangle_{\mathbf{G}}
	-\llangle\mathbf{U},\mathbf{U}\rrangle_{\mathbf{G}}
	-\llangle\mathbf{V},\mathbf{V}\rrangle_{\mathbf{G}}\Big]\,.
\end{equation}
This relation is known as the polarization identity. The quantities on the right-hand side are the squared lengths of $\mathbf{U}+\mathbf{V}$, $\mathbf{U}$, and $\mathbf{V}$, respectively. Therefore, if the lengths of all tangent vectors are preserved by the deformation, then, in particular, the lengths of $\mathbf{U}$, $\mathbf{V}$, and $\mathbf{U}+\mathbf{V}$ are preserved. The polarization identity then implies that the inner product of $\mathbf{U}$ and $\mathbf{V}$ is also preserved. Since this holds for every pair of tangent vectors, preservation of the lengths of all tangent vectors implies preservation of all their pairwise inner products and hence preservation of the complete intrinsic metric of the material surface.} this is equivalent to preservation of the inner product of every pair of tangent vectors, $\llangle\mathbf{U},\mathbf{C}\mathbf{V}\rrangle_{\mathbf{G}}=\llangle\mathbf{U},\mathbf{V}\rrangle_{\mathbf{G}}$, for all tangent vectors $\mathbf{U}$ and $\mathbf{V}$. Hence, intrinsic inextensibility preserves not only tangent lengths but also all intrinsic angles and surface area.

Let $\{\mathbf{E}_1,\mathbf{E}_2\}$ be the orthonormal basis of the reference tangent plane introduced above. The intrinsic-inextensibility condition is then equivalent to
\begin{equation}
	\llangle\mathbf{E}_1,\mathbf{C}\mathbf{E}_1\rrangle_{\mathbf{G}}=1\,,\qquad
	\llangle\mathbf{E}_2,\mathbf{C}\mathbf{E}_2\rrangle_{\mathbf{G}}=1\,,\qquad
	\llangle\mathbf{E}_1,\mathbf{C}\mathbf{E}_2\rrangle_{\mathbf{G}}=0\,.
\end{equation}
Let us define the following three constraint functions
\begin{equation}
	\phi_{\mathsf{R}}^1(\mathbf{C})
	=\llangle\mathbf{E}_1,\mathbf{C}\mathbf{E}_1\rrangle_{\mathbf{G}}-1\,,\qquad
	\phi_{\mathsf{R}}^2(\mathbf{C})
	=\llangle\mathbf{E}_2,\mathbf{C}\mathbf{E}_2\rrangle_{\mathbf{G}}-1\,,\qquad
	\phi_{\mathsf{R}}^3(\mathbf{C})
	=\llangle\mathbf{E}_1,\mathbf{C}\mathbf{E}_2\rrangle_{\mathbf{G}}\,.
\end{equation}
Their derivatives with respect to $\mathbf{C}^{\flat}$ are given as
\begin{equation}
	\phi_{\mathsf{R},\mathbf{C}^{\flat}}^1=\mathbf{E}_1\otimes\mathbf{E}_1\,,\qquad
	\phi_{\mathsf{R},\mathbf{C}^{\flat}}^2=\mathbf{E}_2\otimes\mathbf{E}_2\,,\qquad
	\phi_{\mathsf{R},\mathbf{C}^{\flat}}^3
	=\frac{1}{2}\left(\mathbf{E}_1\otimes\mathbf{E}_2+\mathbf{E}_2\otimes\mathbf{E}_1\right)\,.
\end{equation}
The three derivatives $\phi_{\mathsf{R},\mathbf{C}^{\flat}}^1$, $\phi_{\mathsf{R},\mathbf{C}^{\flat}}^2$, and $\phi_{\mathsf{R},\mathbf{C}^{\flat}}^3$ are linearly independent and span the conormal space of the constraint manifold, defined as the subspace of the ambient cotangent space consisting of covectors that annihilate all tangent vectors to the constraint manifold.
The reaction stress space is therefore defined as
\begin{equation}
	\mathcal{N}_{\mathsf{R}}(\mathbf{N})=\operatorname{span}\Big\{
	\mathbf{E}_1\otimes\mathbf{E}_1,\,\mathbf{E}_2\otimes\mathbf{E}_2,\,
	\mathbf{E}_1\otimes\mathbf{E}_2+\mathbf{E}_2\otimes\mathbf{E}_1\Big\}\,.
\end{equation}
Physically, this three-dimensional normal space contains all symmetric reactive-stress directions acting intrinsically within the tangent plane. The first two directions provide the reactive stresses required to prevent extension or contraction along the two tangent directions, while the third provides the reactive shear stress required to prevent a change in the angle between them. The reactive stress can therefore be written as
\begin{equation}
	\mathbf{S}_r=T_1\,\mathbf{E}_1\otimes\mathbf{E}_1
	+T_2\,\mathbf{E}_2\otimes\mathbf{E}_2
	+T_3\left(\mathbf{E}_1\otimes\mathbf{E}_2+\mathbf{E}_2\otimes\mathbf{E}_1\right)\,,
\end{equation}
where $T_1=T_1(X)$, $T_2=T_2(X)$, and $T_3=T_3(X)$ are indeterminate scalar fields determined as part of the solution of the boundary-value problem. 
Hence, surface inextensibility imposes three independent scalar constraints corresponding to preservation of the intrinsic surface metric. Equivalently, these constraints enforce both surface-area preservation and surface conformality, and therefore preserve intrinsic lengths and angles. They, however, do not restrict extrinsic bending of the surface.

\subsubsection{Volume constraint}

The intrinsic scalar measure associated with a three-dimensional material volume element is its volume ratio $J=\sqrt{\det\mathbf{C}}$. Prescribed volume is expressed as $J=J_0$, with incompressibility corresponding to $J_0=1$. Let us define the constraint function
\begin{equation}
	\phi_{\mathsf{V}}(\mathbf{C})=J-J_0=\sqrt{\det\mathbf{C}}-J_0\,.
\end{equation}
Its derivative with respect to $\mathbf{C}^{\flat}$ is
\begin{equation}
	\phi_{\mathsf{V},\mathbf{C}^{\flat}}=\frac{\partial\phi_{\mathsf{V}}}{\partial\mathbf{C}^{\flat}}
	=\frac{J}{2}\mathbf{C}^{-\sharp}\,,\qquad
	\left(\phi_{\mathsf{V},\mathbf{C}^{\flat}}\right)^{AB}=\frac{J}{2}C^{-AB}\,.
\end{equation}
Thus, the reaction stress space associated with the constraint is defined as
$\mathcal{N}_{\mathsf{V}}(\mathbf{C})=\operatorname{span}\left\{\mathbf{C}^{-\sharp}\right\}$.
The reactive stress therefore has the familiar form $\mathbf{S}_r=-p\,\mathbf{C}^{-\sharp}$, where $p=p(X)$ is an indeterminate scalar field determined as part of the solution of the boundary-value problem. Hence, a volume constraint imposes one scalar constraint. For $J_0=1$, this reduces to the classical incompressibility constraint.

\subsubsection{Combinations of internal constraints}
\label{Sec:Combinations-Internal-Constraints}

The internal constraints discussed above may occur simultaneously. A material may contain several families of inextensible fibers and material surfaces subject to different constraints, together with a volume constraint. For example, incompressibility may coexist with fiber inextensibility, surface-area preservation, surface conformality, or surface inextensibility, and fiber and surface constraints may likewise occur together. The number of independent constraints, and hence the dimension of the corresponding reaction stress space, depends on the number and relative orientations of the fibers and surfaces involved and on possible dependencies among the individual constraints. No attempt is made here to classify all possible internal constraints or their combinations. The preceding examples are intended to illustrate the range of constraints that may arise beyond the classical cases of incompressibility and fiber inextensibility, as well as the possibility of imposing several such constraints simultaneously.

Suppose that a material is subject simultaneously to a collection of fiber inextensibility, surface-area preservation, surface-conformality, and surface inextensibility constraints, and possibly a volume constraint.
At an admissible deformation $\mathbf{C}$, each scalar constraint gives rise to a reactive stress proportional to the corresponding tensor obtained by differentiation of the constraint with respect to $\mathbf{C}^{\flat}$. The total reactive stress is therefore the sum of the reactive stresses associated with all the imposed scalar constraints and may be written as
\begin{equation}
	\mathbf{S}_r=\sum_{\alpha=1}^{m}\lambda_\alpha\,	\phi_{\alpha,\mathbf{C}^{\flat}}\,,
\end{equation}
where $\phi_\alpha(\mathbf{C})=0$, $\alpha=1,\ldots,m$, denotes the complete collection of imposed scalar constraints and $\lambda_\alpha$ are the corresponding indeterminate scalar fields. Accordingly, the reaction stress space associated with the complete system of constraints is defined as
\begin{equation}
	\mathcal{N}_{\mathrm{tot}}(\mathbf{C})=\operatorname{span}\left\{
	\phi_{1,\mathbf{C}^{\flat}},\ldots,\phi_{m,\mathbf{C}^{\flat}}	\right\}\,,
\end{equation}
so that $\mathbf{S}_r\in\mathcal{N}_{\mathrm{tot}}(\mathbf{C})$. The dimension of this space is the number of linearly independent tensors among $\phi_{\alpha,\mathbf{C}^{\flat}}$, and therefore need not equal the total number of imposed scalar constraints when some of their associated reactive stresses are linearly dependent. The indeterminate fields $\lambda_\alpha$ are determined, together with the deformation and stress fields, from the governing equations, the imposed constraints, and the boundary conditions.

Although several scalar constraints may be imposed simultaneously, they need not all be independent at a given admissible deformation. For a collection of constraints $\phi_\alpha(\mathbf{C})=0$, $\alpha=1,\ldots,m$, their local independence is determined by the linear independence of the tensors $\phi_{\alpha,\mathbf{C}^{\flat}}$. If these tensors are linearly dependent, the corresponding constraints are locally dependent and some of them do not contribute additional independent reactive stresses. It follows that the number of independent constraints at $\mathbf{C}$ is given by
\begin{equation}
	r(\mathbf{C})=\dim\operatorname{span}\left\{\phi_{1,\mathbf{C}^{\flat}},\ldots,
	\phi_{m,\mathbf{C}^{\flat}}	\right\}=\dim\mathcal{N}_{\mathrm{tot}}(\mathbf{C})	\leq 6\,,
\end{equation}
where the upper bound follows from the six-dimensionality of the space of symmetric second-order tensors in three dimensions. Thus, the number of independent constraints is determined not simply by the number of constraint equations imposed, but by the rank of their derivatives at the deformation under consideration. This rank may depend on the geometry and relative orientations of the constrained fibers and surfaces and, in general, may vary with the deformation.

For example, consider an incompressible solid reinforced by a family of inextensible fibers with unit tangent $\mathbf{N}$. The constraints are $J=1$ and $\llangle\mathbf{N},\mathbf{C}\mathbf{N}\rrangle_{\mathbf{G}}=1$. The corresponding reaction stress tensors are generated by $\mathbf{N}\otimes\mathbf{N}$ and $\mathbf{C}^{-\sharp}$, respectively. To verify their linear independence, suppose that $a\,\mathbf{N}\otimes\mathbf{N}+b\,\mathbf{C}^{-\sharp}=\mathbf{0}$. If $b=0$, then $a\,\mathbf{N}\otimes\mathbf{N}=\mathbf{0}$, and since $\mathbf{N}\neq\mathbf{0}$, one has $a=0$. A nontrivial linear dependence would therefore require $b\neq0$, in which case $\mathbf{C}^{-\sharp}=-\frac{a}{b}\,\mathbf{N}\otimes\mathbf{N}$. This is impossible because $\mathbf{C}^{-\sharp}$ is nonsingular and has rank three, whereas $\mathbf{N}\otimes\mathbf{N}$ has rank one. Hence $a=b=0$, and the two tensors are linearly independent. The reaction stress space is therefore two-dimensional and is given by
\begin{equation}
	\mathcal{N}_{\mathrm{tot}}(\mathbf{C})
	=\operatorname{span}\left\{\mathbf{N}\otimes\mathbf{N},
	\mathbf{C}^{-\sharp}\right\}.
\end{equation}
Consequently, the most general reactive stress associated with the simultaneous imposition of the two constraints is
\begin{equation}
	\mathbf{S}_r=T_0\,\mathbf{N}\otimes\mathbf{N}
	-p\,\mathbf{C}^{-\sharp}\,,
\end{equation}
where $T_0=T_0(X)$ is the indeterminate tension field associated with fiber inextensibility and $p=p(X)$ is the indeterminate scalar field associated with incompressibility.

As a second example, consider a family of conformally constrained material surfaces together with a family of inextensible fibers that is everywhere transverse to the surfaces. Let $\{\mathbf{E}_1,\mathbf{E}_2\}$ be an orthonormal basis of the tangent planes to the material surfaces, and let $\mathbf{L}=\mathbf{E}_1\times\mathbf{E}_2$ denote the corresponding unit normal. Let $\mathbf{N}$ denote the unit tangent to the fibers. Transversality means that the fibers are nowhere tangent to the material surfaces, or equivalently, $\llangle\mathbf{N},\mathbf{L}\rrangle_{\mathbf{G}}\neq0$. The material is simultaneously subject to the surface-conformality constraint and the fiber-inextensibility constraint. The three corresponding reaction stress generators are
\begin{equation}
	\mathbf{A}_1=\mathbf{E}_1\otimes\mathbf{E}_1-\mathbf{E}_2\otimes\mathbf{E}_2\,,	\qquad
	\mathbf{A}_2=\mathbf{E}_1\otimes\mathbf{E}_2+\mathbf{E}_2\otimes\mathbf{E}_1\,,	\qquad
	\mathbf{A}_3=\mathbf{N}\otimes\mathbf{N}\,.
\end{equation}
These three tensors are linearly independent whenever the fibers are transverse to the surfaces. To see this, suppose that $a\,\mathbf{A}_1+b\,\mathbf{A}_2+c\,\mathbf{A}_3=\mathbf{0}$. Contracting this relation twice with $\mathbf{L}$ and using $\llangle\mathbf{E}_1,\mathbf{L}\rrangle_{\mathbf{G}}=\llangle\mathbf{E}_2,\mathbf{L}\rrangle_{\mathbf{G}}=0$ gives us $c\,\llangle\mathbf{N},\mathbf{L}\rrangle_{\mathbf{G}}^2=0$. Transversality implies $\llangle\mathbf{N},\mathbf{L}\rrangle_{\mathbf{G}}\neq0$, and hence $c=0$. The remaining relation is $a\,\mathbf{A}_1+b\,\mathbf{A}_2=\mathbf{0}$. Contracting this relation twice with $\mathbf{E}_1$ and using the orthonormality of $\{\mathbf{E}_1,\mathbf{E}_2\}$ gives $a=0$, since $\mathbf{A}_1(\mathbf{E}_1,\mathbf{E}_1)=1$ and $\mathbf{A}_2(\mathbf{E}_1,\mathbf{E}_1)=0$. The remaining relation is therefore $b\,\mathbf{A}_2=\mathbf{0}$. Contracting it with $\mathbf{E}_1$ and $\mathbf{E}_2$ gives $b=0$, since $\mathbf{A}_2(\mathbf{E}_1,\mathbf{E}_2)=1$. Thus, $\mathbf{A}_1$, $\mathbf{A}_2$, and $\mathbf{A}_3$ are linearly independent, and the reaction stress space is three-dimensional:
\begin{equation}
	\mathcal{N}_{\mathrm{tot}}=\operatorname{span}\left\{
	\mathbf{E}_1\otimes\mathbf{E}_1-\mathbf{E}_2\otimes\mathbf{E}_2,\,
	\mathbf{E}_1\otimes\mathbf{E}_2+\mathbf{E}_2\otimes\mathbf{E}_1,\,
	\mathbf{N}\otimes\mathbf{N} \right\}\,.
\end{equation}
Consequently, the most general reactive stress associated with the simultaneous imposition of surface conformality and fiber inextensibility is
\begin{equation}
	\mathbf{S}_r
	=T_1\left(\mathbf{E}_1\otimes\mathbf{E}_1-\mathbf{E}_2\otimes\mathbf{E}_2\right)
	+T_2\left(\mathbf{E}_1\otimes\mathbf{E}_2+\mathbf{E}_2\otimes\mathbf{E}_1\right)
	+T_0\,\mathbf{N}\otimes\mathbf{N}\,,
\end{equation}
where $T_1=T_1(X)$ and $T_2=T_2(X)$ are the indeterminate scalar fields associated with surface conformality, while $T_0=T_0(X)$ is the indeterminate tension field associated with fiber inextensibility. If the transversality condition is dropped, the above argument no longer guarantees independence: when $\mathbf{N}$ lies in the tangent plane, $\mathbf{N}\otimes\mathbf{N}$ may have a component in the surface-conformality reaction stress space, and the independence of the three generators must then be examined separately.

\subsection{Stress constraints}

Degenerate elastic materials and mechanical metamaterials may have one or more soft deformation modes, along which deformation can occur without an associated restoring stress in the ideal limit \citep{Milton1995,KadicEtAl2012}. In the linear theory, these modes correspond to zero eigenvalues of the elasticity tensor \citep{Milton1995}. Consequently, the stress response is insensitive to the corresponding strain components. Knowledge of the stress therefore does not determine those strain components: they are constitutively indeterminate. Thus, a soft mode is not merely a deformation mode with a small stiffness. In the ideal zero-stiffness limit, the corresponding strain component cannot be recovered from the constitutive stress relation.

Such materials are usually described in linear elasticity through a positive-semidefinite, rank-deficient elasticity tensor. If $\boldsymbol{\varepsilon}$ denotes the infinitesimal strain and $\boldsymbol{\sigma}=\mathbb{C}\boldsymbol{\varepsilon}$, a soft mode $\mathbf{Z}$ belongs to the null space of $\mathbb{C}$ and hence $\mathbb{C}\mathbf{Z}=\mathbf{0}$. Consequently, $\boldsymbol{\varepsilon}$ and $\boldsymbol{\varepsilon}+a\mathbf{Z}$ generate the same stress for arbitrary $a\in\mathbb{R}$. More generally, if $\{\mathbf{Z}_1,\ldots,\mathbf{Z}_s\}$ is a basis for the null space of $\mathbb{C}$, then
\begin{equation}
	\boldsymbol{\sigma}=\mathbb{C}\boldsymbol{\varepsilon}
	=\mathbb{C}\Big(\boldsymbol{\varepsilon}
	+\sum_{\alpha=1}^{s}a_\alpha\mathbf{Z}_\alpha\Big)\,,
\end{equation}
for arbitrary $a_\alpha\in\mathbb{R}$. Thus, knowledge of $\boldsymbol{\sigma}$ cannot determine the components of the strain associated with the null space of $\mathbb{C}$. Equivalently, an $s$-dimensional null space of the elasticity tensor gives rise to $s$ independent soft modes and an $s$-dimensional constitutive indeterminacy in the strain.

The same constitutive structure may be described from the opposite direction. Rather than beginning with a rank-deficient elasticity tensor and determining its null space, one may prescribe directly the set of stresses that the material is capable of supporting. This leads naturally to the introduction of constraints directly on the stress. Such constraints define the admissible stress states without requiring the prior construction of a singular elasticity tensor whose null space has the desired dimension and structure. The soft deformation modes then emerge from the constitutive indeterminacy associated with these stress restrictions: those components of the deformation that cannot be recovered from an admissible stress are precisely the soft modes. This reverses the usual viewpoint in the theory of degenerate elastic materials. Instead of specifying stiffness, identifying its degeneracies, and then determining the resulting admissible stresses and soft modes, one specifies the admissible stress states themselves and determines the corresponding constitutive indeterminacy on the strain. This formulation is particularly well suited to complementary Cauchy elasticity, in which stress is the constitutive independent variable and strain is determined from stress. Stress constraints can therefore be imposed directly on the domain of the constitutive relation, providing a simple framework for describing degenerate elastic materials and mechanical metamaterials in terms of the stresses they are capable of supporting.

Let us consider $s\leq 5$ independent scalar constraints on the second Piola--Kirchhoff stress,
\begin{equation}
	\psi_\alpha(\mathbf{S})=0\,,\qquad \alpha=1,\ldots,s\,.
\end{equation}
These constraints define locally a stress-constraint manifold $\mathcal{M}_{\mathbf{S}}\subset\mathbb{S}_X$. At an admissible stress $\mathbf{S}\in\mathcal{M}_{\mathbf{S}}$, the tangent space $T_{\mathbf{S}}\mathcal{M}_{\mathbf{S}}$ consists of the stress variations that preserve the constraints to first order. Thus, $\mathbf{H}\in T_{\mathbf{S}}\mathcal{M}_{\mathbf{S}}$ if and only if
\begin{equation}
	D\psi_\alpha\big|_{\mathbf{S}}(\mathbf{H})=0\,,\qquad \alpha=1,\ldots,s\,.
\end{equation}
The derivative $D\psi_\alpha\big|_{\mathbf{S}}$ is a covector on the stress space and therefore belongs to the dual space $\mathbb{S}_X^*$. The annihilator of the tangent space is defined by
\begin{equation}
	\left(T_{\mathbf{S}}\mathcal{M}_{\mathbf{S}}\right)^\circ
	=
	\Big\{
	\boldsymbol{\eta}\in\mathbb{S}_X^*:
	\boldsymbol{\eta}(\mathbf{H})=0
	\quad\text{for every}\quad
	\mathbf{H}\in T_{\mathbf{S}}\mathcal{M}_{\mathbf{S}}
	\Big\}.
\end{equation}
For regular independent constraints, the derivatives $D\psi_\alpha\big|_{\mathbf{S}}$ are linearly independent and span this annihilator:
\begin{equation}
	\left(T_{\mathbf{S}}\mathcal{M}_{\mathbf{S}}\right)^\circ
	=
	\operatorname{span}\left\{
	D\psi_1\big|_{\mathbf{S}},\ldots,D\psi_s\big|_{\mathbf{S}}
	\right\}.
\end{equation}
It follows that $\dim\left(T_{\mathbf{S}}\mathcal{M}_{\mathbf{S}}\right)^\circ=s$ and $\dim T_{\mathbf{S}}\mathcal{M}_{\mathbf{S}}=6-s$. Hence, locally, the admissible stress has only $6-s$ independent components. The $s$ independent stress constraints leave $s$ conjugate strain components constitutively undetermined. These constitutively indeterminate deformation modes are represented by the elements of the annihilator $\left(T_{\mathbf{S}}\mathcal{M}_{\mathbf{S}}\right)^\circ$. Thus, $s$ independent regular stress constraints give rise locally to $s$ independent soft deformation modes.

\begin{remark}[Classification by the number of stress constraints]
Degenerate elastic materials can be classified by the number of independent stress constraints. If $s$ independent stress constraints are imposed, the admissible stress manifold has dimension $6-s$, while the associated space of soft deformation modes has dimension $s$. Accordingly, unimode, bimode, trimode, quadramode, and pentamode materials correspond to $s=1,2,3,4,5$, respectively. They therefore admit five, four, three, two, and one independent stress modes, respectively, together with one, two, three, four, and five independent soft deformation modes. In particular, a pentamode material is characterized by five independent stress constraints, leaving only one independent admissible stress mode and five soft deformation modes.
The complementary stress-work $1$-form restricted to the admissible stress manifold depends only on the strain components conjugate to the admissible stress modes. The constitutively indeterminate soft deformation modes make no contribution to this restricted $1$-form.
\end{remark}

\begin{remark}[Asymmetry between kinematic and stress constraints in nonlinear elasticity]
In nonlinear elasticity, whether a constraint is purely stress-based depends, in general, on the choice of stress measure. Different stress measures are related through the deformation mapping, and consequently a constraint expressed solely in terms of one stress measure may become a mixed stress-strain constraint when rewritten in terms of another. For example, the Cauchy stress and the second Piola--Kirchhoff stress are related by $\boldsymbol{\sigma}=J^{-1}\mathbf{F}\mathbf{S}\mathbf{F}^{\star}$. Hence, a constraint of the form $\widehat{\psi}(\boldsymbol{\sigma})=0$ becomes, when expressed in terms of $\mathbf{S}$, $\widehat{\psi}\!\left(J^{-1}\mathbf{F}\mathbf{S}\mathbf{F}^{\star}\right)=0$, which depends on both $\mathbf{S}$ and the strain. Conversely, a constraint imposed purely on $\mathbf{S}$ will generally acquire strain dependence when expressed in terms of $\boldsymbol{\sigma}$. Thus, in nonlinear elasticity, the designation ``stress constraint'' should always be understood relative to a specified stress measure. A constraint may define a submanifold in the space of one stress measure while its representation in terms of another stress measure defines a mixed stress-strain constraint. 
A simple example is provided by an elastic fluid: the requirement that the Cauchy stress be hydrostatic, $\boldsymbol{\sigma}=-p\,\mathbf{g}^{\sharp}$, is a constraint expressed purely in terms of the Cauchy stress, whereas its pull-back to the reference configuration gives $\mathbf{S}=-pJ\,\mathbf{C}^{-\sharp}$ and therefore involves the strain $\mathbf{C}$.
This distinction is absent in linearized elasticity, where the different stress measures coincide to leading order.

More generally, this observation suggests that, in nonlinear elasticity, an internal constraint may be viewed as a restriction on the constitutive manifold in the joint space of stress and strain, rather than necessarily as a constraint on either stress or strain separately. Purely stress-based and purely kinematic constraints then arise as special cases in which the constraint is independent of the strain or the stress, respectively. This viewpoint is naturally accommodated within implicit elasticity, where stress and strain enter the constitutive relation on an equal footing.

There is, however, a fundamental asymmetry between kinematic and stress constraints in nonlinear elasticity. Although stress and strain may enter an implicit constitutive relation on an equal footing as constitutive variables, they are not equivalent with respect to internal constraints. 
A purely kinematic constraint is representation independent: changing from one strain measure to another changes only its mathematical representation, since the different strain measures are related without introducing a stress variable. Thus, a kinematic constraint remains purely kinematic independently of the strain measure used to express it.
A purely stress-based constraint does not have the same invariance. Different stress measures in nonlinear elasticity are related through deformation, and hence a constraint involving only one stress measure will, in general, become a mixed stress-strain constraint when expressed in terms of another stress measure. Therefore, the formal symmetry between stress and strain in implicit elasticity conceals an important asymmetry: kinematic constraints have a representation-independent meaning as restrictions on the strain, whereas the notion of a purely stress-based constraint is generally tied to the particular stress measure employed.
\end{remark}

\paragraph{One-dimensional admissible stress space and pentamode materials.}
The most restrictive nontrivial case is a one-dimensional admissible stress space. Let $\mathbf{M}$ be a prescribed nonzero symmetric tensor specifying the supported stress mode. The admissible stresses are of the form $\mathbf{S}=q\mathbf{M}$, where $q=q(X)$ is an unrestricted scalar field. Thus, $\mathcal{M}_{\mathbf{S}}=\operatorname{span}\{\mathbf{M}\}$. Five independent scalar stress constraints are required to restrict a general symmetric stress tensor to this one-dimensional space. Accordingly, the stress determines only the scalar $q$, while five independent strain modes remain constitutively indeterminate. The isotropic pentamode case corresponds to a single hydrostatic stress mode and is illustrated by the classical example of an elastic fluid discussed next.

\begin{example}[Elastic fluids as stress-constrained materials]
An important classical example of a stress-constrained material is an elastic fluid. In hyperelasticity, an elastic fluid is characterized by a strain-energy density that depends on the deformation only through the local volume ratio, i.e., $W=W(J)$.\footnote{It is known that any Cauchy elastic fluid is hyperelastic \citep{YavariGoriely2025Cauchy}.} Consequently, the material has resistance to changes of volume but no resistance to isochoric distortions. The corresponding Cauchy stress is hydrostatic. Indeed,
\begin{equation}
	\boldsymbol{\sigma}
	=\frac{1}{J}\frac{\partial W}{\partial\mathbf{F}}\mathbf{F}^{\star}
	=W'(J)\,\mathbf{g}^{\sharp}\,.
\end{equation}
Thus, at each material point the admissible Cauchy stresses belong to the one-dimensional space spanned by $\mathbf{g}^{\sharp}$. 
The same structure may be expressed in terms of the second Piola--Kirchhoff stress. Since $\mathbf{S}=J\mathbf{F}^{-1}\boldsymbol{\sigma}\mathbf{F}^{-\star}$, a hydrostatic Cauchy stress $\boldsymbol{\sigma}=q\,\mathbf{g}^{\sharp}$ corresponds to $\mathbf{S}=q_0\mathbf{C}^{-\sharp}$, where $q_0=Jq$.
Hence, unlike the linear isotropic pentamode idealization, the admissible second Piola--Kirchhoff stress is not, in general, restricted to the fixed space $\operatorname{span}\{\mathbf{G}^{\sharp}\}$. Rather, its admissible one-dimensional space depends on the deformation through $\mathbf{C}^{-\sharp}$. Equivalently, the elastic-fluid stress constraint may be written in spatial form as $\operatorname{dev}_{\mathbf g}\boldsymbol{\sigma}=\mathbf{0}$, which represents five independent scalar stress constraints.

In terms of the stress-constraint framework introduced above, these five independent constraints restrict the Cauchy stress to a single hydrostatic stress mode. Correspondingly, there are five independent soft deformation modes associated with isochoric distortions. The constitutive response determines the volume change through the scalar pressure-volume relation, but it does not determine the isochoric part of the deformation. In this sense, an elastic fluid provides a classical finite-deformation example of an isotropic pentamode material: it supports only a hydrostatic stress mode and offers no elastic resistance to isochoric distortions.
This example also illustrates the natural role of complementary Cauchy elasticity in describing stress-constrained materials. In the usual Cauchy-elastic formulation, strain is the constitutive independent variable, and the constitutive relation must map every admissible strain into a stress satisfying the five hydrostatic-stress constraints. In complementary Cauchy elasticity, stress is instead the constitutive independent variable. The stress can therefore be restricted directly to the admissible hydrostatic stress manifold, while the constitutive relation determines the volumetric deformation from the scalar stress parameter. The five isochoric deformation modes remain constitutively indeterminate, exactly as expected for an ideal elastic fluid.
\end{example}

An ideal pentamode material is characterized locally by a one-dimensional admissible stress space. Let $\mathbf{M}=\mathbf{M}(\mathbf{C})$ be a nonzero symmetric tensor that spans this space at the strain state under consideration. The admissible second Piola--Kirchhoff stress is then of the form $\mathbf{S}=q\,\mathbf{M}(\mathbf{C})$, where $q=q(X)$ is a scalar field. The dependence of $\mathbf{M}$ on $\mathbf{C}$ is allowed but is not required. In particular, a fixed $\mathbf{M}$ gives the simplest pentamode idealization, whereas the elastic-fluid example above has $\mathbf{M}(\mathbf{C})=\mathbf{C}^{-\sharp}$, up to an arbitrary scalar normalization that may be absorbed into $q$. At each strain state, restricting $\mathbf{S}$ to a one-dimensional admissible stress space imposes five independent scalar stress constraints and is associated with five constitutively indeterminate strain modes.

An ideal pentamode material is therefore a degenerate elastic material rather than a strictly stable elastic solid. In the ideal limit, the material has no stiffness associated with its five soft modes. Physical pentamode metamaterials may instead be regarded as approximations to this idealization, with small but nonzero stiffnesses associated with these modes. The ideal limit is nevertheless meaningful for the formulation of boundary-value problems. The stress constraint restricts the stress to its single supported mode, while equilibrium and the boundary conditions determine its scalar amplitude together with whatever information they provide about the strain. What is absent is constitutive determination of the five soft strain modes.

To see this in a simple setting, consider a homogeneous body in the absence of inertial and body forces and a homogeneous deformation $\mathbf{F}=\mathbf{F}_0$. Since $\mathbf{M}=\mathbf{M}(\mathbf{C})$ depends on the strain state, the homogeneity of $\mathbf{F}$ implies that $\mathbf{C}$, and hence $\mathbf{M}$, is constant throughout the body. Since $\mathbf{P}=\mathbf{F}\mathbf{S}=q\,\mathbf{F}_0\mathbf{M}$, equilibrium in components reads $P^{aA}{}_{|A}=F_0^a{}_B M^{BA}q_{,A}=0$. Since $\mathbf{F}_0$ is invertible, this is equivalent to $M^{BA}q_{,A}=0$, i.e., $\mathrm{d}_0q=\frac{\partial q}{\partial X^A} dX^A$ belongs to the kernel of $\mathbf{M}$. Thus, equilibrium does not, in general, require $q$ to be constant. In particular, $q=q_0$ is always a solution, although nonconstant solutions may also exist when $\mathbf{M}$ has a nontrivial kernel. For the constant solution, the corresponding nominal traction on a material surface with unit normal $\mathbf{N}$ is $\mathbf{T}=q_0\mathbf{F}_0\mathbf{M}\mathbf{N}^{\flat}$. Thus, a homogeneous strain state compatible with the prescribed boundary conditions can support a traction field of this form. The stress constraint restricts the stress to its one-dimensional admissible space, while equilibrium and the boundary conditions determine the scalar field $q$ subject to $M^{BA}q_{,A}=0$. The constitutive relation alone does not determine the five soft strain modes.

The isotropic pentamode provides a particularly transparent special case and connects directly with the elastic fluid discussed above. Its supported Cauchy stress mode is hydrostatic, so that $\boldsymbol{\sigma}=q\,\mathbf{g}^{\sharp}$. In the absence of body forces, spatial equilibrium reduces to $dq=\mathbf{0}$, and hence $q$ is constant on each connected component of the body. A uniform pressure loading can therefore be supported exactly. A constitutive relation between $q$ and $J$ may determine the corresponding volume ratio, but it does not determine the isochoric part of the deformation. Different deformation fields having the same local volume ratio may therefore correspond to the same hydrostatic stress field, subject, of course, to compatibility and the imposed boundary conditions.

This example makes precise the sense in which an ideal pentamode material can have a well-defined stress response without constitutively determining the complete strain. Its degeneracy does not prevent the formulation of equilibrium boundary-value problems; rather, it means that the stress-to-strain constitutive relation determines fewer independent strain components than in a nondegenerate elastic solid. This structure is particularly natural in complementary Cauchy elasticity: the admissible stress is restricted directly to its one-dimensional stress manifold, while the five soft strain modes appear explicitly as constitutive indeterminacies.

\begin{remark}
An ideal pentamode material is not necessarily hyperelastic. To see this, consider the second Piola--Kirchhoff stress $\mathbf{S}=q(\mathbf{C}^{\flat},\mathbf{G})\,\mathbf{M}(\mathbf{C}^{\flat},\mathbf{G})$. The corresponding stress-work one-form is written as
\begin{equation}
	\boldsymbol{\Omega}=\frac{1}{2}\mathbf{S}:d\mathbf{C}^{\flat}
	=\frac{1}{2}q\,\mathbf{M}\!:\!d\mathbf{C}^{\flat}=q\,\boldsymbol{\alpha}\,, \qquad 
	\boldsymbol{\alpha}=\frac{1}{2}\mathbf{M}\!:\!d\mathbf{C}^{\flat}\,.
\end{equation}
Its exterior derivative is written as $d\boldsymbol{\Omega}=dq\wedge\boldsymbol{\alpha}+q\,d\boldsymbol{\alpha}$. In general, this two-form does not vanish. An ideal pentamode material is hyperelastic only when the additional integrability condition $d\boldsymbol{\Omega}=0$ is satisfied.
\end{remark}

\section{Kinematic Constraints in Implicit Elasticity}
\label{Sec:InternalConstraints}

Internal constraints restrict the admissible strains and introduce reactive stresses through Lagrange multipliers. The purpose of this section is to formulate the implicit constitutive equations for elastic solids subject to internal constraints and to determine how a constraint and its reactive stress enter the constitutive equations. We then specialize the general formulation to complementary Cauchy elasticity and examine the conditions imposed by the internal constraint and by indifference to the corresponding reactive stress.

\subsection{Incompressibility constraint and reactive stress}
\label{Sec:GeometricIncompressibilityReaction}

The origin of the reactive stress associated with the incompressibility constraint is geometric and does not depend on the particular constitutive representation. Recall that the strain space at $X\in\mathcal B$ is $\mathcal C_X=\operatorname{Sym}^+(T_X^*\mathcal B)$. Consider the scalar function $J:\mathcal C_X\times\operatorname{Sym}^+(T_X\mathcal B)\longrightarrow\mathbb R_+$, $J(\mathbf C^\flat,\mathbf G)=\sqrt{\det(\mathbf G^\sharp\mathbf C^\flat)}$. The incompressibility constraint defines the level set
\begin{equation}
	\mathcal M_J=\left\{\mathbf C^\flat\in\mathcal C_X:J(\mathbf C^\flat,\mathbf G)=1\right\}\,.
\end{equation}
Thus, incompressibility restricts the admissible strain states to the codimension-one submanifold $\mathcal M_J$ of $\mathcal C_X$. This construction is purely kinematic and geometric; no constitutive equation has yet been introduced.

Let $\mathbf C^\flat\in\mathcal M_J$ and consider a smooth curve $t\mapsto\mathbf C^\flat(t)\in\mathcal M_J$ with $\mathbf C^\flat(0)=\mathbf C^\flat$. Since $J(\mathbf C^\flat(t))=1$, $\forall~t$, differentiation at $t=0$ gives us
\begin{equation}
	DJ(\mathbf C^\flat)\big[\dot{\mathbf C}^\flat\big]=0\,,\qquad
	\dot{\mathbf C}^\flat=\left.\frac{d\mathbf C^\flat(t)}{dt}\right|_{t=0}\,.
\end{equation}
Consequently, $T_{\mathbf C^\flat}\mathcal M_J=\ker DJ(\mathbf C^\flat)$. Knowing that $DJ(\mathbf C^\flat)\big[\mathbf H^\flat\big]=\frac{1}{2}J\,\mathbf{C}^{-\sharp}\!:\!\mathbf{H}$, 
the tangent space to the incompressibility manifold is equivalently characterized as 
\begin{equation}
	T_{\mathbf C^\flat}\mathcal M_J
	=\left\{\mathbf H^\flat\in\operatorname{Sym}(T_X^*\mathcal B)
	:\quad \mathbf{C}^{-\sharp}\!:\!\mathbf{H}=0\right\}\,.
\end{equation}
Thus, $\mathbf C^{-\sharp}$ is normal to the incompressibility manifold in the sense that it annihilates every tangent direction $\mathbf H^\flat\in T_{\mathbf C^\flat}\mathcal M_J$.

The stress acts on strain increments through the natural stress-strain pairing. Hence, if $\mathbf H^\flat$ is an admissible strain increment, the action of the second Piola-Kirchhoff stress on $\mathbf H^\flat$ is $\mathbf S\!:\!\mathbf H^\flat=S^{AB}H_{AB}$.
For a constrained strain space, the stresses that do no work on any admissible strain increment define the set $\left\{\boldsymbol{\alpha}\in\operatorname{Sym}(T_X\mathcal B):\quad \boldsymbol{\alpha}\!:\!\mathbf H^\flat=0\,,~\forall\,\mathbf H^\flat\in T_{\mathbf C^\flat}\mathcal M_J\right\}$. Since $T_{\mathbf C^\flat}\mathcal M_J=\ker DJ(\mathbf C^\flat)$ and $DJ(\mathbf C^\flat)\neq0$, the tangent space $T_{\mathbf C^\flat}\mathcal M_J$ is the kernel of a nonzero linear functional. Hence, it is a hyperplane and has codimension one in $\operatorname{Sym}(T_X^*\mathcal B)$. Consequently, the linear functionals that vanish on $T_{\mathbf C^\flat}\mathcal M_J$ form a one-dimensional space spanned by $DJ(\mathbf C^\flat)$. From the expression for $DJ(\mathbf C^\flat)$ given above, this functional is represented under the natural stress-strain pairing by $\mathbf C^{-\sharp}$. Therefore,
\begin{equation}
	\left\{\boldsymbol{\alpha}\in\operatorname{Sym}(T_X\mathcal B):
	\quad \boldsymbol{\alpha}\!:\!\mathbf H^\flat=0\,,~
	\forall\,\mathbf H^\flat \in T_{\mathbf C^\flat}\mathcal M_J\right\}=\operatorname{span}
	\left\{\mathbf C^{-\sharp}\right\}\,.
\end{equation}
If $\mathbf S$ is a second Piola--Kirchhoff stress compatible with the constraint, then adding a reactive stress $\mathbf S_{\mathrm r}=\lambda\,\mathbf C^{-\sharp}$ does not alter the stress work on any admissible strain increment.
As a matter of fact, $\mathbf S_{\mathrm r}\!:\!\mathbf H^\flat=\lambda\,C^{-AB}H_{AB}=0$, $\forall\,\mathbf H^\flat\in T_{\mathbf C^\flat}\mathcal M_J$.
Thus, the incompressibility constraint determines a one-dimensional reactive stress space. In the spatial representation, this space is spanned by the spatial metric, and the corresponding scalar multiplier is the pressure.

It is important to note that this conclusion does not depend on the constitutive equation. The argument uses only the strain space $\mathcal C_X$, the constraint function $J$, and the tangent and cotangent spaces associated with the constraint manifold. In particular, it does not require a Cauchy constitutive relation of the form $\mathbf S=\mathbf S(\mathbf C^\flat,\mathbf G)$, a complementary Cauchy constitutive relation $\mathbf C^\flat=\mathbf C^\flat(\mathbf S,\mathbf G)$, or even an implicit constitutive relation $\boldsymbol{\mathcal F}(\mathbf C^\flat,\mathbf S,\mathbf G)=\mathbf0$. The one-dimensional reactive stress space is therefore a geometric consequence of the incompressibility constraint itself, rather than a consequence of any particular constitutive representation. The corresponding scalar multiplier, however, is determined only after the equilibrium equations and boundary conditions are imposed.

The same geometric observation clarifies the meaning of the constitutive indeterminacy associated with the incompressibility constraint. The admissible strain increments belong to $T_{\mathbf C^\flat}\mathcal M_J$, and therefore they cannot distinguish between second Piola--Kirchhoff stress tensors that differ by a tensor which annihilates every admissible strain increment. For incompressibility, the space of such second Piola--Kirchhoff stress tensors is the one-dimensional space $\operatorname{span}\{\mathbf C^{-\sharp}\}$. Thus, the constitutive response to admissible strain increments determines the second Piola--Kirchhoff stress only up to the addition of a multiple of $\mathbf C^{-\sharp}$. The corresponding scalar multiplier is not determined by the constrained strain increments themselves. 
The multiplier is not determined by the constitutive equation; its value is determined by the balance equations and boundary conditions.

\subsection{Pressure orbits in incompressible implicit elasticity}
\label{Sec:ImplicitIncompressibilityPressureOrbit}

The presence of a kinematic internal constraint requires some care in the formulation of an implicit elastic constitutive relation. For an unconstrained elastic solid, an implicit constitutive equation may be written in the form $\boldsymbol{\mathcal F}(\mathbf C^\flat,\mathbf S,\mathbf G)=\mathbf0$
and defines a constitutive manifold in strain-stress space. In this unconstrained setting, both $\mathbf C^\flat$ and $\mathbf S$ may be regarded as constitutive variables entering the implicit relation. A kinematic internal constraint changes this structure because the total stress is no longer entirely constitutively determined. The constraint introduces a reactive stress whose magnitude is indeterminate constitutively and is determined, together with the deformation, through the balance equations and boundary conditions. 
Consequently, an individual total stress contains information that is irrelevant to the constitutive response.

For incompressibility, the admissible strains belong to the constraint manifold $\mathcal M_J$. As discussed in \S\ref{Sec:GeometricIncompressibilityReaction}, at every $\mathbf C^\flat\in\mathcal M_J$, the reactive part of the second Piola--Kirchhoff stress is proportional to $\mathbf C^{-\sharp}$. Thus, if $\mathbf S$ is a total stress compatible with a given constitutive state, then $\mathbf S-p\,\mathbf C^{-\sharp}$, $p\in\mathbb R$, represents the same constitutive stress state. The parameter $p$ changes only the indeterminate reactive pressure and does not change the constitutively determined part of the stress. It follows that, at fixed isochoric strain, the constitutive theory cannot distinguish between total stresses that differ only by such a reactive stress.

This observation is fundamental for implicit elasticity. The constitutive equation of an incompressible implicit elastic solid should not be regarded as a relation between an admissible strain and an individual total stress. Rather, it is a relation between an admissible strain and an equivalence class of total stresses modulo the reactive stress. At fixed $\mathbf C^\flat\in\mathcal M_J$, let us define
\begin{equation}
	\mathbf S_1\sim_{\mathbf C^\flat}\mathbf S_2
	\quad\Longleftrightarrow\quad
	\mathbf S_2-\mathbf S_1\in\operatorname{span}\left\{\mathbf C^{-\sharp}\right\}\,.
\end{equation}
The corresponding equivalence class is defined as
\begin{equation}
	[\mathbf S]_{\mathbf C^\flat}
	=\left\{\mathbf S-p\,\mathbf C^{-\sharp}:p\in\mathbb R\right\}\,.
\end{equation}
Geometrically, this equivalence class defines the pressure orbit through $\mathbf S$ at fixed $\mathbf C^\flat$, denoted by $\mathcal O_{\mathbf C^\flat}(\mathbf S)=[\mathbf S]_{\mathbf C^\flat}$. The sign in this definition is conventional and is consistent with the usual pressure contribution to the stress.

The constitutively relevant stress variable is therefore $[\mathbf S]_{\mathbf C^\flat}$, not $\mathbf S$ itself. Accordingly, the proper constrained implicit constitutive relation has the following schematic form
\begin{equation}
	\boldsymbol{\mathcal F}
	\big(\mathbf C^\flat,[\mathbf S]_{\mathbf C^\flat},\mathbf G \big)=\mathbf0\,,
	\qquad \mathbf C^\flat\in\mathcal M_J\,.
\end{equation}
This formulation incorporates both consequences of the internal constraint from the outset. First, the strain variable is restricted to the incompressibility manifold $\mathcal M_J$. Second, the stress variable is defined only modulo the reactive pressure. These are two aspects of the same constrained constitutive structure: the kinematic constraint restricts the admissible strains and, through its associated reactive stress, identifies total stresses that represent the same constitutive stress state.

This point is particularly important in implicit elasticity because neither stress nor strain is singled out a priori as the dependent constitutive variable. In ordinary Cauchy elasticity, where stress is expressed as a function of strain, the kinematic constraint restricts the admissible strain states and the reactive stress is appended to the constitutively determined stress. In complementary Cauchy elasticity, where strain is expressed as a function of stress, the stress argument must be understood modulo the reactive stress. Implicit elasticity contains both theories as special representations and therefore must incorporate the equivalence of stresses at the level of the constitutive relation itself. The quotient structure is not peculiar to complementary Cauchy elasticity; it is already present in the general implicit theory whenever a kinematic internal constraint generates an indeterminate reactive stress.

One may, of course, choose a representative $\mathbf S$ of the equivalence class $[\mathbf S]_{\mathbf C^\flat}$ and write the implicit constitutive equation using the total stress. Such a representation does not change the underlying constitutive relation, but it introduces a redundancy: every member of a pressure orbit represents the same constitutive stress state. Consequently, any representation in terms of the total stress must satisfy the condition
\begin{equation}
	\boldsymbol{\mathcal F}(\mathbf C^\flat,\mathbf S,\mathbf G)=\mathbf0
	\quad\Longleftrightarrow\quad
	\boldsymbol{\mathcal F}(\mathbf C^\flat,\mathbf S
	-p\,\mathbf C^{-\sharp},\mathbf G)=\mathbf0\,,
	\quad \forall\,p\in\mathbb R\,.
\end{equation}
Thus, the constitutive manifold written in terms of total stress contains the complete pressure orbit associated with every one of its isochoric strain states. Equivalently, introducing the transformation
\begin{equation}
	R_p:(\mathbf C^\flat,\mathbf S)\mapsto
	(\mathbf C^\flat,\mathbf S-p\,\mathbf C^{-\sharp})\,,
\end{equation}
the total-stress representation of the constrained constitutive manifold $\mathcal K_X$ satisfies the condition $R_p(\mathcal K_X)=\mathcal K_X$. This invariance should not be interpreted as an additional constitutive assumption imposed after incompressibility. Nor is it a property that must be proved from the equation $J=1$. It is the necessary consequence of representing, in terms of individual total stresses, a constitutive relation whose stress argument is intrinsically an equivalence class modulo the reactive pressure.

This observation clarifies the meaning of pressure indifference. Pressure indifference is not an independent constitutive principle that follows from incompressibility. Rather, it is the manifestation, in a total-stress representation, of the fact that the reactive pressure carries no constitutive information. Two stresses that differ only by pressure are not two distinct constitutive inputs to which the material happens to give the same response; they are two representatives of the same constitutive stress state. In this sense, pressure indifference is already built into the proper formulation of incompressible implicit elasticity.
The corresponding local statement follows immediately. At fixed $\mathbf C^\flat$, an increment within a pressure orbit has the form $\delta\mathbf S=-\delta p\,\mathbf C^{-\sharp}$.
Therefore, for any smooth total-stress representation of the constrained implicit constitutive relation we have $D_{\mathbf S}\boldsymbol{\mathcal F}\!\left[\mathbf C^{-\sharp}\right]=\mathbf0$.

This distinction resolves an important issue that arises if one starts instead with an arbitrary implicit relation $\boldsymbol{\mathcal F}(\mathbf C^\flat,\mathbf S,\mathbf G)=\mathbf0$ and subsequently requires only that its admissible strains satisfy $J=1$. Such a procedure imposes incompressibility on the strain component of the constitutive manifold, but it does not by itself identify stresses that differ by reactive pressure. One has then started from a constitutive domain that is too large for an internally constrained material. The appropriate construction is the reverse: the equivalence induced by the reactive stress is incorporated into the constitutive domain from the outset, and any representation using total stress must respect that equivalence.

The same structure is reflected in the stress-work $1$-form. On the strain-stress space $\mathcal X_X=\mathcal C_X\times\mathcal S_X$, consider $\boldsymbol{\Omega}=\frac12\,\mathbf S\!:\!d\mathbf C^\flat$. For every admissible isochoric strain increment $\mathbf K^\flat\in T_{\mathbf C^\flat}\mathcal M_J$, $\mathbf C^{-\sharp}\!:\!\mathbf K^\flat=0$. Hence,
\begin{equation}
	\frac12\left(\mathbf S-p\,\mathbf C^{-\sharp}\right)\!:\!\mathbf K^\flat
	=\frac12\,\mathbf S\!:\!\mathbf K^\flat\,.
\end{equation}
Therefore, the restriction of the stress-work $1$-form to admissible strain increments is independent of the representative chosen from the pressure-equivalence class. The reactive pressure performs no work on admissible strain increments, precisely as required for a reactive stress.

The constitutive and work-conjugate viewpoints are therefore consistent. The geometry of the incompressibility constraint determines the reactive stress space and hence the pressure orbits. The constitutive theory regards every pressure orbit as a single constitutive stress state, while the stress-work $1$-form assigns the same work to all representatives of that state on admissible strain increments. This provides the natural formulation of incompressible implicit elasticity.

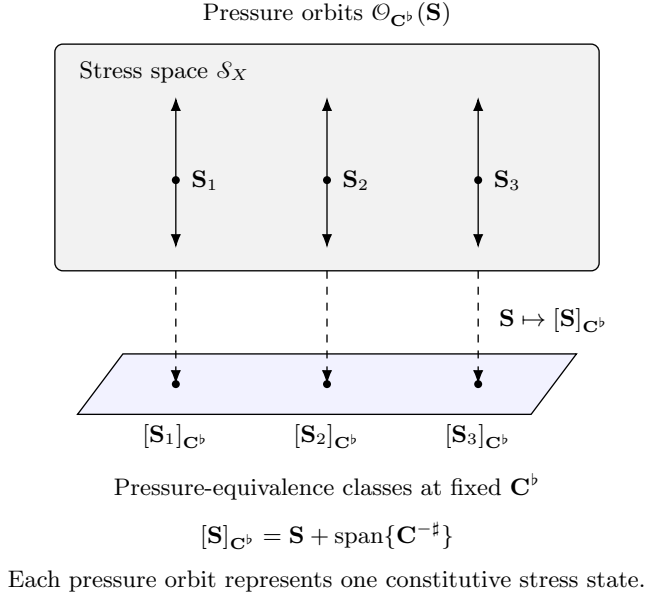
\begin{figure}[htbp]
\centering
\begin{tikzpicture}[x=1cm,y=1cm,font=\small,line width=0.5pt,>=Latex]

\draw[rounded corners=3pt,fill=black!5] (-0.6,1.6) rectangle (6.6,4.6);
\node[anchor=north west] at (-0.4,4.5)
  {Stress space $\mathcal{S}_X$};

\draw[<->] (1,1.9) -- (1,3.9);
\draw[<->] (3,1.9) -- (3,3.9);
\draw[<->] (5,1.9) -- (5,3.9);

\fill (1,2.8) circle (1.5pt);
\fill (3,2.8) circle (1.5pt);
\fill (5,2.8) circle (1.5pt);

\node[right] at (1.08,2.8) {$\mathbf{S}_1$};
\node[right] at (3.08,2.8) {$\mathbf{S}_2$};
\node[right] at (5.08,2.8) {$\mathbf{S}_3$};

\node at (3,5.0)
  {Pressure orbits $\mathcal{O}_{\mathbf{C}^\flat}(\mathbf{S})$};

\draw[fill=blue!5] (-0.3,-0.3) -- (5.7,-0.3)
   -- (6.3,0.5) -- (0.3,0.5) -- cycle;

\draw[dashed,->] (1,1.6) -- (1,0.1);
\draw[dashed,->] (3,1.6) -- (3,0.1);
\draw[dashed,->] (5,1.6) -- (5,0.1);

\node[right] at (5.15,0.95)
  {$\mathbf{S}\mapsto [\mathbf{S}]_{\mathbf{C}^\flat}$};

\fill (1,0.1) circle (1.5pt);
\fill (3,0.1) circle (1.5pt);
\fill (5,0.1) circle (1.5pt);

\node[below] at (1,-0.3) {$[\mathbf{S}_1]_{\mathbf{C}^\flat}$};
\node[below] at (3,-0.3) {$[\mathbf{S}_2]_{\mathbf{C}^\flat}$};
\node[below] at (5,-0.3) {$[\mathbf{S}_3]_{\mathbf{C}^\flat}$};

\node at (3,-1.250)
  {Pressure-equivalence classes at fixed $\mathbf{C}^\flat$};

\node at (3,-1.9)
  {$[\mathbf{S}]_{\mathbf{C}^\flat}
    = \mathbf{S}+\operatorname{span}\{\mathbf{C}^{-\sharp}\}$};

\node at (3,-2.5)
  {Each pressure orbit represents one constitutive stress state.};

\end{tikzpicture}

\caption{Pressure orbits and stress equivalence classes in incompressible implicit elasticity. At fixed strain $\mathbf{C}^\flat$, stresses belonging to the same pressure orbit differ by a reactive stress and represent the same constitutive stress state.}
\label{fig:pressure-orbits}

\end{figure}

\begin{example}[Incompressibility and pressure indifference in complementary Cauchy elasticity.]
The distinction between imposing incompressibility on the image of a stress-to-strain map and requiring pressure indifference can be seen directly from the tangent compliance. Consider a complementary Cauchy elastic relation $\mathbf C^\flat=\widehat{\mathbf C}^\flat(\mathbf S,\mathbf G)$, and let $\mathbb{A}_{ABCD}=\partial C_{AB}/\partial S^{CD}$. 
Suppose first that the image of the constitutive map is incompressible, so that $J(\mathbf C^\flat,\mathbf G)=1$ for every stress in its domain. For an arbitrary stress increment $\delta\mathbf S$, one has $\delta C_{AB}=\mathbb{A}_{ABCD}\,\delta S^{CD}$. Differentiating $J=1$ and using $\delta J=\frac{1}{2}J\,C^{AB}\delta C_{AB}$, one obtains $0=\delta J=\frac{1}{2}C^{AB}\mathbb{A}_{ABCD}\,\delta S^{CD}$. Since $\delta\mathbf S$ is arbitrary, it follows that
\begin{equation}
	C^{AB}\mathbb{A}_{ABCD}=0\,.
\end{equation}
Thus, incompressibility requires the range of the tangent compliance to lie in the tangent space of the incompressibility manifold.

Pressure indifference imposes a different condition. At fixed $\mathbf C^\flat$, an increment along a pressure orbit has the form $\delta\mathbf S=-\delta p\,\mathbf C^{-\sharp}$. Constancy of the strain along this orbit requires $0=\delta C_{AB}=-\delta p\,\mathbb{A}_{ABCD}\,C^{CD}$, and hence
\begin{equation}
	\mathbb{A}_{ABCD}\,C^{CD}=0\,.
\end{equation}
The two kernel conditions are distinct in general because the tangent compliance of a complementary Cauchy elastic material need not have the major symmetry. Therefore, incompressibility of the image of an arbitrary stress-to-strain map does not imply pressure indifference. If the material is complementary hyperelastic, however, the tangent compliance is a Hessian and satisfies $\mathbb{A}_{ABCD}=\mathbb{A}_{CDAB}$; the two conditions then coincide. This illustrates the distinction between merely requiring the image of a stress-to-strain map to satisfy incompressibility and defining an incompressible complementary Cauchy elastic response on the pressure-equivalence classes dictated by the constrained implicit theory.
\end{example}

\begin{example}[An incompressible complementary Cauchy elastic response]
A simple explicit example of an incompressible complementary Cauchy elastic response can be constructed using the Cauchy stress. Let $\boldsymbol{\sigma}$ denote the Cauchy stress and define its deviatoric part by $\boldsymbol{\sigma}_{\mathrm d}=\boldsymbol{\sigma}-\frac{1}{3}(\operatorname{tr}_{\mathbf g}\boldsymbol{\sigma})\,\mathbf g^\sharp$.
Let $\{\hat{\mathbf e}_a\}$ be a $\mathbf g$-orthonormal frame, and denote by $\widehat{\sigma}_{\mathrm d}^{ab}$ the physical components of $\boldsymbol{\sigma}_{\mathrm d}$ with respect to this frame \citep{SadikYavari2026PhysicalComponents}. We denote the matrix of these physical components by $\widehat{\boldsymbol{\sigma}}_{\mathrm d}=\big[\widehat{\sigma}_{\mathrm d}^{ab}\big]$. Let us consider the following constitutive equation
\begin{equation}
	\widehat{b}^{ab}=\left[\exp\left(\frac{\widehat{\boldsymbol{\sigma}}_{\mathrm d}}{\sigma_0}\right)\right]^{ab}\,,
\end{equation}
where $\widehat{b}^{ab}$ are the physical components of the left Cauchy--Green strain $\mathbf b^\sharp$ and $\sigma_0>0$ is a reference stress. In particular, in Cartesian coordinates the coordinate frame is orthonormal and the physical and coordinate components coincide. Thus, the constitutive equation takes the form
\begin{equation}
	b^{ab}=\left[\exp\left(\frac{\boldsymbol{\sigma}_{\mathrm d}}{\sigma_0}\right)\right]^{ab}\,.
\end{equation}
Since $\boldsymbol{\sigma}_{\mathrm d}$ is traceless, in Cartesian coordinates one has
\begin{equation}
	\det\big[b^{ab}\big]=\exp\left(\frac{1}{\sigma_0}\operatorname{tr}\boldsymbol{\sigma}_{\mathrm d}\right)=1\,.
\end{equation}
Thus, the constitutively prescribed strain is isochoric. Moreover, under the addition of an arbitrary hydrostatic stress, $\boldsymbol{\sigma}\mapsto\boldsymbol{\sigma}-p\,\mathbf g^\sharp$, the deviatoric Cauchy stress $\boldsymbol{\sigma}_{\mathrm d}$ is unchanged. Consequently,
\begin{equation}
	\widehat{\mathbf b}^\sharp\left(\boldsymbol{\sigma}
	-p\,\mathbf g^\sharp\right)=\widehat{\mathbf b}^\sharp(\boldsymbol{\sigma})\,.
\end{equation}
Hence, this constitutive response respects both the incompressibility constraint and pressure-indifference. 
\end{example}

\begin{example}[Incompressibility and pressure indifference in complementary hyperelasticity]
\label{ex:pressure-indifference}
\citet{ReynoldsBlume1993} introduced pressure indifference in the complementary formulation of nonlinear elasticity and established its equivalence with incompressibility for complementary hyperelastic solids. In the terminology developed above, pressure indifference expresses the fact that the complementary strain response is well defined on pressure-equivalence classes of stresses.
As shown in the preceding example, for a general complementary Cauchy elastic representation $\mathbf C^\flat=\widehat{\mathbf C}^\flat(\mathbf S,\mathbf G)$, with tangent compliance tensor $\mathbb A_{ABCD}=\partial C_{AB}/\partial S^{CD}$, pressure indifference places $\mathbf C^{-\sharp}$ in the right kernel of the tangent compliance,
\begin{equation}
	\text{pressure indifference:}\qquad \mathbb A_{ABCD}\,C^{CD}=0\,,
\end{equation}
whereas incompressibility places $\mathbf C^{-\sharp}$ in its left kernel,
\begin{equation}
	\text{incompressibility:}\qquad C^{CD}\mathbb A_{CDAB}=0\,.
\end{equation}
For a general complementary Cauchy elastic material these two conditions are distinct because $\mathbb A$ need not possess major symmetry.

Complementary hyperelasticity supplies the additional structure that relates these two conditions. If there exists a complementary strain-energy function $\zeta$ such that $C_{AB}=2\dfrac{\partial\zeta}{\partial S^{AB}}$, then
\begin{equation}
	\mathbb A_{ABCD}=2\frac{\partial^2\zeta}{\partial S^{AB}\partial S^{CD}}\,,
\end{equation}
and consequently $\mathbb A_{ABCD}=\mathbb A_{CDAB}$. The left- and right-kernel conditions therefore coincide. Thus, within complementary hyperelasticity, the local condition associated with incompressibility is equivalent to the local condition associated with pressure indifference.

On a connected stress domain, the local condition $\partial J/\partial S^{AB}=0$ implies that $J$ is constant. If $J=1$ at one constitutive state, incompressibility follows throughout the domain. Thus, with this normalization, incompressibility and pressure indifference are equivalent in complementary hyperelasticity.

This example also illustrates a limitation of restricting attention to complementary hyperelasticity. Within that class, the major symmetry of the tangent compliance makes incompressibility and pressure indifference equivalent, which may suggest that their equivalence is a general feature of stress-based constitutive theories. In general complementary Cauchy elasticity, however, these are distinct constitutive conditions, and neither implies the other.
\end{example}

\subsection{A general scalar internal constraint}

The geometric construction developed above for incompressibility extends directly to a general scalar internal constraint. At every material point $X\in\mathcal B$, let the constraint be defined by $\mu(\mathbf C^\flat)=0$. Assuming that $D\mu(\mathbf C^\flat)\neq0$ at the strain state under consideration, the corresponding constraint manifold in the space of positive-definite strains is the codimension-one submanifold
\begin{equation}
	\mathcal M_\mu=\left\{\mathbf C^\flat\in\operatorname{Sym}^+(T_X^*\mathcal B):\mu(\mathbf C^\flat)=0\right\}\,.
\end{equation}
Let us introduce the notation $\mu_{\mathbf C^\flat}=\frac{\partial\mu}{\partial\mathbf C^\flat}$. For a smooth curve $t\mapsto\mathbf C^\flat(t)\in\mathcal M_\mu$, differentiation of the constraint, one obtains $\mu_{\mathbf C^\flat}\!:\!\dot{\mathbf C}^\flat=0$. Therefore, the tangent space to the constraint manifold is defined as
\begin{equation}
	T_{\mathbf C^\flat}\mathcal M_\mu
	=\left\{\mathbf H^\flat\in\operatorname{Sym}(T_X^*\mathcal B):
	\quad \mu_{\mathbf C^\flat}\!:\!\mathbf H^\flat=0\right\}\,.
\end{equation}
Thus, $\mu_{\mathbf C^\flat}$ represents the normal covector to the constraint manifold under the natural stress-strain pairing. Consequently, the second Piola--Kirchhoff stresses that do no work on any admissible strain increment form the one-dimensional space
\begin{equation}
	\left\{\boldsymbol{\alpha}\in\operatorname{Sym}(T_X\mathcal B):
	\quad \boldsymbol{\alpha}\!:\!\mathbf H^\flat=0\,,~
	\forall\,\mathbf H^\flat\in T_{\mathbf C^\flat}\mathcal M_\mu\right\}
	=\operatorname{span}\big\{\mu_{\mathbf C^\flat}\big\}\,.
\end{equation}
Hence, the reactive part of the second Piola--Kirchhoff stress has the form $\mathbf S_{\mathrm r}=\mathring q\,\mu_{\mathbf C^\flat}$, where $\mathring q$ is an indeterminate scalar multiplier. Accordingly, the total second Piola--Kirchhoff stress can be decomposed as
\begin{equation}
	\mathbf S=\overline{\mathbf S}+\mathring q\,\mu_{\mathbf C^\flat}\,,
\end{equation}
where $\overline{\mathbf S}$ denotes the constitutively determined part of the stress. The multiplier $\mathring q$ is not determined by the constraint itself; its value is determined together with the deformation through the balance equations and boundary conditions.
Using $\boldsymbol{\sigma}=J^{-1}\mathbf F\mathbf S\mathbf F^\star$, the corresponding reactive part of the Cauchy stress is written as $\boldsymbol{\sigma}_{\mathrm r}=q\,\mathbf F\mu_{\mathbf C^\flat}\mathbf F^\star$, where $q=J^{-1}\mathring q$. Thus, the total Cauchy stress admits the decomposition
\begin{equation}
	\boldsymbol{\sigma}
	=\bar{\boldsymbol{\sigma}}+q\,\mathbf F\mu_{\mathbf C^\flat}\mathbf F^\star\,,\qquad
	\bar{\boldsymbol{\sigma}}=J^{-1}\mathbf F\overline{\mathbf S}\mathbf F^\star\,.
\end{equation}

At a fixed admissible strain $\mathbf C^\flat\in\mathcal M_\mu$, stresses that differ only by a reactive stress belong to the same affine reaction-stress orbit. Thus, through a stress state $\mathbf S_0$, the reaction-stress orbit associated with the constraint is defined as
\begin{equation}
	\mathcal O_{\mathbf C^\flat}(\mathbf S_0)
	=\Big\{\mathbf S_0+\lambda\,\mu_{\mathbf C^\flat}:\lambda\in\mathbb R\Big\}\,.
\end{equation}
Every stress on this orbit has the same action on admissible strain increments, since for $\mathbf H^\flat\in T_{\mathbf C^\flat}\mathcal M_\mu$, $\left(\mathbf S_0+\lambda\,\mu_{\mathbf C^\flat}\right)\!:\!\mathbf H^\flat=\mathbf S_0\!:\!\mathbf H^\flat$.
This reaction-stress orbit is determined entirely by the geometry of the internal constraint and does not yet invoke an elastic constitutive relation. As will be formulated below, for an internally constrained material the constitutively relevant stress state is the equivalence class defined by such an orbit, rather than an individual total stress.

\subsection{Reactive stress orbits for a general scalar internal constraint}
\label{Sec:ImplicitConstraintPressureOrbit}

The preceding discussion of incompressibility is a special case of a general structure associated with kinematic internal constraints in implicit elasticity. Consider a regular scalar constraint $\mu(\mathbf C^\flat)=0$, and let
\begin{equation}
	\mathcal M_\mu=\left\{\mathbf C^\flat\in\operatorname{Sym}^+(T_X^*\mathcal B):\mu(\mathbf C^\flat)=0\right\}
\end{equation}
denote the corresponding constraint manifold. At an admissible strain $\mathbf C^\flat\in\mathcal M_\mu$, the admissible infinitesimal strain increments belong to $T_{\mathbf C^\flat}\mathcal M_\mu$. The reactive stresses are those stresses that do no work on every admissible strain increment. For a single regular scalar constraint, the reactive stress space is therefore $\mathcal R_{\mathbf C^\flat}=\operatorname{span}\left\{\mu_{\mathbf C^\flat}\right\}$. Thus, the total stress contains a reactive part of the form $\lambda\,\mu_{\mathbf C^\flat}$, where the scalar multiplier $\lambda$ is not constitutively determined. Its value is determined, together with the deformation, through the balance equations and boundary conditions.

This indeterminacy has a direct consequence for the meaning of an implicit constitutive equation. For an unconstrained elastic solid, one may write $\boldsymbol{\mathcal F}(\mathbf C^\flat,\mathbf S,\mathbf G)=\mathbf0$ with $\mathbf C^\flat$ and $\mathbf S$ entering as constitutive variables. In the presence of the constraint $\mu(\mathbf C^\flat)=0$, however, an individual total stress $\mathbf S$ is not a constitutively distinct stress state. At a fixed admissible strain, changing the total stress by an arbitrary reactive stress changes only the indeterminate reaction associated with the constraint and leaves the constitutively determined part of the stress unchanged. Consequently, stresses that differ only by a reactive stress must represent the same constitutive stress state.
At fixed $\mathbf C^\flat\in\mathcal M_\mu$, we therefore introduce the equivalence relation
\begin{equation}
	\mathbf S_1\sim_{\mathbf C^\flat}\mathbf S_2
	\qquad\Longleftrightarrow\qquad
	\mathbf S_2-\mathbf S_1\in\mathcal R_{\mathbf C^\flat}\,.
\end{equation}
For a single scalar constraint, this is equivalent to $\mathbf S_2-\mathbf S_1=\lambda\,\mu_{\mathbf C^\flat}$ for some $\lambda\in\mathbb R$. The corresponding equivalence class is defined as
\begin{equation}
	[\mathbf S]_{\mathbf C^\flat}
	=\left\{\mathbf S+\lambda\,\mu_{\mathbf C^\flat}:\lambda\in\mathbb R\right\}\,.
\end{equation}
Geometrically, this equivalence class is the reaction-stress orbit through $\mathbf S$, i.e., $[\mathbf S]_{\mathbf C^\flat}=	\mathcal O_{\mathbf C^\flat}(\mathbf S)$.
Every member of this orbit represents the same constitutive stress state; the members differ only in the indeterminate reactive stress.

The proper implicit constitutive relation for the constrained material is therefore not fundamentally a relation between $\mathbf C^\flat$ and an individual total stress $\mathbf S$. It is a relation between an admissible strain and an equivalence class of stresses,
\begin{equation}
	\boldsymbol{\mathcal F} \big(\mathbf C^\flat,[\mathbf S]_{\mathbf C^\flat},
	\mathbf G\big)=\mathbf0\,,	\qquad	\mathbf C^\flat\in\mathcal M_\mu\,.
\end{equation}
The kinematic constraint consequently modifies both aspects of the constitutive description. It restricts the admissible strain states to $\mathcal M_\mu$, and it identifies total stresses that differ only by the reactive stress. The constitutively relevant stress variable is the equivalence class $[\mathbf S]_{\mathbf C^\flat}$, while the reactive multiplier carries no constitutive information.

This structure is intrinsic to constrained implicit elasticity and does not depend on choosing either stress or strain as the dependent constitutive variable. In a Cauchy-elastic representation, the admissible strain is prescribed and the constitutive equation determines the constitutively determined part of the stress, while the reactive stress remains indeterminate. In a complementary Cauchy-elastic representation, the strain is determined from the constitutively relevant stress state, which is necessarily an equivalence class modulo the reactive stress. Both descriptions are special representations of the same constrained implicit constitutive structure.
If one nevertheless chooses to represent the implicit constitutive relation using an individual total stress $\mathbf S$, the representation necessarily contains a redundancy. Every member of the same reaction-stress orbit must represent the same constitutive state. Hence,
\begin{equation}
	\boldsymbol{\mathcal F}(\mathbf C^\flat,\mathbf S,\mathbf G)=\mathbf0
	\quad\Longleftrightarrow\quad
	\boldsymbol{\mathcal F}\big(\mathbf C^\flat,\mathbf S
	+\lambda\,\mu_{\mathbf C^\flat},\mathbf G \big)=\mathbf0\,,
	\quad \forall\,\lambda\in\mathbb R\,.
\end{equation}
This invariance is not an additional constitutive assumption imposed on the constrained material. It follows from representing, by individual total stresses, a constitutive relation that is intrinsically defined on equivalence classes of stresses.
The corresponding local statement follows by considering a variation within a reaction-stress orbit at fixed strain, $\delta\mathbf S=\delta\lambda\,\mu_{\mathbf C^\flat}$. For any smooth total-stress representation of the constrained implicit constitutive relation, differentiation along the reaction-stress orbit, one obtains $D_{\mathbf S}\boldsymbol{\mathcal F}\!\left[\mu_{\mathbf C^\flat}\right]=\mathbf0$. This kernel condition is therefore the local differential expression of the equivalence of stresses modulo the reactive stress. 

The same structure appears in the stress-work $1$-form. If $\mathbf H^\flat\in T_{\mathbf C^\flat}\mathcal M_\mu$ is an admissible strain increment, then $\mu_{\mathbf C^\flat}\!:\!\mathbf H^\flat=0$. Consequently,
\begin{equation}
	\frac12\left(\mathbf S+\lambda\,\mu_{\mathbf C^\flat}\right)\!:\!\mathbf H^\flat
	=\frac12\,\mathbf S\!:\!\mathbf H^\flat\,.
\end{equation}
Thus, the stress work on admissible strain increments is independent of the representative chosen from $[\mathbf S]_{\mathbf C^\flat}$. The reactive stress performs no work on admissible strain increments and carries no constitutive information. The equivalence-class formulation of the constitutive relation and the workless character of the reactive stress are therefore two manifestations of the same constrained structure.
There is an important geometric subtlety. Since $\mu_{\mathbf C^\flat}$ generally depends on $\mathbf C^\flat$, the reactive stress space $\mathcal R_{\mathbf C^\flat}$ generally varies over the constraint manifold. Hence, one cannot in general construct the constitutive stress space by quotienting the entire stress space by a single fixed linear subspace. Instead, each admissible strain $\mathbf C^\flat\in\mathcal M_\mu$ carries its own reaction-stress orbit. The constrained implicit constitutive relation is therefore naturally described in terms of these strain-dependent equivalence classes.

\begin{example}[Fiber inextensibility and tension indifference]
Consider a family of inextensible fibers with unit material tangent $\mathbf N$, for which the constraint is $C_{AB}N^AN^B=1$. The corresponding reactive stress is proportional to $\mathbf N\otimes\mathbf N$, and the associated reaction-stress orbits are tension orbits. For a complementary Cauchy elastic material with tangent compliance $\mathbb A_{ABCD}$, tension indifference requires that $\mathbb A_{ABCD}N^CN^D=0$, whereas fiber inextensibility of the image of the stress-to-strain map requires that $N^CN^D\mathbb A_{CDAB}=0$.
Thus, as in the incompressible case, tension indifference and fiber inextensibility correspond to right- and left-kernel conditions, respectively, and are distinct for a general complementary Cauchy elastic material. For complementary hyperelasticity, the major symmetry $\mathbb A_{ABCD}=\mathbb A_{CDAB}$ makes the two conditions coincide. Hence, subject to the corresponding connectedness and normalization conditions, fiber inextensibility and tension indifference are equivalent in complementary hyperelasticity.

In order to illustrate the distinction between these two conditions, let us consider an orthonormal material basis $\{\mathbf E_1,\mathbf E_2,\mathbf E_3\}$ and a family of fibers with unit material tangent $\mathbf N=\mathbf E_1$. First, consider the complementary Cauchy elastic constitutive relation
\begin{equation}
	C_{11}=1+\alpha S^{22}\,,\qquad
	C_{22}=1+\beta S^{22}\,,\qquad
	C_{33}=1+\gamma S^{33}\,,
\end{equation}
with all off-diagonal components vanishing and $\alpha\neq 0$. The stress domain is restricted so that $\mathbf C$ is positive definite. Since the strain components are independent of $S^{11}$, we have $\mathbb A_{ABCD}N^CN^D=\mathbb A_{AB11}=0$. Thus, tension indifference holds. However, $N^CN^D\mathbb A_{CD22}=\mathbb A_{1122}=\alpha\neq 0$. Hence, fiber inextensibility does not hold. Next, consider the constitutive relation
\begin{equation}
	C_{11}=1\,,\qquad
	C_{22}=1+\alpha S^{11}+\beta S^{22}\,,\qquad
	C_{33}=1\,,
\end{equation}
with $\alpha\neq 0$ and all off-diagonal components vanishing. Again, the stress domain is restricted so that $\mathbf C$ remains positive definite. In this case, $N^CN^D\mathbb A_{CDAB}
=\mathbb A_{11AB}=0$ and $\mathbb A_{2211}=\alpha\neq 0$. Thus, fiber inextensibility holds, but tension indifference fails.
\end{example}

\begin{remark}
The preceding examples are instances of the same structure for a general scalar kinematic constraint $\mu(\mathbf C^\flat)=0$. If $\mathbb A$ denotes the tangent compliance, satisfaction of the constraint by the image of a stress-to-strain map requires $\mathbb A^\star[\mu_{\mathbf C^\flat}]=\mathbf0$, where $\mathbb A^\star$ denotes the adjoint action of the tangent compliance, so that in components $\big(\mathbb A^\star[\mu_{\mathbf C^\flat}]\big)_{AB}=(\mu_{\mathbf C^\flat})^{CD}\mathbb A_{CDAB}$. Indifference to the corresponding reactive stress requires that $\mathbb A[\mu_{\mathbf C^\flat}]=\mathbf0$, where $\big(\mathbb A[\mu_{\mathbf C^\flat}]\big)_{AB}=\mathbb A_{ABCD} (\mu_{\mathbf C^\flat})^{CD}$. Thus, the derivative $\mu_{\mathbf C^\flat}$ of the constraint belongs to the left kernel of the tangent compliance in the first case and to its right kernel in the second. These conditions are distinct in general complementary Cauchy elasticity and coincide under the major symmetry of complementary hyperelasticity. For several independent scalar constraints, the same conditions hold for the derivative of each constraint, and hence for the entire reactive stress space.
\end{remark}

\section{Balance Laws and Thermodynamics of Complementary Cauchy Elasticity}
\label{Sec:BalanceLawsComplementaryCauchyElasticity}

For a general complementary Cauchy elastic solid, a Lagrangian is not available, and hence Hamilton's principle cannot be written in its standard form. The balance laws, however, are independent of constitutive equations and material classes and may simply be postulated. Alternatively, they may be obtained from the balance of energy by imposing an appropriate invariance requirement.\footnote{The use of invariance of the energy balance to obtain the balance laws has a long history in continuum mechanics. In Euclidean ambient spaces, \citet{Green64} required the balance of energy to be invariant under superposed translations and rotations, leading to what is commonly referred to as the Green--Naghdi--Rivlin theorem. A related formulation was given by \citet{Noll1963}, with changes of observer interpreted passively in terms of time-dependent coordinate systems. The geometric extension from invariance under Euclidean motions to covariance under arbitrary spatial diffeomorphisms was developed for hyperelastic bodies in Riemannian ambient spaces by \citet{HuMa1977}. They showed that covariance of the energy balance yields the balance laws together with the Doyle--Ericksen formula \citep{doyle1956nonlineari}; see also \citep{MarsdenHughes1983,YaMaOr2006}. These classical developments assume the existence of an underlying energy function and hence apply to hyperelastic materials. Covariance arguments have also been extended to anelasticity when an energy function is available \citep{Yavari2010}. The connection between covariance of the energy balance and the Lagrangian field-theoretic formulation of elasticity was studied in \citep{YavariMarsden2012}. The principle of virtual work, another standard route to the balance laws \citep{Antman1979,Maugin1980}, can itself be related to covariance of the energy balance; see \citet{MarsdenHughes1983}.} We follow the covariance approach and adapt the argument used for Cauchy elasticity by \citet{YavariGoriely2025Cauchy} to complementary Cauchy elasticity.

\subsection{The first law of thermodynamics: balance of energy} \label{sec:1stLaw}

Let $\mathcal{U}\subset\mathcal{B}$ be an arbitrary material subbody. We write the energy balance in the following form
\begin{equation}
\begin{aligned}
	\frac{d}{dt}\int_{\mathcal{U}}
	\frac{1}{2}\rho_0\lVert\mathbf{V}\rVert_{\mathbf{g}}^2\,dV
	+\int_{\mathcal{U}}\left(\widetilde{\mathbf{P}}\!:\!\nabla^{\varphi_t}\mathbf{V}
	+\rho_0\mathsf{f}\dot{\mathscr{N}}\right)dV
	=\int_{\mathcal{U}}\rho_0\left(
	\llangle\mathbf{B},\mathbf{V}\rrangle_{\mathbf{g}}+R\right)dV
	+\int_{\partial\mathcal{U}}\left(
	\llangle\mathbf{T},\mathbf{V}\rrangle_{\mathbf{g}}+H\right)dA\,.
\end{aligned}
\end{equation}
Here, $\rho_0$ is the referential mass density, $\mathbf{B}$ is the body force per unit mass, $\mathbf{T}$ is the traction per unit reference area, $R$ is the specific heat supply, $H$ is the outward heat flux through the boundary, $\mathscr{N}$ is the specific entropy. 
The two-point tensor $\widetilde{\mathbf{P}}$ and the scalar field $\mathsf{f}$ are initially left unspecified.\footnote{Unlike classical covariance arguments, we deliberately leave these two quantities unspecified: covariance of the energy balance will determine $\widetilde{\mathbf{P}}$, while the second law of thermodynamics will determine $\mathsf{f}$.}

The Cauchy traction theorem does not require the existence of a strain-energy function. Therefore, there exists a first Piola--Kirchhoff stress $\mathbf{P}$ such that $\mathbf{T}=\mathbf{P}\mathbf{N}^{\flat}$, where $\mathbf{N}$ is the outward unit normal to $\partial\mathcal{U}$ in the reference configuration. The divergence theorem implies that
\begin{equation}  \label{Traction-Divergence}
	\int_{\partial\mathcal{U}}
	\llangle\mathbf{T},\mathbf{V}\rrangle_{\mathbf{g}}\,dA
	=\int_{\mathcal{U}}\left[
	\llangle\operatorname{Div}\mathbf{P},\mathbf{V}\rrangle_{\mathbf{g}}
	+\mathbf{P}\!:\!\nabla^{\varphi_t}\mathbf{V}\right]dV\,.
\end{equation}
Using this identity in the energy balance and denoting the material acceleration by $\mathbf{A}=\frac{D\mathbf{V}}{Dt}$, one obtains
\begin{equation}
\begin{aligned}
	\int_{\mathcal{U}}\Big[
	\llangle\operatorname{Div}\mathbf{P}
	+\rho_0(\mathbf{B}-\mathbf{A}),\mathbf{V}\rrangle_{\mathbf{g}}
	+\left(\mathbf{P}-\widetilde{\mathbf{P}}\right)\!:\!\nabla^{\varphi_t}\mathbf{V}
	-\frac{1}{2}\dot{\rho}_0\lVert\mathbf{V}\rVert_{\mathbf{g}}^2
	+\rho_0R-\rho_0\mathsf{f}\dot{\mathscr{N}} \Big]dV
	+\int_{\partial\mathcal{U}}H\,dA=0\,.
\end{aligned}
\end{equation}

Next, let us consider an arbitrary time-dependent diffeomorphism $\xi_t:\mathcal{S}\rightarrow\mathcal{S}$ such that $\xi_0=\operatorname{id}_{\mathcal{S}}$. Its Eulerian velocity field is defined by $\mathbf{w}_t=\frac{\partial\xi_t}{\partial t}\circ\xi_t^{-1}$. The transformed motion and spatial metric are $\varphi'_t=\xi_t\circ\varphi_t$ and $\mathbf{g}'=\xi_{t*}\mathbf{g}$, respectively. Let us define the vector field along the transformed motion by $\mathbf{W}_t=\mathbf{w}_t\circ\varphi'_t$, or equivalently $\mathbf{W}_t=\frac{\partial\xi_t}{\partial t}\circ\varphi_t$.
Let $\mathbf{v}_t=\mathbf{V}\circ\varphi_t^{-1}$ and $\mathbf{v}'_t=\mathbf{V}'\circ(\varphi'_t)^{-1}$ denote the spatial velocity fields of the original and transformed motions, respectively. We also introduce the push-forward $\mathbf{z}_t=\xi_{t*}\mathbf{v}_t$. The transformed spatial velocity is $\mathbf{v}'_t=\mathbf{w}_t+\mathbf{z}_t$, and composition with $\varphi'_t$ gives us
\begin{equation} \label{Velocity-Transformed}
	\mathbf{V}'=T\xi_t\cdot\mathbf{V}+\mathbf{W}_t\,.
\end{equation}
At $t=0$, one has $\xi_0=\operatorname{id}_{\mathcal{S}}$ and $T\xi_0=\operatorname{id}_{T\mathcal{S}}$. Thus, writing $\mathbf{W}=\mathbf{W}_0$, the transformed velocity and its covariant derivative read
\begin{equation}
	\mathbf{V}'_0=\mathbf{V}_0+\mathbf{W}\,,\qquad
	\nabla^{\varphi_0}\mathbf{V}'_0=\nabla^{\varphi_0}\mathbf{V}_0
	+\nabla^{\varphi_0}\mathbf{W}\,.
\end{equation}

We next determine how the material acceleration transforms. Let $\mathbf{A}'=\frac{D'\mathbf{V}'}{Dt}$ denote the material acceleration of the transformed motion, where $\frac{D'}{Dt}$ is the covariant time derivative induced by the Levi--Civita connection of the transformed spatial metric $\mathbf{g}'=\xi_{t*}\mathbf{g}$. We compute separately the covariant time derivatives of the two terms in the preceding velocity transformation.
Since $\mathbf{W}_t=\mathbf{w}_t\circ\varphi'_t$, its value at a material point $X\in\mathcal{B}$ is $\mathbf{W}_t(X)=\mathbf{w}_t(\varphi'_t(X))$. Thus, $\mathbf{W}_t(X)$ is obtained by evaluating the time-dependent spatial vector field $\mathbf{w}_t$ along the transformed trajectory $t\mapsto\varphi'_t(X)$. Its covariant time derivative contains two contributions: one from the explicit time dependence of $\mathbf{w}_t$ at a fixed spatial point and another from the motion $\varphi'_t(X)$ through the spatially varying field $\mathbf{w}_t$. By the covariant chain rule, one writes
\begin{equation}
	\frac{D'\mathbf{W}_t}{Dt}=\left(\frac{\partial\mathbf{w}_t}{\partial t}
	+\nabla^{\mathbf{g}'}_{\mathbf{v}'_t}\mathbf{w}_t\right)\circ\varphi'_t\,.
\end{equation}
Let $\{{x'}^a\}$ be spatial coordinates and let ${\gamma'}^a{}_{bc}$ denote the Christoffel symbols of the Levi--Civita connection of $\mathbf{g}'$. Since $W_t^a(X)=w_t^a(\varphi'_t(X))$ and $\partial_t{\varphi'}^a_t(X)={V'}^a(X)={v'}^a_t(\varphi'_t(X))$, the components of the covariant time derivative are written as
\begin{equation}
	\left(\frac{D'\mathbf{W}_t}{Dt}\right)^a=\frac{dW_t^a}{dt}
	+{\gamma'}^a{}_{bc}(\varphi'_t)\,{V'}^b\,W_t^c\,.
\end{equation}
The ordinary chain rule gives us
\begin{equation}
	\frac{dW_t^a}{dt}=\left(\frac{\partial w_t^a}{\partial t}
	+{v'}^b_t\frac{\partial w_t^a}{\partial {x'}^b}\right)\circ\varphi'_t\,.
\end{equation}
Consequently,
\begin{equation}
	\left(\frac{D'\mathbf{W}_t}{Dt}\right)^a
	=\left[\frac{\partial w_t^a}{\partial t}+{v'}^b_t\left(\frac{\partial w_t^a}{\partial {x'}^b}
	+{\gamma'}^a{}_{bc}\,w_t^c\right)\right]\circ\varphi'_t
	=\left(\frac{\partial w_t^a}{\partial t}+{v'}^b_t\,\nabla'_b w_t^a\right)\circ\varphi'_t\,.
\end{equation}
The term $\frac{\partial\mathbf{w}_t}{\partial t}$ measures the change of the spatial vector field $\mathbf{w}_t$ at a fixed spatial point, whereas $\nabla^{\mathbf{g}'}_{\mathbf{v}'_t}\mathbf{w}_t$ measures its spatial variation along the transformed trajectory.

We let us differentiate the push-forward $T\xi_t\cdot\mathbf{V}=\mathbf{z}_t\circ\varphi'_t$. Because both $\xi_t$ and $\mathbf{V}$ depend on time, this derivative contains the push-forward of the original acceleration and an additional term generated by the spatial variation of $\mathbf{w}_t$ in the direction $\mathbf{z}_t$. More specifically,
\begin{equation}
	\frac{D'}{Dt}\left(T\xi_t\cdot\mathbf{V}\right)
	=T\xi_t\cdot\mathbf{A}
	+\left(\nabla^{\mathbf{g}'}_{\mathbf{z}_t}\mathbf{w}_t\right)\circ\varphi'_t\,.
\end{equation}
Adding the preceding two identities, from \eqref{Velocity-Transformed} one obtains
\begin{equation}
	\mathbf{A}'=T\xi_t\cdot\mathbf{A}+\left(\frac{\partial\mathbf{w}_t}{\partial t}
	+\nabla^{\mathbf{g}'}_{\mathbf{v}'_t}\mathbf{w}_t
	+\nabla^{\mathbf{g}'}_{\mathbf{z}_t}\mathbf{w}_t\right)\circ\varphi'_t\,.
\end{equation}
Using $\mathbf{v}'_t=\mathbf{w}_t+\mathbf{z}_t$ and the linearity of the covariant derivative in its lower argument, we obtain
\begin{equation}
	\mathbf{A}'=T\xi_t\cdot\mathbf{A}+\left(\frac{\partial\mathbf{w}_t}{\partial t}
	+\nabla^{\mathbf{g}'}_{\mathbf{w}_t}\mathbf{w}_t
	+2\nabla^{\mathbf{g}'}_{\mathbf{z}_t}\mathbf{w}_t\right)\circ\varphi'_t\,.
\end{equation}
Thus, in addition to the push-forward of the original acceleration, the transformed acceleration contains the local acceleration of the spatial diffeomorphism, its convective acceleration, and a velocity-dependent term.\footnote{For a rigid change of observer, these additional terms reduce to the familiar translational, Euler, centrifugal, and Coriolis accelerations. In particular, the term that depends on the material velocity contains the Coriolis acceleration. For a general spatial diffeomorphism, they represent the corresponding generalized inertial accelerations.} Recalling that $\mathbf{z}_t=\xi_{t*}\mathbf{v}_t$, the result may equivalently be written as
\begin{equation}
	\mathbf{A}'=T\xi_t\cdot\mathbf{A}+\left(\frac{\partial\mathbf{w}_t}{\partial t}
	+\nabla^{\mathbf{g}'}_{\mathbf{w}_t}\mathbf{w}_t
	+2\nabla^{\mathbf{g}'}_{\xi_{t*}\mathbf{v}_t}\mathbf{w}_t\right)\circ\varphi'_t\,.
\end{equation}
At $t=0$, one has $\xi_0=\operatorname{id}_{\mathcal{S}}$, $T\xi_0=\operatorname{id}_{T\mathcal{S}}$, $\varphi'_0=\varphi_0$, $\mathbf{g}'=\mathbf{g}$, and $\mathbf{z}_0=\mathbf{v}_0$. Consequently,
\begin{equation}
	\mathbf{A}'_0=\mathbf{A}_0+\left(\left.\frac{\partial\mathbf{w}_t}{\partial t}\right|_{t=0}
	+\nabla^{\mathbf{g}}_{\mathbf{w}_0}\mathbf{w}_0
	+2\nabla^{\mathbf{g}}_{\mathbf{v}_0}\mathbf{w}_0\right)\circ\varphi_0\,.
\end{equation}

The transformation of the acceleration vector follows kinematically from the transformed motion. The transformation of the body force, however, does not follow from this calculation or from the energy balance. We postulate covariance of the energy balance, i.e., the energy balance retains its form under every such spatial diffeomorphism. The material scalar fields and the referential volume and area elements are unchanged, while spatial vector fields and the spatial indices of two-point tensor fields are pushed forward by $\xi_t$. In particular, $\mathbf{P}'=T\xi_t\cdot\mathbf{P}$ and $\widetilde{\mathbf{P}}'=T\xi_t\cdot\widetilde{\mathbf{P}}$. We further postulate that the body force transforms in such a way that the effective body force $\mathbf{B}-\mathbf{A}$ is a spatial vector \citep{MarsdenHughes1983}, namely
\begin{equation}
	\mathbf{B}'-\mathbf{A}'=T\xi_t\cdot(\mathbf{B}-\mathbf{A})\,.
\end{equation}
At $t=0$, since $\xi_0=\operatorname{id}_{\mathcal{S}}$ and $T\xi_0=\operatorname{id}_{T\mathcal{S}}$, this transformation law reduces to
\begin{equation}
	\mathbf{B}'_0-\mathbf{A}'_0=\mathbf{B}_0-\mathbf{A}_0\,.
\end{equation}
For rigid changes of observer, the additional terms in the transformed body force have the usual interpretation as inertial forces. For a general spatial diffeomorphism, this transformation rule is a covariance assumption rather than a consequence of the energy balance.

Evaluating the transformed energy balance at $t=0$ and subtracting the original balance one obtains
\begin{equation}
\begin{aligned}
	\int_{\mathcal{U}}\Big[
	\llangle\operatorname{Div}\mathbf{P}
	+\rho_0(\mathbf{B}-\mathbf{A}),\mathbf{W}\rrangle_{\mathbf{g}}
	+\left(\mathbf{P}-\widetilde{\mathbf{P}}\right)\!:\!\nabla^{\varphi_t}\mathbf{W}
	-\dot{\rho}_0\left(\frac{1}{2}\lVert\mathbf{W}\rVert_{\mathbf{g}}^2
	+\llangle\mathbf{V},\mathbf{W}\rrangle_{\mathbf{g}}\right)	\Big]dV=0\,.
\end{aligned}
\end{equation}
This identity holds for every vector field $\mathbf{W}$. Substituting $\mathbf{W}=\beta\widehat{\mathbf{W}}$, where $\widehat{\mathbf{W}}$ is a $\mathbf{g}$-unit vector field, and differentiating twice with respect to $\beta$, one obatins
\begin{equation}
	\int_{\mathcal{U}}\dot{\rho}_0\,dV=0\,.
\end{equation}
Since $\mathcal{U}$ is arbitrary, one obtains the referential form of conservation of mass: $\dot{\rho}_0=0$. The covariance identity therefore reduces to
\begin{equation}
	\int_{\mathcal{U}}\left[
	\llangle\operatorname{Div}\mathbf{P}
	+\rho_0(\mathbf{B}-\mathbf{A}),\mathbf{W}\rrangle_{\mathbf{g}}
	+\left(\mathbf{P}-\widetilde{\mathbf{P}}\right)
	\!:\!\nabla^{\varphi_t}\mathbf{W}\right]dV=0\,.
\end{equation}
At any point, the values of $\mathbf{W}$ and $\nabla^{\varphi_t}\mathbf{W}$ may be prescribed independently. Their arbitrariness implies that
\begin{equation}
	\operatorname{Div}\mathbf{P}+\rho_0\mathbf{B}=\rho_0\mathbf{A}\,,
	\qquad \widetilde{\mathbf{P}}=\mathbf{P}\,.
\end{equation}
Thus, covariance of the energy balance gives the balance of linear momentum and identifies the tensor entering the internal stress power with the first Piola--Kirchhoff stress.
Using conservation of mass, the balance of linear momentum, and $\widetilde{\mathbf{P}}=\mathbf{P}$, the energy balance reduces to
\begin{equation}
	\int_{\mathcal{U}}\rho_0\left(R-\mathsf{f}\dot{\mathscr{N}}\right)dV
	+\int_{\partial\mathcal{U}}H\,dA=0\,.
\end{equation}
By the Cauchy heat-flux theorem, the heat transfer across a material surface is represented by a referential heat-flux vector $\mathbf{Q}$. Writing $H=-\llangle\mathbf{Q},\mathbf{N}\rrangle_{\mathbf{G}}$, where $\mathbf{N}$ is the outward $\mathbf{G}$-unit normal to $\partial\mathcal{U}$, the divergence theorem gives us
\begin{equation}
	\int_{\mathcal{U}}\left(\rho_0R-\operatorname{Div}\mathbf{Q}
	-\rho_0\mathsf{f}\dot{\mathscr{N}}\right)dV=0\,.
\end{equation}
Since $\mathcal{U}$ is arbitrary, the local form of the reduced energy balance is written as
\begin{equation} \label{LocalEnergyBalance}
	\rho_0R-\operatorname{Div}\mathbf{Q}-\rho_0\mathsf{f}\dot{\mathscr{N}}=0\,.
\end{equation}
This is identical to the reduced energy balance in Cauchy elasticity; the distinction between Cauchy and complementary Cauchy elasticity lies in the constitutive relation, not in the balance laws.

\subsection{Covariance and the balance of angular momentum in Cauchy elasticity}

The stress-work $1$-form is defined on the space of right Cauchy--Green strains as $\boldsymbol{\Omega}=\frac{1}{2}\mathbf{S}\!:\!d\mathbf{C}^{\flat}$. Since $\mathbf{C}^{\flat}$ is symmetric, $d\mathbf{C}^{\flat}$ is symmetric, and hence the antisymmetric part of $\mathbf{S}$ does not contribute to $\boldsymbol{\Omega}$. Therefore,
\begin{equation}
	\boldsymbol{\Omega}
	=\frac{1}{2}\operatorname{sym}(\mathbf{S})\!:\!d\mathbf{C}^{\flat}
	=\frac{1}{4}\left(\mathbf{S}+\mathbf{S}^{\star}\right)\!:\!d\mathbf{C}^{\flat}\,.
\end{equation}
This identity does not assume that $\mathbf{S}$ is symmetric; it only shows that the stress-work $1$-form is independent of the antisymmetric part of $\mathbf{S}$. Differentiating $\mathbf{C}^{\flat}=\mathbf{F}^{\star}\mathbf{g}\mathbf{F}$, we have $d\mathbf{C}^{\flat}=d\mathbf{F}^{\star}\mathbf{g}\mathbf{F}+\mathbf{F}^{\star}\mathbf{g}\,d\mathbf{F}$. It follows that
\begin{equation}
	\boldsymbol{\Omega}
	=\frac{1}{2}\mathbf{gF}
	\left(\mathbf{S}+\mathbf{S}^{\star}\right)\!:\!d\mathbf{F}
	=\mathbf{gF}\operatorname{sym}(\mathbf{S})\!:\!d\mathbf{F}\,.
\end{equation}
Using the relation $\mathbf{P}=\mathbf{F}\mathbf{S}$, one has $\mathbf{S}=\mathbf{F}^{-1}\mathbf{P}$ and $\mathbf{S}^{\star}=\mathbf{P}^{\star}\mathbf{F}^{-\star}$. Therefore,
\begin{equation} \label{Stress-Work-Form-In-Terms-of-P}
	\boldsymbol{\Omega}
	=\frac{1}{2}\mathbf{g}
	\left(\mathbf{P}
	+\mathbf{F}\mathbf{P}^{\star}\mathbf{F}^{-\star}\right)
	\!:\!d\mathbf{F}\,.
\end{equation}
The same calculation can be written explicitly in components. Since $C_{AB}=g_{ab}F^a{}_A F^b{}_B$, one has $dC_{AB}=g_{ab}\left(dF^a{}_A F^b{}_B+F^a{}_A dF^b{}_B\right)$, and hence
\begin{equation}
\begin{aligned}
	\boldsymbol{\Omega} &=\frac{1}{2}S^{AB}dC_{AB}
	=\frac{1}{2}g_{ab}F^b{}_B \left(S^{AB}+S^{BA}\right)dF^a{}_A
	=\frac{1}{2}g_{ab}\left[P^{bA}+F^b{}_B\,F^{-A}{}_c\,P^{cB}\right]dF^a{}_A\,.
\end{aligned}
\end{equation}
Equation \eqref{Stress-Work-Form-In-Terms-of-P} reduces to
$\boldsymbol{\Omega}=\mathbf{gP}\!:\!d\mathbf{F}$ if and only if $\mathbf{P}=\mathbf{F}\mathbf{P}^{\star}\mathbf{F}^{-\star}$, or equivalently, $\mathbf{P}\mathbf{F}^{\star}=\mathbf{F}\mathbf{P}^{\star}$, which is equivalent to $\mathbf{S}=\mathbf{S}^{\star}$.

The preceding calculation shows that the identification $\boldsymbol{\Omega}=\mathbf{gP}\!:\!d\mathbf{F}$ cannot be made before the balance of angular momentum has been established. We now give a derivation of the balance of angular momentum.\footnote{In our earlier work \citep{YavariGoriely2025Cauchy}, the identity $\boldsymbol{\Omega}=\mathbf{gP}\!:\!d\mathbf{F}$ was used before the balance of angular momentum had been established. As the preceding calculation shows, this identity already requires $\mathbf{S}=\mathbf{S}^{\star}$. Consequently, using it to derive the symmetry of $\mathbf{S}$ makes that particular argument circular. The derivation presented here corrects this issue. This correction does not affect the other constitutive developments or results of that paper.}
We know that the stress-work $1$-form has the Darboux representation
\begin{equation}
	\boldsymbol{\Omega}=\phi_1\,d\psi_1+\phi_2\,d\psi_2+\phi_3\,d\psi_3\,.
\end{equation}
Initially, the generalized energy functions are written as functions of the deformation gradient and the two metrics: $\phi_i=\phi_i(X,\mathbf{F},\mathbf{G},\mathbf{g})$ and $\psi_i=\psi_i(X,\mathbf{F},\mathbf{G},\mathbf{g})$, $i=1,2,3$. Evaluating the stress-work $1$-form along a motion gives us
\begin{equation}
	\dot{\boldsymbol{\Omega}}=\sum_{i=1}^{3}\phi_i\dot{\psi}_i
	=\left(\sum_{i=1}^{3}
	\phi_i\frac{\partial\psi_i}{\partial\mathbf{F}}\right)\!:\!\nabla^{\varphi_t}\mathbf{V}\,.
\end{equation}
At this stage, $\dot{\boldsymbol{\Omega}}$ must not be identified with
$\mathbf{P}\!:\!\nabla^{\varphi_t}\mathbf{V}$. Instead, the expression obtained from the Darboux representation is used as the internal stress-work rate in the balance of energy:
\begin{equation}
\begin{aligned} 
	\frac{d}{dt}\int_{\mathcal{U}}\frac{1}{2}\rho_0\lVert\mathbf{V}\rVert_{\mathbf{g}}^2\,dV
	+\int_{\mathcal{U}}\left[\left(\sum_{i=1}^{3}\phi_i
	\frac{\partial\psi_i}{\partial\mathbf{F}}\right)\!:\!\nabla^{\varphi_t}\mathbf{V}
	+\rho_0\mathsf{f}\dot{\mathscr{N}}\right]dV
	&=\int_{\mathcal{U}}\rho_0\left(\llangle\mathbf{B},\mathbf{V}\rrangle_{\mathbf{g}}+R\right)dV\\
	&\quad+\int_{\partial\mathcal{U}}\left(\llangle\mathbf{T},\mathbf{V}\rrangle_{\mathbf{g}}	
	+H\right)dA\,.
\end{aligned}
\end{equation}
We now postulate covariance of the energy balance under arbitrary time-dependent spatial diffeomorphisms. Evaluating the transformed balance at $t=0$, using \eqref{Traction-Divergence} and subtracting the original balance, and using conservation of mass and the balance of linear momentum, one obtains
\begin{equation}
	\int_{\mathcal{U}}\left[\mathbf{P}-\sum_{i=1}^{3}
	\phi_i\frac{\partial\psi_i}{\partial\mathbf{F}}\right]\!:\!\nabla^{\varphi_0}\mathbf{W}\,dV=0\,.
\end{equation}
At a fixed material point, the value of $\nabla^{\varphi_0}\mathbf{W}$ can be prescribed arbitrarily. Since $\mathcal{U}\subset\mathcal{B}$ is arbitrary, localization yields the generalized Doyle--Ericksen formula
\begin{equation} \label{Generalized-Doyle-Ericksen-Cauchy-Corrected}
	\mathbf{P}=\sum_{i=1}^{3}\phi_i\frac{\partial\psi_i}{\partial\mathbf{F}}\,.
\end{equation}

Up to this point, no assumption has been made concerning the transformation properties of the generalized energy functions $\phi_i$ and $\psi_i$. Covariance of the energy balance, which led to \eqref{Generalized-Doyle-Ericksen-Cauchy-Corrected}, does not by itself impose covariance on these constitutive functions. We now introduce this as an additional and independent postulate. Namely, we require the generalized energy functions to be covariant under arbitrary spatial diffeomorphisms. This requirement may be viewed as the natural geometric extension of the classical principle of material frame indifference to a general Riemannian ambient space. It is also the analogue, for the generalized energy functions in the Darboux representation, of the covariance postulate imposed on the strain-energy function in hyperelasticity \citep{MarsdenHughes1983}.

In order to accommodate material anisotropy, we include the referential structural tensors $\boldsymbol{\zeta}_1,\ldots,\boldsymbol{\zeta}_N$ introduced in \S\ref{Sec:Anisotropic-complementary-Cauchy-elasticity} among the constitutive arguments. Since these tensors are referential, they remain unchanged under spatial diffeomorphisms. More precisely, for every spatial diffeomorphism $\xi:\mathcal{S}\rightarrow\mathcal{S}$,
\begin{equation}
\begin{aligned}
    &\phi_i\left(X,T\xi\cdot\mathbf{F},\mathbf{G},
    \xi_*\mathbf{g},\boldsymbol{\zeta}_1,\ldots,\boldsymbol{\zeta}_N\right)
    =\phi_i\left(X,\mathbf{F},\mathbf{G},\mathbf{g},
    \boldsymbol{\zeta}_1,\ldots,\boldsymbol{\zeta}_N\right)\,,\\
    &\psi_i\left(X,T\xi\cdot\mathbf{F},\mathbf{G},
    \xi_*\mathbf{g},\boldsymbol{\zeta}_1,\ldots,\boldsymbol{\zeta}_N\right)
    =\psi_i\left(X,\mathbf{F},\mathbf{G},\mathbf{g},
    \boldsymbol{\zeta}_1,\ldots,\boldsymbol{\zeta}_N\right)\,,
    \qquad i=1,2,3\,.
\end{aligned}
\end{equation}
Note that the Edelen--Darboux potentials are not unique. In particular, for arbitrary differentiable functions $g_i$ satisfying $g_i'\neq0$, the transformations $\widetilde{\psi}_i=g_i(\psi_i)$ and $\widetilde{\phi}_i=\phi_i/g_i'(\psi_i)$ leave each term unchanged, since $\widetilde{\phi}_i\,d\widetilde{\psi}_i=\phi_i\,d\psi_i$. If $\phi_i$ and $\psi_i$ are covariant, then $\widetilde{\phi}_i$ and $\widetilde{\psi}_i$ are covariant as well.

Covariance of the generalized energy functions implies that their dependence on $\mathbf{F}$ and $\mathbf{g}$ can be expressed through $\mathbf{C}^{\flat}=\mathbf{F}^{\star}\mathbf{g}\mathbf{F}$, with the referential structural tensors retained as additional constitutive arguments.
Therefore, $\phi_i=\widehat{\phi}_i(X,\mathbf{C}^{\flat},\mathbf{G})$ and
$\psi_i=\widehat{\psi}_i(X,\mathbf{C}^{\flat},\mathbf{G})$, $i=1,2,3$. The chain rule gives us
\begin{equation}
	\frac{\partial\psi_i}{\partial\mathbf{F}}=2\mathbf{F}
	\frac{\partial\widehat{\psi}_i}{\partial\mathbf{C}^{\flat}}\,,\qquad i=1,2,3\,.
\end{equation}
Substituting these relations into
\eqref{Generalized-Doyle-Ericksen-Cauchy-Corrected} yields
\begin{equation}
	\mathbf{P}=2\mathbf{F}\sum_{i=1}^{3}\widehat{\phi}_i
	\frac{\partial\widehat{\psi}_i}{\partial\mathbf{C}^{\flat}}\,.
\end{equation}
Using $\mathbf{P}=\mathbf{F}\mathbf{S}$, one obtains
\begin{equation}
	\mathbf{S}=2\sum_{i=1}^{3}\widehat{\phi}_i
	\frac{\partial\widehat{\psi}_i}{\partial\mathbf{C}^{\flat}}\,.
\end{equation}
Because $\mathbf{C}^{\flat}$ is symmetric, each derivative $\partial\widehat{\psi}_i/\partial\mathbf{C}^{\flat}$ is symmetric. Consequently,
\begin{equation}
	\mathbf{S}=\mathbf{S}^{\star}\,,\quad \text{or}\qquad  
	\mathbf{P}\mathbf{F}^{\star}=\mathbf{F}\mathbf{P}^{\star}\,.
\end{equation}
This is the balance of angular momentum. Its derivation requires two separate covariance postulates: covariance of the energy balance and covariance of the generalized energy functions in the Darboux representation. The former identifies the first Piola--Kirchhoff stress with the coefficient obtained from the Darboux representation, while the latter implies that the generalized energy functions depend on $\mathbf{F}$ and $\mathbf{g}$ through $\mathbf{C}^{\flat}$ and consequently implies symmetry of the second Piola--Kirchhoff stress.

\subsection{Balance of angular momentum in complementary Cauchy elasticity} \label{sec:AngularMomentum}

We now derive the balance of angular momentum for complementary Cauchy elastic solids. Recall that the complementary stress-work $1$-form is defined as $\boldsymbol{\Lambda}
=\frac{1}{2}\mathbf{C}^{\flat}\!:\!d\mathbf{S}$. The stress-work and complementary stress-work $1$-forms satisfy the identity
\begin{equation} \label{Stress-Complementary-Stress-Work-Relation}
    \boldsymbol{\Omega}
    =d\left(\frac{1}{2}\mathbf{S}\!:\!\mathbf{C}^{\flat}\right)
    -\boldsymbol{\Lambda}\,.
\end{equation}
Using the product rule, one obtains $\boldsymbol{\Omega}=\frac{1}{2}d\mathbf{S}\!:\!\mathbf{C}^{\flat}+\frac{1}{2}\mathbf{S}\!:\!d\mathbf{C}^{\flat}-\frac{1}{2}\mathbf{C}^{\flat}\!:\!d\mathbf{S}
=\frac{1}{2}\mathbf{S}\!:\!d\mathbf{C}^{\flat}$. Since $\mathbf{C}^{\flat}$ is symmetric, $d\mathbf{C}^{\flat}$ is symmetric, and hence the antisymmetric part of $\mathbf{S}$ does not contribute to the stress-work $1$-form. Therefore,
\begin{equation} \label{Complementary-Stress-Work-Rate}
    \boldsymbol{\Omega}
    =\frac{1}{2}\operatorname{sym}(\mathbf{S})\!:\!d\mathbf{C}^{\flat}
    =\frac{1}{4}\left(\mathbf{S}+\mathbf{S}^{\star}\right)\!:\!d\mathbf{C}^{\flat}\,.
\end{equation}
This identity does not assume that $\mathbf{S}$ is symmetric; it only shows that the stress-work $1$-form is independent of the antisymmetric part of $\mathbf{S}$.
Evaluating \eqref{Complementary-Stress-Work-Rate} along a motion, one obtains
\begin{equation}
    \dot{\boldsymbol{\Omega}}
    =\frac{1}{2}\operatorname{sym}(\mathbf{S})
    \!:\!\dot{\mathbf{C}}^{\flat}\,.
\end{equation}
It is important that \eqref{Complementary-Stress-Work-Rate} has been obtained without assuming symmetry of $\mathbf{S}$. Thus, this representation of the stress-work $1$-form is valid before the balance of angular momentum has been established and may therefore be used in its derivation.

Since $\mathbf{C}^{\flat}=\mathbf{F}^{\star}\mathbf{g}\mathbf{F}$, metric compatibility implies that $\dot{\mathbf{C}}^{\flat}=\left(\nabla^{\varphi_t}\mathbf{V}\right)^{\star}\mathbf{gF}+\mathbf{F}^{\star}\mathbf{g}\nabla^{\varphi_t}\mathbf{V}$.
Substituting this identity into the stress-work rate and using the symmetry of $\operatorname{sym}(\mathbf{S})$, one obtains
\begin{equation} \label{Complementary-Stress-Work-Velocity-Gradient}
    \dot{\boldsymbol{\Omega}}
    =\mathbf{gF}\operatorname{sym}(\mathbf{S})
    \!:\!\nabla^{\varphi_t}\mathbf{V}\,.
\end{equation}

Therefore, the balance of energy is written as
\begin{equation}
	\begin{aligned}\frac{d}{dt}\int_{\mathcal{U}}
	\frac{1}{2}\rho_0\lVert\mathbf{V}\rVert_{\mathbf{g}}^2\,dV
	+\int_{\mathcal{U}}\left[\mathbf{gF}\operatorname{sym}(\mathbf{S})
	\!:\!\nabla^{\varphi_t}\mathbf{V}+\rho_0\mathsf{f}\dot{\mathscr{N}}\right]dV
	&=\int_{\mathcal{U}}\rho_0\left(\llangle\mathbf{B},\mathbf{V}\rrangle_{\mathbf{g}}+R\right)dV\\
	&\quad+\int_{\partial\mathcal{U}}\left(\llangle\mathbf{T},\mathbf{V}\rrangle_{\mathbf{g}}	
	+H\right)dA\,.
\end{aligned}
\end{equation}
The traction is related to the first Piola--Kirchhoff stress by $\mathbf{T}=\mathbf{P}\mathbf{N}^{\flat}$. Applying the divergence theorem to the traction-power term gives us \eqref{Traction-Divergence}. We now postulate covariance of the energy balance under arbitrary time-dependent spatial diffeomorphisms. Evaluating the transformed balance at $t=0$, subtracting the original balance, and using conservation of mass and the balance of linear momentum, one obtains
\begin{equation}
	\int_{\mathcal{U}}\left[\mathbf{P}
	-\mathbf{F}\operatorname{sym}(\mathbf{S})\right]\!:\!\nabla^{\varphi_0}\mathbf{W}\,dV=0\,.
\end{equation}
The value of $\nabla^{\varphi_0}\mathbf{W}$ may be prescribed arbitrarily at a material point. Since $\mathcal{U}$ is arbitrary, localization gives us
\begin{equation} \label{Complementary-Cauchy-Generalized-Doyle-Ericksen}
	\mathbf{P}=\mathbf{F}\operatorname{sym}(\mathbf{S})\,.
\end{equation}
On the other hand, the first and second Piola--Kirchhoff stresses are related by $\mathbf{P}=\mathbf{F}\mathbf{S}$. Comparing this relation with \eqref{Complementary-Cauchy-Generalized-Doyle-Ericksen} and using the invertibility of $\mathbf{F}$, one concludes that
\begin{equation}
	\mathbf{S}=\operatorname{sym}(\mathbf{S})=\mathbf{S}^{\star}\,,\quad
	\text{or~equivalently}\qquad
	\mathbf{P}\mathbf{F}^{\star}=\mathbf{F}\mathbf{P}^{\star}\,,
\end{equation}
which is the material representation of the balance of angular momentum. Thus, covariance of the energy balance implies the balance of angular momentum without any prior assumption of symmetry of the second Piola--Kirchhoff stress.

\begin{remark}
There is a fundamental difference between the preceding derivation and its counterpart in Cauchy elasticity. In Cauchy elasticity, the independent constitutive variable is the right Cauchy--Green strain $\mathbf{C}^{\flat}$, which is kinematically symmetric. Thus, the stress-work $1$-form is defined on a six-dimensional constitutive manifold, independently of the balance of angular momentum. The Darboux classification provides a representation of this $1$-form in terms of generalized energy functions. Covariance of the energy balance then yields the generalized Doyle--Ericksen formula \eqref{Generalized-Doyle-Ericksen-Cauchy-Corrected}, but this relation alone does not imply symmetry of $\mathbf{S}$. An additional postulate---covariance of the generalized energy functions---is needed to restrict their dependence on $\mathbf{F}$ and $\mathbf{g}$ to dependence through $\mathbf{C}^{\flat}$ (and any referential structural tensors). The resulting constitutive relation implies that $\mathbf{S}$ is symmetric.

In complementary Cauchy elasticity, by contrast, the independent constitutive variable is $\mathbf{S}$, which cannot initially be assumed symmetric. Nevertheless, its work-conjugate variable $\mathbf{C}^{\flat}$ is kinematically symmetric. Consequently, the stress-work rate depends only on $\operatorname{sym}(\mathbf{S})$, and covariance of the energy balance directly yields $\mathbf{P}=\mathbf{F}\operatorname{sym}(\mathbf{S})$. Comparing this with $\mathbf{P}=\mathbf{F}\mathbf{S}$ implies that $\mathbf{S}=\mathbf{S}^{\star}$. Thus, unlike in Cauchy elasticity, neither the Darboux classification nor an additional covariance postulate on the generalized energy functions is needed to derive the balance of angular momentum. However, the six-dimensional Darboux classification of the complementary stress-work $1$-form is justified only after symmetry of $\mathbf{S}$ has been established.
\end{remark}

\subsection{The second law of thermodynamics} \label{sec:2ndLaw}

The covariance of the energy balance identifies the tensor entering the internal mechanical power with the first Piola--Kirchhoff stress. The scalar field $\mathsf{f}$ conjugate to the entropy rate remains to be determined. This identification follows from the second law of thermodynamics.
For every material subbody $\mathcal{U}\subset\mathcal{B}$, the Clausius--Duhem inequality is written as \citep{Truesdell1952,gurtin1974modern,MarsdenHughes1983}
\begin{equation}
	\frac{d}{dt}\int_{\mathcal{U}}\rho_0\mathscr{N}\,dV
	-\int_{\mathcal{U}}\rho_0\frac{R}{T}\,dV
	-\int_{\partial\mathcal{U}}\frac{H}{T}\,dA
	\geq0\,,
\end{equation}
where $T$ is the absolute temperature. Let $H=-\llangle\mathbf{Q},\mathbf{N}\rrangle_{\mathbf{G}}$, where $\mathbf{Q}$ is the material heat-flux vector and $\mathbf{N}$ is the outward $\mathbf{G}$-unit normal to $\partial\mathcal{U}$. Localizing the Clausius--Duhem inequality and multiplying it by $T$ gives the material dissipation inequality
\begin{equation}
	\mathscr{D}=\rho_0T\dot{\mathscr{N}}+\operatorname{Div}\mathbf{Q}
	-\rho_0R -\frac{1}{T}\langle dT,\mathbf{Q}\rangle \geq0\,,
\end{equation}
where $\mathscr{D}$ is the material dissipation density and $\langle dT,\mathbf{Q}\rangle$ denotes the natural pairing of the temperature gradient and the material heat flux.
Substitution of \eqref{LocalEnergyBalance} into the dissipation inequality yields
\begin{equation}
	\mathscr{D}
	=\rho_0(T-\mathsf{f})\dot{\mathscr{N}}
	-\frac{1}{T}\langle dT,\mathbf{Q}\rangle
	\geq0\,.
\end{equation}
For a spatially uniform temperature, $dT=0$. Requiring the inequality to hold for arbitrary entropy rates then implies that $\mathsf{f}=T$. Consequently, the dissipation inequality reduces to the heat-conduction restriction
\begin{equation}
	\mathscr{D}
	=-\frac{1}{T}\langle dT,\mathbf{Q}\rangle
	\geq0\,,
\end{equation}
and the entropy balance is written as
\begin{equation}
	\rho_0T\dot{\mathscr{N}}
	=\rho_0R-\operatorname{Div}\mathbf{Q}\,.
\end{equation}
The local first law may therefore be written as
\begin{equation}
	\rho_0\dot{\mathscr{E}}
	=\mathbf{P}\!:\!\nabla^{\varphi_t}\mathbf{V}
	+\rho_0T\dot{\mathscr{N}}
	=\mathbf{P}\!:\!\nabla^{\varphi_t}\mathbf{V}
	+\rho_0R-\operatorname{Div}\mathbf{Q}\,.
\end{equation}
Using the work-conjugacy relation between $\mathbf{S}$ and $\frac{1}{2}\mathbf{C}^{\flat}$ one writes
\begin{equation}
	\rho_0\dot{\mathscr{E}}=\frac{1}{2}\mathbf{S}\!:\!\dot{\mathbf{C}}^{\flat}
	+\rho_0T\dot{\mathscr{N}}\,.
\end{equation}
For a complementary Cauchy elastic response, the strain is prescribed by $\mathbf{C}^{\flat}
=\mathbf{C}^{\flat}(\mathbf{S},\mathbf{G})$, and therefore
\begin{equation}
	\dot{\mathbf{C}}^{\flat}
	=\mathbb{A}[\dot{\mathbf{S}}]\,,
	\qquad
	\mathbb{A}_{ABCD}
	=\frac{\partial C_{AB}}{\partial S^{CD}}\,.
\end{equation}
Thus, the first law may be expressed entirely in terms of the stress-space variables as
\begin{equation}
	\rho_0\dot{\mathscr{E}}
	=\frac{1}{2}\mathbf{S}\!:\!\mathbb{A}[\dot{\mathbf{S}}]
	+\rho_0T\dot{\mathscr{N}}\,.
\end{equation}
The mechanical term in this identity is the stress-work rate. It must not be confused with the complementary stress-work rate $\boldsymbol{\Lambda}(\dot{\mathbf{S}})	=\frac{1}{2}\mathbf{C}^{\flat}\!:\!\dot{\mathbf{S}}$. The two rates are related as
\begin{equation}
	\frac{d}{dt}\left(\frac{1}{2}\mathbf{S}\!:\!\mathbf{C}^{\flat}\right)
	=\frac{1}{2}\mathbf{S}\!:\!\mathbb{A}[\dot{\mathbf{S}}]
	+\boldsymbol{\Lambda}(\dot{\mathbf{S}})\,.
\end{equation}
Accordingly, the pullback of the stress-work $1$-form to stress space is given as
\begin{equation}
	\boldsymbol{\Omega}
	=\frac{1}{2}\mathbf{S}\!:\!\mathbb{A}[d\mathbf{S}]
	=d\left(\frac{1}{2}\mathbf{S}\!:\!\mathbf{C}^{\flat}\right)
	-\boldsymbol{\Lambda}\,.
\end{equation}

\begin{remark}[Cyclic stress histories]
Suppose that the process is adiabatic and has neither a heat supply nor a heat flux. The first law then reduces to
\begin{equation}
	\rho_0\dot{\mathscr{E}}\,dt
	=\frac{1}{2}\mathbf{S}\!:\!\mathbb{A}[d\mathbf{S}]
	=\boldsymbol{\Omega}\,.
\end{equation}
Let $\gamma$ be a closed stress path. Since the strain response is a single-valued function of stress, the corresponding strain path is also closed. Therefore,
\begin{equation}
	\int_{\gamma}
	d\left(\frac{1}{2}\mathbf{S}\!:\!\mathbf{C}^{\flat}\right)=0\,,
\end{equation}
and hence
\begin{equation}
	\int_{\gamma}\boldsymbol{\Omega}=-\int_{\gamma}\boldsymbol{\Lambda}\,.
\end{equation}
Thus, nonzero complementary stress-work around a closed stress path is compatible with the first and second laws of thermodynamics. It is the negative of the stress work around the corresponding constitutive cycle and is not an entropy-production term. In particular, the second law does not require $\boldsymbol{\Lambda}$ to be closed, does not impose the major symmetry of the tangent compliance tensor, and does not reduce complementary Cauchy elasticity to complementary hyperelasticity.
\end{remark}

\section{Linearized Implicit, Cauchy, and Complementary Cauchy Elasticity}
\label{Sec:LinearizedElasticity}

Linear elasticity may be obtained by linearizing a nonlinear constitutive theory about a given finitely deformed configuration. The resulting linear theory depends, in general, on both the deformation and the stress in the configuration about which the linearization is carried out. The classical theory of small deformations superposed on finite deformations goes back at least to \citet{GreenRivlinShield1952}; a systematic geometric treatment of incremental elasticity was later developed by \citet{MarsdenHughes1983}. Our interest here is in the linearization of implicit elasticity. The kinematic and balance-law linearizations are the same as those of classical nonlinear elasticity; what changes is the constitutive equation. We therefore recall only the kinematic relations needed for the linearization of the implicit constitutive equation.

We first linearize an implicit elastic constitutive relation about an arbitrary finitely deformed and prestressed configuration. Linearized Cauchy elasticity and linearized complementary Cauchy elasticity are then obtained as the two explicit cases in which the tangent constitutive relation can be solved locally for the incremental stress and incremental strain, respectively. The configuration about which the linearization is carried out is not assumed to be undeformed or stress free.

\subsection{Linearized kinematics about a finitely deformed configuration}

Let $\mathring{\varphi}:\mathcal{B}\rightarrow\mathcal{S}$ be a fixed deformation, and let $\mathring{\mathbf{F}}=T\mathring{\varphi}$ and $\mathring{\mathbf{C}}^{\flat}=\mathring{\varphi}^{*}\mathbf{g}=\mathring{\mathbf{F}}^{\star}\mathbf{g}\mathring{\mathbf{F}}$. We consider a smooth one-parameter family of deformations $\varphi_{\epsilon}$ such that $\varphi_0=\mathring{\varphi}$ and define the variation field
\begin{equation}
	\mathbf{U}=\frac{d}{d\epsilon}\Big|_{\epsilon=0}\varphi_{\epsilon}\,.
\end{equation}
Its spatial representation on the deformed configuration $\mathring{\varphi}(\mathcal{B})$ is $\mathbf{u}=\mathbf{U}\circ\mathring{\varphi}^{-1}$. The infinitesimal strain associated with this incremental displacement is written as
\begin{equation}
	\boldsymbol{\epsilon}=\frac{1}{2}\big(\nabla^{\mathbf g}\mathbf{u}^{\flat}
	+[\nabla^{\mathbf g}\mathbf{u}^{\flat}]^{\star}\big)\,.
\end{equation}
Linearization of the right Cauchy--Green strain about $\mathring{\varphi}$ is computed as
\begin{equation}\label{Linearized-Cauchy-Green}
	\delta\mathbf{C}^{\flat}\coloneq\frac{d}{d\epsilon}\Big|_{\epsilon=0}\mathbf{C}_{\epsilon}^{\flat}	
	=\mathring{\varphi}^{*}\mathcal{L}_{\mathbf{u}}\mathbf{g}
	=2\mathring{\varphi}^{*}\boldsymbol{\epsilon}=2\boldsymbol{\varepsilon}\,,
\end{equation}
where $\boldsymbol{\varepsilon}=\mathring{\varphi}^{*}\boldsymbol{\epsilon}$ is the linearized strain. Notice that \eqref{Linearized-Cauchy-Green} is valid for an arbitrary finite base deformation $\mathring{\varphi}$; in particular, $\mathring{\mathbf{C}}^{\flat}$ need not coincide with the material metric $\mathbf{G}$.

In what follows, $\mathring{\mathbf{S}}$ denotes the second Piola--Kirchhoff stress associated with the initial state. No assumption is made that $\mathring{\mathbf{S}}=\mathbf{0}$. Thus, the constitutive equations are linearized about the finitely deformed state $\big(\mathring{\mathbf{C}}^{\flat},\mathring{\mathbf{S}}\big)$.

\subsection{Linearized implicit elasticity}

Consider a smooth implicit elastic constitutive relation $\boldsymbol{\mathcal F}\big(\mathbf{C}^{\flat},\mathbf{S},\mathbf{G}\big)=\mathbf{0}$ and let $\big(\mathring{\mathbf{C}}^{\flat},\mathring{\mathbf{S}}\big)$ be a regular constitutive state, so that $\boldsymbol{\mathcal F}\big(\mathring{\mathbf{C}}^{\flat},\mathring{\mathbf{S}},\mathbf{G}\big)=\mathbf{0}$. Let $\big(\mathbf{C}_{\epsilon}^{\flat},\mathbf{S}_{\epsilon}\big)$ be a smooth one-parameter family of constitutive states passing through the base state at $\epsilon=0$. Since the material metric is fixed, differentiation of the constitutive equation gives us
\begin{equation}\label{Linearized-Implicit-General}
	D_{\mathbf{C}^{\flat}}\boldsymbol{\mathcal F}\Big|_0[\delta\mathbf{C}^{\flat}]
	+D_{\mathbf{S}}\boldsymbol{\mathcal F}\Big|_0[\delta\mathbf{S}]=\mathbf{0}\,,
\end{equation}
Here and throughout this subsection, the notation $|_0$ indicates evaluation at the base constitutive state $\big(\mathring{\mathbf{C}}^{\flat},\mathring{\mathbf{S}},\mathbf{G}\big)$, corresponding to $\epsilon=0$. Using \eqref{Linearized-Cauchy-Green}, the incremental constitutive relation is written as
\begin{equation}\label{Linearized-Implicit-Strain}
	2D_{\mathbf{C}^{\flat}}\boldsymbol{\mathcal F}\Big|_0[\boldsymbol{\varepsilon}]
	+D_{\mathbf{S}}\boldsymbol{\mathcal F}\Big|_0[\delta\mathbf{S}]=\mathbf{0}\,.
\end{equation}

Equation \eqref{Linearized-Implicit-Strain} is the linearized constitutive equation of implicit elasticity. Geometrically, it characterizes the tangent space to the constitutive manifold at the initial state:
\begin{equation}
	\big(2\boldsymbol{\varepsilon},\delta\mathbf{S}\big)\in 
	T_{\big(\mathring{\mathbf{C}}^{\flat},\mathring{\mathbf{S}}\big)}\mathcal{K}_X
	=\ker D\boldsymbol{\mathcal F}\Big|_0\,.
\end{equation}
Thus, the linearized implicit constitutive equation defines a six-dimensional linear subspace of the product of the incremental strain and stress spaces. It need not be the graph of stress over strain or of strain over stress.

Suppose first that $D_{\mathbf{S}}\boldsymbol{\mathcal F}\Big|_0$ is nonsingular. Equation \eqref{Linearized-Implicit-Strain} can then be solved uniquely for the incremental stress:
\begin{equation}\label{Implicit-Linearized-Stiffness}
	\delta\mathbf{S}=-2\left(D_{\mathbf{S}}\boldsymbol{\mathcal F}\Big|_0\right)^{-1}
	D_{\mathbf{C}^{\flat}}\boldsymbol{\mathcal F}\Big|_0[\boldsymbol{\varepsilon}]
	\eqcolon \boldsymbol{\mathbb C}[\boldsymbol{\varepsilon}]\,.
\end{equation}
Hence, the incremental stiffness operator associated with the implicit constitutive relation is
\begin{equation}\label{Implicit-Stiffness-Operator}
	\boldsymbol{\mathbb C}=-2\left(D_{\mathbf{S}}\boldsymbol{\mathcal F}\Big|_0\right)^{-1}
	D_{\mathbf{C}^{\flat}}\boldsymbol{\mathcal F}\Big|_0\,.
\end{equation}
Alternatively, suppose that $D_{\mathbf{C}^{\flat}}\boldsymbol{\mathcal F}\Big|_0$ is nonsingular. The same linearized implicit relation may then be solved uniquely for the incremental strain:
\begin{equation}\label{Implicit-Linearized-Compliance}
	\boldsymbol{\varepsilon}
	=-\frac{1}{2}\left(D_{\mathbf{C}^{\flat}}\boldsymbol{\mathcal F}\Big|_0\right)^{-1}
	D_{\mathbf{S}}\boldsymbol{\mathcal F}\Big|_0[\delta\mathbf{S}]
	\eqcolon \boldsymbol{\mathbb A}[\delta\mathbf{S}]\,.
\end{equation}
Thus, the incremental compliance operator is
\begin{equation}\label{Implicit-Compliance-Operator}
	\boldsymbol{\mathbb A}
	=-\frac{1}{2}\left(D_{\mathbf{C}^{\flat}}\boldsymbol{\mathcal F}\Big|_0\right)^{-1}
	D_{\mathbf{S}}\boldsymbol{\mathcal F}\Big|_0\,.
\end{equation}
If both partial derivatives are nonsingular, both explicit incremental descriptions exist, and $	\boldsymbol{\mathbb A}=\boldsymbol{\mathbb C}^{-1}$.
If one of the two partial derivatives is singular, only one of these explicit linearized representations may exist. If both are singular, the tangent constitutive relation \eqref{Linearized-Implicit-Strain} remains meaningful even though it cannot be written locally in either stiffness or compliance form. This is the linearized counterpart of the distinction among implicit, Cauchy, and complementary Cauchy elasticity established in \S\ref{Sec:DifferentialGeometricStructure}.

\subsection{Linearized Cauchy elasticity}

Let us consider now a Cauchy elastic constitutive equation $\mathbf{S}=\widehat{\mathbf{S}}\big(\mathbf{C}^{\flat},\mathbf{G}\big)$. Linearization about the finitely strained and prestressed state $\big(\mathring{\mathbf{C}}^{\flat},\mathring{\mathbf{S}}\big)$ gives us
\begin{equation}
	\delta\mathbf{S}
	=D_{\mathbf{C}^{\flat}}\widehat{\mathbf{S}}\Big|_0[\delta\mathbf{C}^{\flat}]
	=2D_{\mathbf{C}^{\flat}}\widehat{\mathbf{S}}\Big|_0[\boldsymbol{\varepsilon}]\,.
\end{equation}
Let us define the incremental elasticity tensor by
\begin{equation}\label{Linearized-Cauchy-Elasticity-Tensor}
	\boldsymbol{\mathbb C}=2D_{\mathbf{C}^{\flat}}\widehat{\mathbf{S}}\Big|_0\,.
\end{equation}
The linearized constitutive equation takes the form $\delta\mathbf{S}=\boldsymbol{\mathbb C}[\boldsymbol{\varepsilon}]$. In components,
\begin{equation}
	\delta S^{AB}=\mathbb C^{ABCD}\varepsilon_{CD}\,,\qquad 
	\mathbb C^{ABCD}=2\frac{\partial\widehat S^{AB}}{\partial C_{CD}}\Bigg|_0\,.
\end{equation}
The minor symmetries of $\boldsymbol{\mathbb C}$ follow from the symmetry of the stress and strain tensors. Major symmetry, $\mathbb C^{ABCD}=\mathbb C^{CDAB}$ is an additional constitutive restriction. It holds when the stress-work $1$-form is locally closed at the base state and, in particular, for a hyperelastic response. A general Cauchy elastic solid need not possess this symmetry.

\subsection{Linearized complementary Cauchy elasticity}

Consider next a complementary Cauchy elastic constitutive equation $\mathbf{C}^{\flat}=\widehat{\mathbf{C}}^{\flat}\big(\mathbf{S},\mathbf{G}\big)$. Linearization about the finitely strained and prestressed constitutive state gives us
\begin{equation}
	\delta\mathbf{C}^{\flat}=D_{\mathbf{S}}\widehat{\mathbf{C}}^{\flat}\Big|_0[\delta\mathbf{S}]\,.
\end{equation}
Since $\delta\mathbf{C}^{\flat}=2\boldsymbol{\varepsilon}$, one obtains
\begin{equation}
	\boldsymbol{\varepsilon}
	=\frac{1}{2}D_{\mathbf{S}}\widehat{\mathbf{C}}^{\flat}\Big|_0[\delta\mathbf{S}]\,.
\end{equation}
We therefore define the incremental compliance tensor by
\begin{equation}\label{Linearized-Complementary-Compliance}
	\boldsymbol{\mathbb A}=\frac{1}{2}D_{\mathbf{S}}\widehat{\mathbf{C}}^{\flat}\Big|_0\,,
\end{equation}
so that $\boldsymbol{\varepsilon}=\boldsymbol{\mathbb A}[\delta\mathbf{S}]$. In components,
\begin{equation}
	\varepsilon_{AB}=\mathbb A_{ABCD}\,\delta S^{CD}\,,\qquad 
	\mathbb A_{ABCD}=\frac{1}{2}\frac{\partial\widehat C_{AB}}{\partial S^{CD}}\Bigg|_0\,.
\end{equation}
The minor symmetries of $\boldsymbol{\mathbb A}$ follow from the symmetry of strain and stress. Complementary hyperelasticity imposes the additional major symmetry $\mathbb A_{ABCD}=\mathbb A_{CDAB}$. A general complementary Cauchy elastic response need not satisfy this condition.

\subsection{Hyperelasticity of the linearized constitutive equations}

An interesting question arises when a non-hyperelastic elastic solid is linearized about a given constitutive state. One might expect the lack of hyperelasticity of the nonlinear constitutive equation to be inherited by its linearization. This is not necessarily the case. The obstruction to hyperelasticity may disappear at first order, so that the linearized constitutive equation is hyperelastic even though the underlying nonlinear solid is not. The non-hyperelastic character of the material may become visible only at second or higher order. This phenomenon has been demonstrated for nonlinear Cauchy elasticity by \citet{GillandersYavariGoriely2026}. Here, we extend the analysis to general implicit elasticity and examine Cauchy and complementary Cauchy elasticity as special cases.

Let us consider a smooth implicit elastic constitutive equation $\boldsymbol{\mathcal F}
\big(\mathbf{C}^{\flat},\mathbf{S},\mathbf{G}\big)=\mathbf{0}$, and let $\big(\mathring{\mathbf{C}}^{\flat},\mathring{\mathbf{S}}\big)$ be a regular constitutive state. As established in the preceding subsection, its linearization is given in \eqref{Linearized-Implicit-Strain}. The admissible pairs $\big(2\boldsymbol{\varepsilon},\delta\mathbf{S}\big)$ form the tangent space to the constitutive manifold at the base state. This tangent space is six-dimensional, but it need not be the graph of a linear map from strain increments to stress increments or vice versa. Recall that the stress-work $1$-form and its exterior derivative read $\boldsymbol{\Omega}=\frac{1}{2}\mathbf{S}\!:\!\mathrm{d}\mathbf{C}^{\flat}$, and $\mathrm{d}\boldsymbol{\Omega}=\frac{1}{2}\mathrm{d}\mathbf{S}
\wedge\mathrm{d}\mathbf{C}^{\flat}$, respectively, where the wedge product includes contraction of the tensor indices. Consequently, for two tangent vectors $\big(2\boldsymbol{\varepsilon}_1,\delta\mathbf{S}_1\big),  \big(2\boldsymbol{\varepsilon}_2,\delta\mathbf{S}_2\big)  \in T_{\big(\mathring{\mathbf{C}}^{\flat}, \mathring{\mathbf{S}}\big)}\mathcal{K}_X$, one has
\begin{equation}
    \mathrm{d}\boldsymbol{\Omega}\Big|_0
    \Big[ \big(2\boldsymbol{\varepsilon}_1,\delta\mathbf{S}_1\big),
    \big(2\boldsymbol{\varepsilon}_2,\delta\mathbf{S}_2\big) \Big]
    = \delta\mathbf{S}_1\!:\!\boldsymbol{\varepsilon}_2
    -\delta\mathbf{S}_2\!:\!\boldsymbol{\varepsilon}_1\,.
\end{equation}
Thus, the linearized implicit constitutive relation is hyperelastic if and only if
\begin{equation}\label{Linearized-Implicit-Hyperelasticity}
    \delta\mathbf{S}_1\!:\!\boldsymbol{\varepsilon}_2
    =  \delta\mathbf{S}_2\!:\!\boldsymbol{\varepsilon}_1\,,
\end{equation}
for every pair of admissible incremental constitutive states. Geometrically, this condition means that the tangent constitutive space is a Lagrangian subspace of the product of the strain and stress spaces equipped with the canonical symplectic structure induced by $\mathrm{d}\boldsymbol{\Omega}$. This characterization does not require the existence of either an incremental stiffness or an incremental compliance operator.

If the linearized implicit relation admits a stiffness representation $\delta\mathbf{S}    =\boldsymbol{\mathbb C}[\boldsymbol{\varepsilon}]$, condition \eqref{Linearized-Implicit-Hyperelasticity} reduces to
\begin{equation}
    \boldsymbol{\varepsilon}_1\!:\!
    \boldsymbol{\mathbb C}[\boldsymbol{\varepsilon}_2]
    = \boldsymbol{\varepsilon}_2\!:\!
    \boldsymbol{\mathbb C}[\boldsymbol{\varepsilon}_1]\,.
\end{equation}
Hence, the incremental stiffness operator must have the major symmetry. Similarly, when a compliance representation exists, i.e., $\boldsymbol{\varepsilon}=\boldsymbol{\mathbb A}[\delta\mathbf{S}]$, the corresponding condition is
\begin{equation}
    \delta\mathbf{S}_1\!:\!
    \boldsymbol{\mathbb A}[\delta\mathbf{S}_2]
    =\delta\mathbf{S}_2\!:\!
    \boldsymbol{\mathbb A}[\delta\mathbf{S}_1]\,.
\end{equation}
Thus, major symmetry of the stiffness and compliance operators represents two special cases of the same geometric condition.

The nonlinear implicit constitutive relation is locally hyperelastic when the restriction of $\mathrm{d}\boldsymbol{\Omega}$ to its constitutive manifold vanishes. For its linearization to be hyperelastic, however, it is sufficient that this restriction vanish at the particular constitutive state about which the linearization is performed. Consequently, a nonlinear implicit constitutive relation may fail to be hyperelastic even though its linearization at a given state is hyperelastic. This observation applies regardless of whether the constitutive relation admits an explicit stress or strain representation.

Let us consider first a Cauchy elastic solid with constitutive equation $\mathbf{S}=\widehat{\mathbf{S}}(\mathbf{C}^{\flat},\mathbf{G})$, and suppose that the solid is not hyperelastic. Let $\boldsymbol{\Omega}=\frac{1}{2}\mathbf{S}\!:\!\mathrm{d}\mathbf{C}^{\flat}$ denote the stress-work one-form. Non-hyperelasticity means that $\mathrm{d}\boldsymbol{\Omega}$ does not vanish identically on the constitutive domain. This, however, does not require $\mathrm{d}\boldsymbol{\Omega}$ to be nonzero at every deformation state.
Linearization about $\mathring{\mathbf{C}}^{\flat}$ gives us $\delta\mathbf{S}=\mathbb{C}_0\!:\!\boldsymbol{\varepsilon}$, where $\mathbb{C}_0=2D_{\mathbf{C}^{\flat}}\widehat{\mathbf{S}}\Big|_0$. The linearized constitutive equation is hyperelastic if and only if $\mathbb{C}_0$ has the major symmetry. Equivalently, for arbitrary symmetric covariant tensors $\mathbf{A}$ and $\mathbf{B}$, $\mathbf{A}\!:\!\mathbb{C}_0[\mathbf{B}]=\mathbf{B}\!:\!\mathbb{C}_0[\mathbf{A}]$.
This condition is equivalent to $\mathrm{d}\boldsymbol{\Omega}\big|_0=\mathbf{0}$. Thus, although $\mathrm{d}\boldsymbol{\Omega}$ does not vanish identically for a non-hyperelastic Cauchy elastic solid, its value may vanish at the state about which the constitutive equation is linearized. In that case, the first-order theory is hyperelastic.
More specifically, suppose that $\mathrm{d}\boldsymbol{\Omega}\big|_0=\mathbf{0}$, while $D_{\mathbf{C}^{\flat}}(\mathrm{d}\boldsymbol{\Omega})\big|_0\neq\mathbf{0}$. Then the obstruction to hyperelasticity is absent from the first-order theory but may appear at the next order. Hence, non-hyperelasticity of a nonlinear Cauchy elastic solid need not be detectable from its linearization about a single deformation state.

An entirely analogous situation occurs in complementary Cauchy elasticity. Consider a complementary Cauchy elastic solid with constitutive equation $\mathbf{C}^{\flat}=\widehat{\mathbf{C}}^{\flat}(\mathbf{S},\mathbf{G})$, and suppose that the solid is not complementary hyperelastic. Let $\boldsymbol{\Lambda}=\frac{1}{2}\mathbf{C}^{\flat}\!:\!\mathrm{d}\mathbf{S}$ denote the complementary stress-work one-form. Non-complementary-hyperelasticity means that $\mathrm{d}\boldsymbol{\Lambda}$ does not vanish identically on the constitutive domain, although it may vanish at particular stress states. Linearization about $\mathring{\mathbf{S}}$ gives us $\boldsymbol{\varepsilon}=\mathbb{K}_0\!:\!\delta\mathbf{S}$, where $\mathbb{K}_0=\frac{1}{2}D_{\mathbf{S}}\widehat{\mathbf{C}}^{\flat}\Big|_0$. The linearized constitutive equation is complementary hyperelastic if and only if $\mathbb{K}_0$ has the major symmetry. Equivalently, for arbitrary symmetric contravariant tensors $\mathbf{A}$ and $\mathbf{B}$, $\mathbf{A}\!:\!\mathbb{K}_0[\mathbf{B}]=\mathbf{B}\!:\!\mathbb{K}_0[\mathbf{A}]$. This condition is equivalent to $\mathrm{d}\boldsymbol{\Lambda}\big|_0=\mathbf{0}$. Thus, although $\mathrm{d}\boldsymbol{\Lambda}$ does not vanish identically for a non-complementary-hyperelastic solid, its value may vanish at the state about which the constitutive equation is linearized. In that case, the first-order theory is complementary hyperelastic. More specifically, suppose that $\mathrm{d}\boldsymbol{\Lambda}\big|_0=\mathbf{0}$, while $D_{\mathbf{S}}(\mathrm{d}\boldsymbol{\Lambda})\big|_0\neq\mathbf{0}$. Then the obstruction to complementary hyperelasticity is absent from the first-order theory but may appear at the next order. Hence, non-complementary-hyperelasticity of a nonlinear complementary Cauchy elastic solid need not be detectable from its linearization about a single stress state.

\begin{example}
Let $\mathbf{M}$ be a fixed symmetric material tensor that is not proportional to $\mathbf{G}^{\sharp}$, and consider, in a neighborhood of $\mathbf{C}^{\flat}=\mathbf{G}$, the following Cauchy elastic constitutive equation
\begin{equation}
    \mathbf{S}
    =\lambda \operatorname{tr}(\mathbf{C}^{\flat}-\mathbf{G})\,\mathbf{G}^{\sharp}
    +2\mu\,(\mathbf{C}^{\flat}-\mathbf{G})^{\sharp}
    +\alpha\left[\mathbf{M}\!:\!(\mathbf{C}^{\flat}-\mathbf{G})\right]^2\mathbf{G}^{\sharp}\,.
\end{equation}
Writing $2\mathbf{E}^{\flat}=\mathbf{C}^{\flat}-\mathbf{G}$, differentiation in the direction of an arbitrary symmetric tensor $\mathbf{A}$, one obtains
\begin{equation}
    D_{\mathbf{C}^{\flat}}\widehat{\mathbf{S}}[\mathbf{A}]
    =\lambda (\operatorname{tr}\mathbf{A})\,\mathbf{G}^{\sharp}
    +2\mu\,\mathbf{A}^{\sharp}
    +4\alpha(\mathbf{M}\!:\!\mathbf{E}^{\flat})(\mathbf{M}\!:\!\mathbf{A})\,\mathbf{G}^{\sharp}\,.
\end{equation}
At $\mathbf{C}^{\flat}=\mathbf{G}$, one has $\mathbf{E}^{\flat}=\mathbf{0}$, and hence
\begin{equation}
    D_{\mathbf{C}^{\flat}}\widehat{\mathbf{S}}\Big|_{\mathbf{C}^{\flat}=\mathbf{G}}\,[\mathbf{A}]
    =\lambda (\operatorname{tr}\mathbf{A})\,\mathbf{G}^{\sharp}
    +2\mu\,\mathbf{A}^{\sharp}\,.
\end{equation}
This linear map is self-adjoint, and hence the linearized constitutive equation about $\mathbf{C}^{\flat}=\mathbf{G}$ is hyperelastic.
The nonlinear constitutive equation, however, is not hyperelastic. For arbitrary symmetric tensors $\mathbf{A}$ and $\mathbf{B}$, the antisymmetric part of the constitutive derivative is given as
\begin{equation}
    \mathbf{B}\!:\!D_{\mathbf{C}^{\flat}}\widehat{\mathbf{S}}[\mathbf{A}]
    -\mathbf{A}\!:\!D_{\mathbf{C}^{\flat}}\widehat{\mathbf{S}}[\mathbf{B}]
    =4\alpha(\mathbf{M}\!:\!\mathbf{E}^{\flat})
    \left[
    (\mathbf{M}\!:\!\mathbf{A})\operatorname{tr}\mathbf{B}
    -(\mathbf{M}\!:\!\mathbf{B})\operatorname{tr}\mathbf{A}
    \right]\,.
\end{equation}
For $\alpha\neq0$ and $\mathbf{M}$ not proportional to $\mathbf{G}^{\sharp}$, the right-hand side does not vanish identically. Thus, the nonlinear Cauchy elastic constitutive equation is not hyperelastic, even though its linearization about $\mathbf{C}^{\flat}=\mathbf{G}$ is hyperelastic. In this example, the obstruction to hyperelasticity first appears at quadratic order in the constitutive response.
\end{example}

\begin{example}
Let $\mathbf{M}$ be a fixed symmetric material tensor that is not proportional to $\mathbf{G}$, and consider, in a neighborhood of $\mathbf{S}=\mathbf{0}$, the following complementary Cauchy elastic constitutive equation
\begin{equation}
    \mathbf{C}^{\flat}
    =\mathbf{G}
    +a(\mathbf{G}\!:\!\mathbf{S})\mathbf{G}
    +b\,\mathbf{S}^{\flat}
    +\beta(\mathbf{M}\!:\!\mathbf{S})^2\mathbf{G}\,,
\end{equation}
where $a$, $b$, and $\beta$ are material constants. Its derivative in the direction of an arbitrary symmetric tensor $\mathbf{A}$ is
\begin{equation}
    D_{\mathbf{S}}\widehat{\mathbf{C}}^{\flat}[\mathbf{A}]
    =a(\mathbf{G}\!:\!\mathbf{A})\mathbf{G}+b\,\mathbf{A}^{\flat}
    +2\beta(\mathbf{M}\!:\!\mathbf{S})(\mathbf{M}\!:\!\mathbf{A})\mathbf{G}\,.
\end{equation}
At $\mathbf{S}=\mathbf{0}$, the quadratic term vanishes, and hence
\begin{equation}
    D_{\mathbf{S}}\widehat{\mathbf{C}}^{\flat}\Big|_{\mathbf{S}=\mathbf{0}}[\mathbf{A}]
    =a(\mathbf{G}\!:\!\mathbf{A})\,\mathbf{G}+b\,\mathbf{A}^{\flat}\,.
\end{equation}
This linear map is self-adjoint, and hence the linearized complementary Cauchy elastic constitutive equation about $\mathbf{S}=\mathbf{0}$ is complementary hyperelastic.
The nonlinear constitutive equation, however, is not complementary hyperelastic. For arbitrary symmetric tensors $\mathbf{A}$ and $\mathbf{B}$, the antisymmetric part of the constitutive derivative is written as
\begin{equation}
    \mathbf{B}\!:\!D_{\mathbf{S}}\widehat{\mathbf{C}}^{\flat}[\mathbf{A}]
    -\mathbf{A}\!:\!D_{\mathbf{S}}\widehat{\mathbf{C}}^{\flat}[\mathbf{B}]
    =2\beta(\mathbf{M}\!:\!\mathbf{S})
    \left[
    (\mathbf{M}\!:\!\mathbf{A})(\mathbf{G}\!:\!\mathbf{B})
    -(\mathbf{M}\!:\!\mathbf{B})(\mathbf{G}\!:\!\mathbf{A})
    \right]\,.
\end{equation}
For $\beta\neq0$ and $\mathbf{M}$ not proportional to $\mathbf{G}$, the right-hand side does not vanish identically. Thus, the nonlinear complementary Cauchy elastic constitutive equation is not complementary hyperelastic, even though its linearization about $\mathbf{S}=\mathbf{0}$ is complementary hyperelastic. In this example, the obstruction to complementary hyperelasticity first appears at quadratic order in the constitutive response.
\end{example}

These observations show that hyperelasticity of a linearized constitutive equation is determined by the tangent constitutive space at the state about which the linearization is performed. For general implicit elasticity, the relevant condition is the vanishing of the restriction of $\mathrm{d}\boldsymbol{\Omega}$ to this tangent space. When an explicit stiffness or compliance representation exists, this condition reduces to the familiar major symmetry of the corresponding operator. A nonlinear implicit constitutive relation may therefore fail to be hyperelastic while its linearization at a particular state is hyperelastic. The Cauchy and complementary Cauchy examples demonstrate this phenomenon explicitly: in both cases, the obstruction to integrability vanishes at the state of linearization but becomes apparent at higher order. Thus, hyperelasticity of the tangent constitutive space at a single state does not imply hyperelasticity of the underlying nonlinear constitutive relation in a neighborhood of that state.

\section{Conclusions}
\label{Sec:Conclusions}

We have developed a nonlinear theory of complementary Cauchy elasticity in which the right Cauchy--Green strain is prescribed as a function of the second Piola--Kirchhoff stress, without assuming constitutive invertibility or the existence of a complementary strain-energy function. Complementary Cauchy elasticity is a distinct class of explicit constitutive responses within implicit elasticity, rather than an algebraic reformulation of conventional Cauchy elasticity. The two theories admit locally equivalent descriptions only when their constitutive response functions are mutually inverse. More generally, implicit elasticity provides a common constitutive framework in which neither stress nor strain need be uniquely determined by the other.

The complementary stress-work $1$-form provides the geometric foundation of the theory. Its exterior derivative characterizes the obstruction to complementary hyperelasticity and is determined by the major-antisymmetric part of the tangent compliance tensor. For general implicit elasticity, the stress-work and complementary stress-work $1$-forms are naturally defined on the constitutive manifold, where their exterior derivatives are negatives of one another. Consequently, hyperelasticity and complementary hyperelasticity are equivalent intrinsic properties of this manifold, characterized locally by its Lagrangian structure with respect to the canonical symplectic form on stress-strain space. This characterization remains meaningful even when neither an explicit stress response nor an explicit strain response exists.

The Darboux classification provides local canonical representations of complementary Cauchy elastic solids without requiring complementary hyperelasticity. For compressible anisotropic materials, the six-dimensional stress space admits six nontrivial canonical classes. For isotropic materials, the reduction to invariant stress space yields the complementary Green, complementary Ericksen, and complementary Edelen classes. Explicit isotropic constitutive responses were constructed for the non-hyperelastic Ericksen and Edelen classes, establishing that these classes correspond to genuine constitutive behavior rather than merely formal possibilities. The failure of complementary hyperelasticity in these examples is demonstrated both geometrically, through the complementary stress-work $1$-form, and constitutively, through the lack of major symmetry of the tangent compliance.

The complementary formulation also provides a natural description of stress-constrained elastic materials. Restrictions on the admissible stress states are associated with constitutively indeterminate strain modes, leading to a stress-based characterization of extremal composite materials, including pentamode solids. For materials subject to internal kinematic constraints, a more fundamental structure emerges: the constitutively relevant stress variable is not an individual total stress but an equivalence class of stresses modulo the reactive stress associated with the constraint. The constitutive response is therefore naturally defined on a quotient of stress space. A central consequence of this quotient structure is the distinction between the left and right kernels of the tangent compliance. 
The left-kernel condition expresses satisfaction of the kinematic constraint, whereas the right-kernel condition expresses the constancy of the constitutive strain response along reactive-stress orbits.
These conditions are generally independent in complementary Cauchy elasticity and coincide when the tangent compliance has major symmetry. Their distinction is therefore essential for internally constrained elastic materials that are not complementary hyperelastic. The analysis of incompressibility, fiber inextensibility, and general scalar internal constraints establishes the corresponding constitutive conditions without imposing major symmetry.

The covariant formulation of the balance laws reveals a further distinction between Cauchy and complementary Cauchy elasticity. In the complementary theory, the symmetry of the right Cauchy--Green strain ensures that the stress-work rate depends only on the symmetric part of the second Piola--Kirchhoff stress. Covariance of the energy balance then yields the balance of angular momentum without requiring an additional covariance postulate on the generalized energy functions. This differs from the corresponding derivation in Cauchy elasticity, where covariance of the generalized energy functions is needed to establish stress symmetry. 

Finally, the linearization of general implicit elasticity yields a tangent constitutive relation that need not admit either a stiffness or a compliance representation. Its hyperelasticity is characterized intrinsically by the Lagrangian property of the tangent constitutive space, reducing to major symmetry when an explicit incremental stiffness or compliance operator exists. Importantly, a nonlinear constitutive relation that is not hyperelastic may nevertheless have a hyperelastic linearization at a particular constitutive state. The obstruction to hyperelasticity may vanish at that state and become apparent only at higher order, as demonstrated by explicit Cauchy and complementary Cauchy elastic examples.

These results establish complementary Cauchy elasticity as a general stress-based constitutive theory whose geometric and mechanical structure extends beyond the existence of complementary energy and beyond the invertibility of conventional stress-strain relations. In particular, the Darboux classification, the intrinsic symplectic geometry of implicit elasticity, and the quotient structure associated with reactive stresses provide a unified framework for investigating non-hyperelastic, noninvertible, and constrained elastic responses.

An important direction for future research is a deeper understanding of the mathematical and geometric structure of implicit elasticity. The present work represents a step toward such a theory by treating constitutive relations as geometric objects independently of their explicit representations in terms of stress or strain. Fundamental questions remain concerning the geometry and classification of general implicit constitutive relations, the structure of their singularities and constitutive branches, and the role of material symmetry. Addressing these questions may provide a more comprehensive geometric framework for implicit elasticity and clarify the relationships among its different constitutive classes. Another important direction is the characterization and classification of universal deformations in complementary Cauchy elasticity, and their relationship to the known universal deformations of Cauchy elastic solids. The development of computational methods for complementary Cauchy elasticity is also an important direction. Since strain is prescribed as a function of stress, conventional displacement-based finite element formulations are not directly applicable unless the constitutive relation can be inverted. This motivates the development of mixed stress--deformation formulations, as well as stress-based formulations in which the compatibility of the constitutively prescribed strain must be enforced. More generally, the development of computational methods for implicit elasticity presents a fundamental challenge. Since neither stress nor strain need be uniquely determined by the other, mixed formulations that retain stress and deformation as independent unknowns provide a natural approach. Unlike conventional mixed formulations for internally constrained materials, such formulations may be necessary even for compressible materials without internal constraints. Their mathematical well-posedness and numerical stability remain important questions for future investigation.

\section*{Acknowledgements}

The author is grateful to Amit Acharya for discussions on complementary potential energy in nonlinear elasticity that motivated the initial exploration of stress-based constitutive formulations and ultimately led to the developments presented in this paper.

\bibliographystyle{abbrvnat}
\bibliography{ref}

\appendix

\section{Darboux Classification of Differential Forms and Pfaffian Equations} 
\label{App:DarbouxClassification}

A related review of the Darboux classification and Pfaffian equations was given in \citep{YavariGoriely2025Cauchy} in connection with the stress-work $1$-form of Cauchy elasticity. We summarize the required material here to make the present paper self-contained and to establish the notation used for the complementary stress-work $1$-form. The application is different: the relevant constitutive manifold is now stress space, and the coefficients of the $1$-form are determined by the strain response.

Consider a differential form $\boldsymbol{\alpha}$ defined on an $n$-dimensional manifold $\mathcal{M}$. We use the following notation for its repeated exterior product:\footnote{Notice that although the wedge product of a $1$-form with itself always vanishes, the same is not generally true for a $2$-form. Indeed, interchanging two differential forms of degrees $p$ and $q$ introduces the sign $(-1)^{pq}$. Thus, two $1$-forms anticommute, whereas two $2$-forms commute. Consequently, since $d\boldsymbol{\Lambda}$ is a $2$-form, $(d\boldsymbol{\Lambda})^2$ need not vanish. For example, if $\boldsymbol{\omega}=dx\wedge dy+dz\wedge dw$, then $\boldsymbol{\omega}\wedge\boldsymbol{\omega}=2\,dx\wedge dy\wedge dz\wedge dw\neq 0$.}
\begin{equation}
	(\boldsymbol{\alpha})^{k}=\overbrace{\boldsymbol{\alpha} \wedge \hdots 
	\wedge \boldsymbol{\alpha}}^{k-\text{factors}}\,.
\end{equation}
Let $\Lambda^1(\mathcal{M})$ denote the space of differential $1$-forms on $\mathcal{M}$. 
For $\boldsymbol{\Lambda}\in\Lambda^1(\mathcal{M})$, the nonnegative integer $k$ is called the \textit{rank} of $\boldsymbol{\Lambda}$ when $(d\boldsymbol{\Lambda})^k\neq0$ and $(d\boldsymbol{\Lambda})^{k+1}=0$.
Since $(d\boldsymbol{\Lambda})^k$ is a $2k$-form, its non-vanishing requires $n\geq2k$, or equivalently $n-k\geq k$. The following theorem gives the local canonical forms of a differential $1$-form.

\begin{thm}[Darboux \citep{Darboux1882,Slebodzinski1970,Sternberg1999,Bryant2013,Suhubi2013}]
Suppose that $\boldsymbol{\Lambda}\in\Lambda^1(\mathcal{M})$ has rank $k$. If $\boldsymbol{\Lambda}\wedge(d\boldsymbol{\Lambda})^k=0$ everywhere on $\mathcal{M}$, then, in a neighborhood of each point, there exist coordinates $\{y^1,\hdots,y^{n-k},z^1,\hdots,z^k\}$ in which
\begin{equation}
    \boldsymbol{\Lambda}=y^1dz^1+\hdots+y^k dz^k    \,.
\end{equation}
If, instead, $\boldsymbol{\Lambda}\wedge(d\boldsymbol{\Lambda})^k\neq 0$ everywhere on $\mathcal{M}$, then, in a neighbourhood of each point, there exist coordinates $\{y^1,\hdots,y^{n-k},$ $z^1,\hdots,z^k\}$ in which
\begin{equation}
	\boldsymbol{\Lambda}=y^1dz^1+\hdots+y^kdz^k+dy^{k+1}  \,.
\end{equation}
\end{thm}
\noindent The representations appearing in the theorem are referred to as the \textit{canonical (normal) forms} of $\boldsymbol{\Lambda}$.

These normal forms will be used in complementary Cauchy elasticity. For this purpose, we need the classification of differential $1$-forms on manifolds of dimensions one, two, three, and six. These dimensions correspond, respectively, to the constitutive spaces arising in incompressible and compressible isotropic complementary Cauchy elasticity and in anisotropic complementary Cauchy elasticity. Here, balance of angular momentum has already been imposed, so that the second Piola--Kirchhoff stress tensor is symmetric and the corresponding stress space is six-dimensional.

On a one-dimensional manifold, $d\boldsymbol{\Lambda}=0$, and therefore $k=0$. It follows that there is a scalar field $\zeta$ for which $\boldsymbol{\Lambda}=d\zeta$. On a two-dimensional manifold, the rank can only be zero or one. Moreover, $\boldsymbol{\Lambda}\wedge d\boldsymbol{\Lambda}$ is a $3$-form and thus vanishes identically. The possible normal forms are
\begin{equation} \label{Darboux-2D}
\begin{aligned}
	k=0:& \quad d\boldsymbol{\Lambda}=0\,,~ \boldsymbol{\Lambda}\wedge(d\boldsymbol{\Lambda})^0
	=\boldsymbol{\Lambda}\neq 0 && \Rightarrow ~ \boldsymbol{\Lambda}=d\zeta_1\,, \\
	k=1:& \quad d\boldsymbol{\Lambda}\wedge d\boldsymbol{\Lambda}=0, ~ 
	\boldsymbol{\Lambda}\wedge d\boldsymbol{\Lambda} =0  && \Rightarrow ~ \boldsymbol{\Lambda}	
	=\kappa_1 d\zeta_1\,.
\end{aligned}
\end{equation}
For a three-dimensional manifold, the rank is likewise restricted to zero or one. In this case, however, $\boldsymbol{\Lambda}\wedge d\boldsymbol{\Lambda}$ need not vanish, and the following alternatives are possible:
\begin{equation}
\begin{aligned}
	k=0:& \quad d\boldsymbol{\Lambda}=0\,,~ \boldsymbol{\Lambda}\wedge (d\boldsymbol{\Lambda})^0
	=\boldsymbol{\Lambda}\neq 0 ~ \Rightarrow ~ \boldsymbol{\Lambda}=d\zeta_1\,,\\
	k=1:& \quad d\boldsymbol{\Lambda}\wedge d\boldsymbol{\Lambda}=0, ~ 
	\begin{dcases}
	\boldsymbol{\Lambda}\wedge d\boldsymbol{\Lambda} =0    ~ \Rightarrow ~ \boldsymbol{\Lambda}	
	=\kappa_1 d\zeta_1\,,\\
	\boldsymbol{\Lambda}\wedge d\boldsymbol{\Lambda} \neq 0    ~ \Rightarrow ~ \boldsymbol{\Lambda}	=\kappa_1 d\zeta_1+d\zeta_2 \,.
	\end{dcases}
\end{aligned}
\end{equation}
For a six-dimensional manifold, the rank of a differential $1$-form may be $0,1,2$, or $3$. Notice that $\boldsymbol{\Lambda}\wedge d\boldsymbol{\Lambda}\wedge d\boldsymbol{\Lambda}\wedge d\boldsymbol{\Lambda}$ is a $7$-form and hence vanishes identically on every $6$-manifold. The possible normal forms are therefore
\begin{equation}
\begin{aligned}
	k=0:& \quad d\boldsymbol{\Lambda}=0\,,~ \boldsymbol{\Lambda}\wedge 	
	(d\boldsymbol{\Lambda})^0=\boldsymbol{\Lambda}\neq 0 ~ \Rightarrow ~ \boldsymbol{\Lambda}	
	=d\zeta_1\,, \\
	k=1:& \quad d\boldsymbol{\Lambda}\wedge d\boldsymbol{\Lambda}=0, ~ 
	\begin{dcases}
	\boldsymbol{\Lambda}\wedge d\boldsymbol{\Lambda} =0    ~ \Rightarrow ~ \boldsymbol{\Lambda}	
	=\kappa_1 d\zeta_1\,, \\
	\boldsymbol{\Lambda}\wedge d\boldsymbol{\Lambda} \neq 0    ~ \Rightarrow ~ \boldsymbol{\Lambda}	=\kappa_1 d\zeta_1+d\zeta_2\,,
	\end{dcases}\\
	k=2:& \quad d\boldsymbol{\Lambda}\wedge d\boldsymbol{\Lambda}\wedge d\boldsymbol{\Lambda}=0, ~ 
	\begin{dcases}
	\boldsymbol{\Lambda}\wedge d\boldsymbol{\Lambda}\wedge d\boldsymbol{\Lambda} =0    ~ \Rightarrow 	~ \boldsymbol{\Lambda}=\kappa_1 d\zeta_1+\kappa_2 d\zeta_2\,,\\
	\boldsymbol{\Lambda}\wedge d\boldsymbol{\Lambda}\wedge d\boldsymbol{\Lambda} \neq 0    ~ 	
	\Rightarrow ~ \boldsymbol{\Lambda}=\kappa_1 d\zeta_1+\kappa_2 d\zeta_2+d\zeta_3\,,
	\end{dcases}\\ 
	k=3:& \quad d\boldsymbol{\Lambda} \wedge d\boldsymbol{\Lambda}\wedge d\boldsymbol{\Lambda}	
	\wedge d\boldsymbol{\Lambda}=0, ~ 
	\boldsymbol{\Lambda}\wedge d\boldsymbol{\Lambda}\wedge d\boldsymbol{\Lambda}\wedge 	
	d\boldsymbol{\Lambda} =0    ~ \Rightarrow ~ \boldsymbol{\Lambda}=\kappa_1 d\zeta_1+\kappa_2 	
	d\zeta_2+\kappa_3 d\zeta_3\,.
\end{aligned}
\end{equation}

Given $\boldsymbol{\Lambda}\in\Lambda^1(\mathcal{M})$, consider the first-order partial differential equation 
\begin{equation} \label{Pfaff}
	\boldsymbol{\Lambda}=0\,,
\end{equation}
which is known as a \textit{Pfaffian equation} and is the simplest example of an exterior differential system. A $p$-dimensional integral manifold of \eqref{Pfaff} is an immersion $f:\mathcal{W}\to\mathcal{M}$, not necessarily an embedding, such that $f^*\boldsymbol{\Lambda}=0$, where $f^*$ is the pullback by $f$. An integral curve of $\boldsymbol{\Lambda}$ is a curve $c:I\to\mathcal{M}$, with $I$ an interval of the real line, satisfying $\boldsymbol{\Lambda}(\dot{c}(t))=0$ for every $t\in I$.

The Pfaffian equation $\boldsymbol{\Lambda}=0$ is said to have the \emph{local accessibility} property if, for every $x\in\mathcal{M}$, there is a neighborhood $U$ of $x$ such that each $y\in U$ can be connected to $x$ by an integral curve of \eqref{Pfaff}. Conversely, \eqref{Pfaff} has the \emph{local inaccessibility} property if every neighborhood $U$ of every $x\in\mathcal{M}$ contains at least one point $y\in U$ that cannot be connected to $x$ by an integral curve of \eqref{Pfaff}.

The rank of the Pfaffian equation \eqref{Pfaff} is the integer $r$ determined by $\boldsymbol{\Lambda}\wedge(d\boldsymbol{\Lambda})^r\neq0$ and $\boldsymbol{\Lambda}\wedge(d\boldsymbol{\Lambda})^{r+1}=0$.
According to Carath\'eodory's theorem, if this rank is constant, the Pfaffian equation has the local accessibility property if and only if $r\geq2$. Hence, local inaccessibility can occur only for $r=0$ or $r=1$. These two cases have the normal forms $\boldsymbol{\Lambda}=d\zeta_1$ and $\boldsymbol{\Lambda}=\kappa_1d\zeta_1$, respectively. In either case, the equation $\boldsymbol{\Lambda}=0$ yields $\zeta_1=\text{constant}$. These level sets are hypersurfaces and constitute integral manifolds of the largest possible dimension, namely $n-1$.

\end{document}